\documentclass[runningheads]{llncs}

\PassOptionsToPackage{pagebackref=true}{hyperref}

\usepackage[mobile]{eccv}

\usepackage{eccvabbrv}

\usepackage{graphicx,verbatim}
\usepackage{booktabs}
\usepackage{lmodern}     % scalable Type1 fonts (enables microtype font expansion on any TeX system)
\usepackage{microtype}   % camera-ready micro-typography: keeps every line within the column width
\usepackage{multirow}
\usepackage{amsmath}
\usepackage{amssymb}
\usepackage{tikz}
\usepackage{array}
\usetikzlibrary{arrows.meta,positioning,backgrounds,fit,calc,shapes.geometric,shapes.multipart}
\usepackage{xcolor}
\definecolor{flblue}{RGB}{44,123,182}
\definecolor{fllightblue}{RGB}{214,234,248}
\definecolor{kdred}{RGB}{215,25,28}
\definecolor{kdlightred}{RGB}{253,232,230}
\definecolor{datagreen}{RGB}{26,150,65}
\definecolor{datalightgreen}{RGB}{212,239,215}
\definecolor{modelorange}{RGB}{230,159,0}
\definecolor{modellightorange}{RGB}{253,238,210}
\definecolor{arrowgrey}{RGB}{80,80,80}

\usepackage[accsupp]{axessibility}  % Improves PDF readability for those with disabilities.
\usepackage{subcaption}
\usepackage{longtable}
\newcommand{\sd}[1]{{\scriptsize$\,\pm#1$}}

\usepackage{etoolbox}
\AtBeginEnvironment{thebibliography}{\interlinepenalty=10000}

\usepackage{float}
\newfloat{algorithm}{tbp}{loa}
\floatname{algorithm}{Algorithm}
\newcounter{algln}
\newenvironment{algsteps}{%
  \par\footnotesize
  \begin{list}{\textup{\arabic{algln}:}}{%
    \usecounter{algln}%
    \setlength{\leftmargin}{2.0em}\setlength{\labelwidth}{1.4em}%
    \setlength{\labelsep}{0.55em}\setlength{\itemsep}{1.2pt}%
    \setlength{\parsep}{0pt}\setlength{\topsep}{2pt}}%
}{\end{list}}
\newcommand{\algkw}[1]{\textbf{#1}}        % pseudocode keyword
\newcommand{\algind}{\hspace*{1.2em}}       % one indentation level
\newcommand{\algcom}[1]{\hfill{\footnotesize\itshape$\triangleright$\,#1}}  % trailing comment

\usepackage{orcidlink}

\makeatletter
\begingroup\catcode`\&=12\relax\gdef\flkdamp{&}\endgroup
\@ifpackageloaded{hyperref}{%
  \hypersetup{%
    colorlinks=true, % Enables colored links instead of boxes
    linkcolor=red,   % Colors the page back-references red
    citecolor=green, % Colors the citation numbers green (optional, adjust to preference)
    urlcolor=blue,   % Colors URLs blue (optional)
    pdftitle={Benchmarking Federated Learning \& Knowledge Distillation for Point Cloud Classification},%
    pdfauthor={Aizierjiang Aiersilan},%
    pdfsubject={A multi-seed benchmark that jointly evaluates federated learning and knowledge distillation for 3D point cloud classification across thirteen federated algorithms, ten distillation objectives, and their full 130-pair cross-product, on ModelNet40 and a real-world clinical craniosynostosis dataset. It quantifies how federated learning degrades under extreme non-IID label skew, how distillation compresses the teacher into a student about 74 percent smaller and roughly twice as fast, and how a hard-label cross-entropy term in the combined pipeline masks federated failure by reusing the private labels that federation was meant to protect. The paper recommends evaluating federated-distillation pipelines with label-free distillation so that the reported accuracy reflects the federated teacher.},%
    pdfkeywords={Federated learning, Knowledge distillation, 3D point cloud, Model compression, Non-IID, Privacy-preserving learning, Medical imaging, Benchmark}%
  }%
}{%
  \pdfinfo{%
    /Title (Benchmarking Federated Learning \flkdamp\ Knowledge Distillation for Point Cloud Classification)%
    /Author (Aizierjiang Aiersilan)%
    /Subject (A multi-seed benchmark that jointly evaluates federated learning and knowledge distillation for 3D point cloud classification across thirteen federated algorithms, ten distillation objectives, and their full 130-pair cross-product, on ModelNet40 and a real-world clinical craniosynostosis dataset. It quantifies how federated learning degrades under extreme non-IID label skew, how distillation compresses the teacher into a student about 74 percent smaller and roughly twice as fast, and how a hard-label cross-entropy term in the combined pipeline masks federated failure by reusing the private labels that federation was meant to protect. The paper recommends evaluating federated-distillation pipelines with label-free distillation so that the reported accuracy reflects the federated teacher.)%
    /Keywords (Federated learning, Knowledge distillation, 3D point cloud, Model compression, Non-IID, Privacy-preserving learning, Medical imaging, Benchmark)%
    /Creator (LaTeX with pdfTeX)%
  }%
}%
\makeatother

\begin{document}

% ---------------------------------------------------------------
\title{Benchmarking Federated Learning \& Knowledge Distillation for Point Cloud Classification}

\titlerunning{FL and KD for Point Cloud Classification}

% Author list for review version
\author{Aizierjiang Aiersilan}

% Abbreviated list of authors.
\authorrunning{Aizierjiang Aiersilan}

% Institution list.
\institute{University of Macau \\
    \email{ezharjan@outlook.com}}

\maketitle

\begin{abstract}
Deploying 3D point cloud analysis in privacy-sensitive and resource-constrained settings faces two coupled barriers: data cannot be centralized for training, and the trained model must run on limited edge hardware. We present a multi-seed benchmark that jointly evaluates federated learning (FL) and knowledge distillation (KD) for 3D point cloud classification. It spans thirteen FL algorithms and ten KD objectives, supporting their full $130$-pair teacher--objective cross-product per dataset; every standardized configuration is repeated over three random seeds for $504$ training runs in total, with the complete combined grid evaluated at multi-seed scale on the clinical dataset. We characterize federated degradation and the combined-pipeline pitfall on ModelNet40, then validate them on a real-world clinical craniosynostosis dataset of patient head shapes, where the privacy and edge-deployment stakes are concrete. We report three findings. First, under extreme non-independent and identically distributed (non-IID) label skew, standalone FL degrades sharply: on ModelNet40 the strongest method reaches only $76.32\%$ against a $92.26\%$ centralized reference, on the clinical data the best reaches $75.83\%$ against $100\%$, and the four server-side optimizers collapse to near the chance level; the best algorithm differs by dataset, so none is universally robust. Second, distillation compresses the teacher into a student $74.51\%$ smaller and roughly twice as fast at inference, with five of the seven objectives evaluated on ModelNet40 matching or surpassing the $92.44\%$ teacher. Third, the combined pipeline exposes an evaluation pitfall: when distillation keeps a hard-label cross-entropy term on a labeled proxy split, a collapsed federated teacher at $8.50\%$ paired with Logit-MSE still yields a $92.94\%$ student. This $84.4$-point gap reflects the proxy labels rather than the federated model, and the hard-label term reuses the very labels whose privacy motivated federation. Objectives without a hard-label term instead track teacher quality ($r\approx0.99$ on the clinical grid) and collapse when the teacher does. We therefore recommend evaluating FL-KD pipelines with label-free distillation, so that the reported accuracy reflects the federated teacher rather than the proxy.
\end{abstract}

\smallskip
\noindent\textbf{Benchmark \& Code:} \url{https://ezharjan.github.io/FLKD3DBenchmark}

\section{Introduction}
\label{sec:intro}

3D point cloud analysis underpins applications including autonomous driving, robotics, augmented reality, and medical imaging, where sensor data are naturally represented as irregular point sets in Euclidean space. Architectures such as PointNet~\cite{qi2017pointnet} and PointNet++~\cite{pointnet2} achieve strong classification performance on standard benchmarks, yet their deployment in privacy-sensitive, resource-constrained settings faces two challenges that have not been studied together in one benchmark.

The first challenge is data privacy. In medical~\cite{rieke2020future,sheller2020federated,liu2022privacy}, industrial~\cite{nguyen2021federated,lu2019blockchain,hao2019efficient}, and multi-institutional settings~\cite{yang2019federated,li2020federateddirections,long2020federated}, raw sensor data cannot be centralized because of regulatory or competitive constraints. Federated learning (FL)~\cite{mcmahan2017communication} addresses this by coordinating model training across clients without exchanging raw data, using iterative local updates and global aggregation. Even so, gradient inversion attacks~\cite{zhu2019deep,geiping2020inverting} show that transmitted gradients can still leak private data, and the broader open problems in FL are surveyed in~\cite{kairouz2021advances}. The statistical heterogeneity of data across clients, the non-IID problem~\cite{hsu2019measuring,lu2024federated,jimenez2024non}, degrades convergence and final accuracy even for well-established aggregation algorithms. These effects are well documented for image classification, but their impact on hierarchical 3D feature extraction remains largely uncharacterized.

The second challenge is computational efficiency at deployment. A full PointNet++ model is often unsuitable for resource-constrained edge platforms such as embedded robotic controllers or portable clinical scanners. Knowledge distillation (KD)~\cite{hinton2015distilling} transfers the capacity of a large teacher into a compact student, producing a smaller and faster network that retains much of the teacher's accuracy. Several distillation objectives exist for image recognition, including feature hints~\cite{romero2014fitnets}, attention transfer~\cite{zagoruyko2016paying}, and self-distillation~\cite{furlanello2018born}, but their effectiveness for 3D point cloud models has not been systematically benchmarked.

Combining FL and KD is appealing: FL trains a shared model without centralizing data, and KD then compresses that model for edge deployment. A recent survey~\cite{qin2025knowledge} identifies the absence of domain-specific benchmarks as a critical gap. In practice, the two-stage pipeline is usually evaluated end to end, by reporting the compressed student's accuracy, rather than by asking whether that accuracy actually reflects the federated teacher. To the best of our knowledge, no prior work provides a systematic, reproducible evaluation of FL and KD strategies for 3D point cloud classification across the full cross-product of federated teachers and distillation objectives, let alone on clinical 3D shape data; 3DFFL~\cite{khan20243dffl} addresses federated few-shot learning for point clouds but does not evaluate KD losses or the combined pipeline.

We study two complementary datasets. On ModelNet40~\cite{wu20153d}, a $40$-class benchmark, we characterize federated degradation and isolate the combined-pipeline pitfall against a strong centralized reference. We then validate that pitfall on a real-world clinical craniosynostosis dataset~\cite{schaufelberger2021statistical} of patient head shapes, where privacy mandates federation~\cite{rieke2020future,sheller2020federated} and edge compression is a deployment necessity; there the full multi-seed cross-product confirms that hard-label masking conceals federated failure on a privacy-critical task.
We make four contributions:
\begin{itemize}
\item \textbf{Benchmark design.} The benchmark provides thirteen FL algorithms, a centralized baseline, and ten KD objectives, and supports the complete two-stage cross-product of all $130$ teacher--objective pairs per dataset. Our standardized multi-seed evaluation comprises $504$ training runs over three seeds: on ModelNet40 the centralized baseline and the thirteen FL teachers, and on the clinical dataset the centralized baseline, the thirteen FL teachers, the ten KD objectives, and the full $130$-pair combined grid; the ModelNet40 distillation and combined results come from a separate focused round (\Cref{sec:kd_results}) selected for its wide spread of teacher quality. All results are reported as mean and standard deviation. The framework is open-source and is built to be extended with further FL algorithms, KD objectives, and datasets through one shared interface.
\item \textbf{Quantified degradation under label skew.} On ModelNet40 the strongest FL method trails the centralized reference by about $16$ points (FedNova at $76.32\%$ against $92.26\%$), on the clinical data by about $24$ points (FedProx at $75.83\%$ against $100\%$), and the four server-side optimizers (FedAvgM, FedAdam, FedYogi, FedAdagrad) collapse to near the chance level on both. The method ranking is dataset-dependent, so no single algorithm is robust to label skew.
\item \textbf{Effective compression through distillation.} On ModelNet40 distillation transfers the teacher's accuracy into a student that is $74.51\%$ smaller and roughly twice as fast at inference (about $50\%$ lower latency), with five of seven objectives matching or surpassing the $92.44\%$ teacher; the clinical task exhibits the same near-ceiling compression.
\item \textbf{A controlled diagnosis of the combined pipeline.} The teacher-by-objective grid separates the distillation objectives by their reliance on hard labels. Objectives that keep a hard-label cross-entropy term recover the student to near-centralized accuracy regardless of teacher quality, so that on ModelNet40 a collapsed federated teacher at $8.50\%$ paired with Logit-MSE still yields a $92.94\%$ student, an $84.4$-point gap driven by the proxy labels, not by transfer. The feature- and attention-based objectives whose cross-entropy weight is exactly zero instead track the teacher and collapse when it collapses; the full multi-seed cross-product on the clinical data confirms the same split across all thirteen teachers, with the four pure-transfer objectives at $r\approx0.99$.
Hard-label distillation thus masks federated failure while reusing exactly the labels that federation was meant to keep private, and we accordingly recommend two controls (distilling with the cross-entropy term removed, or on an unlabeled proxy split) that distinguish genuine knowledge transfer from this artifact.
\end{itemize}

% =====================================================================
\section{Related Work}
\label{sec:related}

\subsection{3D Point Cloud Learning}

Early volumetric methods discretized shapes onto occupancy grids and applied 3D convolutions~\cite{wu20153d}, but their memory cost grows cubically with resolution. PointNet~\cite{qi2017pointnet} introduced a permutation-invariant architecture that processes raw point sets via shared multi-layer perceptrons and global max pooling. PointNet++~\cite{pointnet2} extended this with hierarchical set abstraction layers that capture local geometry at progressively coarser scales through farthest point sampling and ball-query grouping. Later architectures incorporate graph-based aggregation as in DGCNN~\cite{wang2019dynamic} and self-attention as in Point Transformer~\cite{zhao2021point}. Further advances include kernel-point convolutions (KPConv)~\cite{thomas2019kpconv}, random-sampling segmentation with RandLA-Net~\cite{hu2020randla}, the Point Cloud Transformer (PCT)~\cite{guo2021pct}, PointNeXt~\cite{qian2022pointnext}, and Point Transformer V2~\cite{wu2022pointtransv2}. Sarker \etal~\cite{sarker2024comprehensive} survey deep learning for point cloud understanding. We use PointNet++ single-scale grouping (SSG) as the teacher backbone because its hierarchical, locality-sensitive feature extraction provides a demanding setting for federated aggregation under non-IID label skew.

\subsection{Federated Learning}

We benchmark thirteen FL algorithms spanning the major families of aggregation strategy. FedAvg~\cite{mcmahan2017communication} averages client models weighted by dataset size after local gradient descent. To counter client drift under heterogeneity, FedProx~\cite{li2020federated} adds a proximal term, SCAFFOLD~\cite{karimireddy2020scaffold} introduces client-level control variates that reduce gradient variance, FedNova~\cite{wang2020tackling} normalizes the accumulated local updates to remove the objective inconsistency caused by differing local-step counts, and FedDyn~\cite{acar2021federated} adds a dynamic per-client regularizer that aligns local and global stationary points. A second family modifies the server-side update: FedAvgM applies server momentum~\cite{hsu2019measuring}, while FedAdam, FedYogi, and FedAdagrad apply adaptive server optimizers~\cite{reddi2020adaptive}. FedMedian~\cite{yin2018byzantine} replaces averaging with coordinate-wise median aggregation for robustness, and FedBN~\cite{li2021fedbn} keeps batch-normalization statistics client-local to mitigate feature shift. Two methods tailored to non-IID data are MOON~\cite{li2021model}, which adds a model-contrastive term aligning client and global representations, and Ditto~\cite{li2021ditto}, which learns a personalized model per client alongside the global one. The effects of non-IID partitions on several of these strategies are documented for image classification~\cite{hsu2019measuring,lu2024federated,jimenez2024non}; we characterize them for hierarchical 3D point cloud architectures, including the methods most often recommended for label skew.

\subsection{Knowledge Distillation}

We benchmark ten KD objectives spanning the major families of distillation loss. Hinton \etal~\cite{hinton2015distilling} showed that a student can match the soft class-probability outputs of a larger teacher via temperature-scaled Kullback--Leibler divergence. FitNets~\cite{romero2014fitnets} aligned intermediate layer representations through hint-based training. Zagoruyko and Komodakis~\cite{zagoruyko2016paying} transferred attention maps derived from intermediate feature tensors. Born-Again Networks~\cite{furlanello2018born} showed that self-distillation can yield a student that surpasses its teacher. Our benchmark also evaluates four representative relational and decoupled objectives: contrastive representation distillation (CRD)~\cite{tian2019contrastive}, relational knowledge distillation (RKD)~\cite{park2019relational}, similarity-preserving distillation (SP)~\cite{tung2019similarity}, and decoupled knowledge distillation (DKD)~\cite{zhao2022decoupled}, alongside simpler logit-matching variants~\cite{ba2014deep,luo2018cosine}. Other objectives such as the comprehensive feature overhaul~\cite{heo2019comprehensive} and the label-smoothing reinterpretation of KD~\cite{yuan2020revisiting} are not evaluated here. A general survey is given by Gou \etal~\cite{gou2021knowledge}. For 3D data, structured distillation for detection~\cite{zhang2023pointdistiller} and adversarial distillation~\cite{sanjay2025adversarial} have been explored, but neither benchmarks multiple loss objectives nor connects KD with federated pre-training. Qin \etal~\cite{qin2025knowledge} survey the combination of FL and KD and identify the absence of domain-specific evaluations as a key open problem. KD has been integrated into FL to mitigate non-IID degradation and catastrophic forgetting in clinical settings~\cite{kim2024federated,yu2022cyclic}, yet systematic benchmarking for 3D hierarchical models remains unexplored. A complementary line distills on unlabeled or server-side proxy data~\cite{lin2020ensemble}, the approach our diagnosis ultimately motivates.

\subsection{Clinical Shape Analysis}

Craniosynostosis is the premature fusion of cranial sutures, which alters head shape. Schaufelberger \etal~\cite{schaufelberger2021statistical} released a statistical shape model of patients, with instances for the principal suture-fusion types and a control group, giving a privacy-conscious source of patient-like head shapes. Because such scans are distributed across clinics and cannot be pooled freely~\cite{rieke2020future,sheller2020federated,liu2022privacy}, the setting motivates federated training and compact models for point-of-care devices.

% =====================================================================
\section{Benchmark Design}
\label{sec:benchmark}

\subsection{Datasets and Non-IID Partitioning}

We evaluate on two complementary datasets. ModelNet40~\cite{wu20153d} provides point clouds across $40$ shape categories; the standard pre-processed normal-resampled split provides $9{,}843$ training and $2{,}468$ test shapes. It serves as a widely used benchmark on which to measure federated degradation at scale. The craniosynostosis dataset~\cite{schaufelberger2021statistical} provides 3D head-surface instances sampled from a statistical shape model, with four classes (a healthy control group and the coronal, metopic, and sagittal suture-fusion pathologies) and $100$ instances per class; we use a deterministic stratified $80/20$ split, giving $320$ training and $80$ test shapes. Each shape is represented by $1{,}024$ points with three spatial coordinates and no surface normals, an input tensor of shape $3 \times 1{,}024$. The two datasets are deliberately contrasting: a large-vocabulary collection where the centralized task is moderately hard, and a small four-class clinical collection where the centralized task is essentially solved. We report the primary empirical results on ModelNet40, the more demanding benchmark, and use the clinical collection as a corroborating study; its near-perfect centralized ceiling lets us read off teacher quality unambiguously.
Beyond these two, the data loader is dataset-agnostic and currently supports ModelNet10~\cite{wu20153d}, OmniObject3D~\cite{wu2023omniobject3d}, YCB~\cite{calli2015ycb}, and GazeboSim~\cite{downs2022google} under the same interface, so the protocol can be transferred to other point-cloud sources.

To model federated deployment, the training set of each dataset is partitioned among five clients with a label-skew non-IID strategy: all training samples are sorted by class label and the ordered sequence is divided into five consecutive equal slices. Each client therefore concentrates on a disjoint segment of the label space, reflecting the class imbalance that arises when different facilities collect data for different purposes. This is a deliberately severe, worst-case form of heterogeneity; we adopt it because it most directly isolates the failure modes the benchmark is designed to surface, and we return to its implications in \Cref{sec:discussion}.

\smallskip\noindent\textbf{Multi-seed protocol.}
To make the comparison statistically interpretable rather than dependent on a single run, every configuration (baseline, FL, KD, and combined) is trained independently with three random seeds, $\{7, 42, 123\}$. These seeds control the model initialization and the stochastic elements of training, namely mini-batch ordering and point-cloud augmentation; the label-skew partition is deterministic and held fixed across seeds, so the reported variance isolates training-time randomness rather than partition luck. Unless stated otherwise, all reported numbers are the mean over the three seeds, and tables give the corresponding standard deviation. 
% In total the benchmark comprises $504$ training runs.

\subsection{Model Architectures}
\label{sec:architectures}

\noindent\textbf{Teacher: PointNet++ SSG.}
The teacher follows the PointNet++ single-scale grouping architecture~\cite{pointnet2} on raw coordinates without surface normals, with three hierarchical set-abstraction levels followed by a two-layer classification head.
It occupies $5.65$\,MB on ModelNet40 and $5.62$\,MB on the four-class clinical data, measured from its parameter and buffer tensors.

\smallskip\noindent\textbf{Student: a compact PointNet++ SSG.}
The student, SmallPointNet2, keeps the teacher's hierarchical structure with narrower set-abstraction and classification layers.
It occupies $1.44$\,MB on ModelNet40 (a $74.51\%$ reduction from the teacher) and $1.42$\,MB on the clinical data, and exposes a penultimate feature so that feature-based distillation has a signal.

\subsection{Federated Learning Strategies}
\label{sec:fl_strategies}

We evaluate thirteen FL aggregation algorithms on the teacher architecture, chosen to span the major mechanism families: drift correction, server-side optimization, robust aggregation, and personalization. All configurations share the same federated setting of five clients, five local epochs per round, and twenty communication rounds, with local training using the Adam optimizer at learning rate $0.001$ and batch size $24$. Each strategy is listed with its distinguishing mechanism alongside its accuracy in \Cref{tab:fl_results}.

\subsection{Knowledge Distillation Objectives}
\label{sec:kd_strategies}

We evaluate ten distillation objectives, all transferring knowledge from the PointNet++ SSG teacher to the compact student, spanning the major families of KD loss. Logit matching is represented by Vanilla KD~\cite{hinton2015distilling}, Logit-MSE~\cite{ba2014deep}, Cosine~\cite{luo2018cosine}, and DKD~\cite{zhao2022decoupled}; intermediate-feature and attention alignment by Feature KD~\cite{romero2014fitnets}, Attention Transfer~\cite{zagoruyko2016paying}, and SP~\cite{tung2019similarity}; relational and contrastive transfer by RKD~\cite{park2019relational} and CRD~\cite{tian2019contrastive}; and self-distillation by the Born-Again protocol~\cite{furlanello2018born}. All but DKD share a soft-target weight $\alpha{=}0.5$, and the feature, attention, SP, and RKD objectives add an intermediate-representation weight $\beta{=}0.5$; the soft-target objectives use temperature $T{=}2.0$. For Self-Distillation, an initial compact student trained from the teacher then supervises the final reported student.

Central to our combined-pipeline diagnosis is the weight on the hard-label cross-entropy term, $\mathcal{L}_{\text{CE}}$. Vanilla KD, Logit-MSE, Cosine, CRD, and Self-Distillation weight it by $(1-\alpha){=}0.5$, and DKD adds it at full weight, so all six retain a direct supervised signal from the ground-truth labels. The feature, attention, SP, and RKD objectives weight it by $(1-\alpha-\beta)$, which is exactly $0$ at $\alpha{=}\beta{=}0.5$, so these four are pure teacher-transfer objectives with no hard-label term.

\subsection{Combined Pipeline}
\label{sec:combined}

\begin{figure}[tb]
    \centering
    \includegraphics[width=\textwidth]{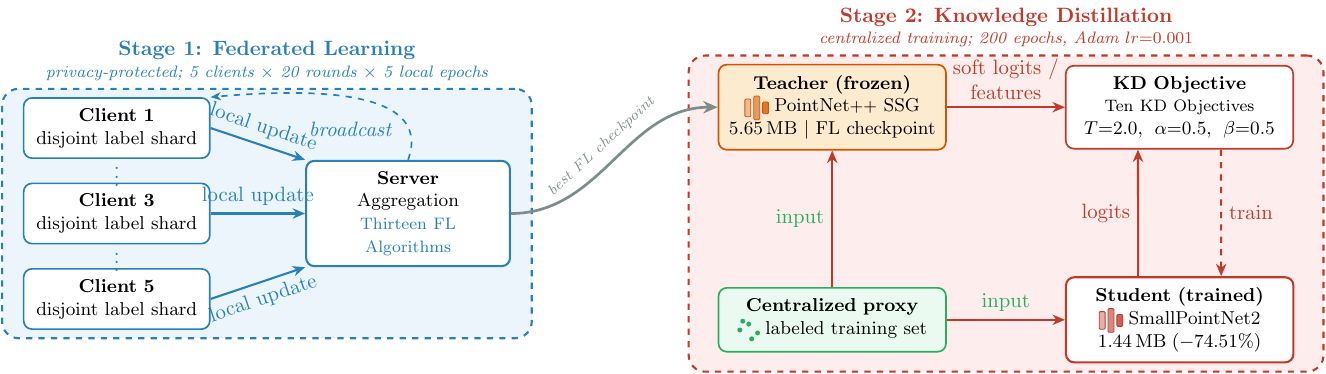}
    \caption{Overview of the two-stage FL-KD benchmark pipeline.
\textbf{Stage~1} (blue, privacy-protected): Five non-IID clients train on
label-skewed slices of the dataset; the server applies one of the thirteen benchmarked FL algorithms to produce a shared PointNet++ SSG teacher.
\textbf{Stage~2} (red): The best FL teacher checkpoint supervises a
compact student on the centralized training set with a KD objective, yielding a smaller and faster student model.}
    \label{fig:pipeline}
\end{figure}

The two-stage combined pipeline first trains a federated teacher (PointNet++ SSG) for twenty rounds under the non-IID partition; the best checkpoint then supervises the compact student in a KD run on the centralized training set.
\Cref{fig:pipeline} illustrates the workflow.
This setup mirrors deployments where a periodically aggregated federated model is compressed for edge hardware using a server-side proxy dataset. We run Stage~2 on the fully labeled centralized data precisely to test the pitfall it can create, so that the spread of teacher quality and the role of the hard-label term can both be measured directly. The pipeline is dataset-agnostic; on ModelNet40 we pair federated teachers spanning the full quality range with the distillation objectives. The same two stages apply unchanged to the clinical dataset, where we run the complete multi-seed cross-product of all thirteen FL teachers with all ten KD objectives ($130$ teacher--objective pairs) and reproduce the same recovery illusion.

% =====================================================================
\section{Experiments}
\label{sec:experiments}

\subsection{Centralized Baseline}

We train the teacher on the full training split of each dataset over three seeds, using the common $200$-epoch training protocol.
The teacher reaches $\mathbf{92.26\pm0.04\%}$ instance accuracy on ModelNet40 ($89.58\pm0.10\%$ mean class accuracy) and $\mathbf{100.00\pm0.00\%}$ on the four-class clinical data, using the best checkpoint per run averaged over seeds. The clinical task is essentially solved when trained centrally, which is useful for our purposes: it fixes an unambiguous reference for teacher quality against which the FL degradation and the combined-pipeline behavior can be read. We compare all FL and KD results against these centralized references.

\subsection{Federated Learning Results}

\Cref{tab:fl_results} reports the best instance accuracy of each FL strategy over twenty communication rounds under the non-IID partition, for both datasets.
Per-round convergence trajectories appear in the supplementary material (Appendix~C, Figure~A7).

\begin{table}[tb]
\caption{Standalone FL instance accuracy (\%, best per run, mean$\pm$std over three seeds) on both datasets, with each method's distinguishing mechanism (5 clients, non-IID label-skew, 20 rounds, 5 local epochs per round). Methods are ordered by ModelNet40 accuracy. The ranking is dataset-dependent: FedNova leads on ModelNet40 but drops to mid-pack on the clinical data, where FedProx leads, while the four server-side optimizers fail on both. On the balanced clinical test set the mean-class accuracy equals the instance accuracy.}
\label{tab:fl_results}
\centering
\small
\setlength{\tabcolsep}{4pt}
\begin{tabular}{@{} l >{\raggedright\arraybackslash}p{4.3cm} cc @{}}
\toprule
Method & Distinguishing Mechanism & ModelNet40 $\uparrow$ & Craniosynostosis $\uparrow$ \\
\midrule
Centralized & Non-federated reference & $92.26$\sd{0.04} & $100.00$\sd{0.00} \\
\midrule
FedNova~\cite{wang2020tackling}         & Normalized local averaging & $\mathbf{76.32}$\sd{1.55} & $50.83$\sd{8.04} \\
FedProx~\cite{li2020federated}          & Proximal regularization & $71.04$\sd{0.87} & $\mathbf{75.83}$\sd{10.63} \\
FedDyn~\cite{acar2021federated}         & Dynamic regularization & $66.00$\sd{1.28} & $58.33$\sd{8.13} \\
SCAFFOLD~\cite{karimireddy2020scaffold} & Control-variate correction & $64.26$\sd{3.75} & $36.67$\sd{1.44} \\
Ditto~\cite{li2021ditto}                & Personalized plus global model & $61.36$\sd{4.11} & $71.25$\sd{9.92} \\
FedAvg~\cite{mcmahan2017communication}  & Weighted parameter averaging & $58.51$\sd{0.25} & $69.58$\sd{2.60} \\
FedBN~\cite{li2021fedbn}                & Client-local batch-norm & $55.28$\sd{2.37} & $65.42$\sd{4.73} \\
MOON~\cite{li2021model}                 & Model-contrastive term & $45.84$\sd{7.51} & $69.17$\sd{8.78} \\
FedMedian~\cite{yin2018byzantine}       & Coordinate-wise median & $39.13$\sd{5.79} & $64.58$\sd{7.53} \\
FedAdagrad~\cite{reddi2020adaptive}     & Server-side Adagrad & $\phantom{0}6.28$\sd{1.09} & $26.25$\sd{2.17} \\
FedAdam~\cite{reddi2020adaptive}        & Server-side Adam & $\phantom{0}4.71$\sd{0.98} & $25.00$\sd{0.00} \\
FedAvgM~\cite{hsu2019measuring}         & Server-side momentum & $\phantom{0}4.05$\sd{0.00} & $25.00$\sd{0.00} \\
FedYogi~\cite{reddi2020adaptive}        & Server-side Yogi & $\phantom{0}4.05$\sd{0.00} & $25.00$\sd{0.00} \\
\bottomrule
\end{tabular}
\end{table}

Federated training degrades severely under the label-skew partition on both datasets (\Cref{tab:fl_results}). On ModelNet40 the strongest method, FedNova, reaches only $76.32\%$, $15.94$ points below centralized; the drift-correction methods (FedProx $71.04\%$, FedDyn $66.00\%$, SCAFFOLD $64.26\%$) improve on plain FedAvg ($58.51\%$), whereas MOON ($45.84\%$) falls below it and Ditto ($61.36\%$) exceeds it only modestly. FedMedian reaches $39.13\%$, and the four server-side optimizers fall to near-trivial accuracy.

The ranking is dataset-dependent, itself a finding: no single algorithm is robust. FedNova, the ModelNet40 leader, drops to mid-pack on the clinical data ($50.83\%$), where FedProx leads at $75.83\%$ and is the most consistent across both; SCAFFOLD falls from strong to $36.67\%$ on the four-class problem, while Ditto and MOON rise into the upper group. The key result is constant: the best method trails centralized by $15.94$ points on ModelNet40 and $24.17$ on the clinical data, and the four server-side optimizers collapse to near-chance accuracy on both ($4.05$--$6.28\%$ near $2.5\%$ ModelNet40 floor, $25.00$--$26.25\%$ near the four-class floor).

These results point to a vulnerability of this hierarchical 3D extractor to label skew. When each client sees only a few categories, the early set-abstraction layers likely calibrate their geometric priors to locally dominant shapes and transfer poorly after aggregation. The server-side optimizers are the most fragile: for the three adaptive variants, even at a tuned server rate of $0.05$ the averaged client updates form an unreliable pseudo-gradient whose second-moment estimates become ill-conditioned, likely amplifying rather than correcting divergence~\cite{reddi2020adaptive}. Adaptive server aggregation is thus a practical risk in label-skewed edge settings, where validation data for tuning is rarely available. This is a worst-case regime; softer partitions would narrow the gap (\Cref{sec:discussion}).

Communication is identical across strategies: the full model occupies $5.65$\,MB per transmission, and each communication round sends and receives across five clients, totaling $56.5$\,MB per round, or approximately $1.13$\,GB over twenty rounds. Wall-clock cost varies with the local objective: MOON and Ditto roughly double FedAvg's per-round time by training an auxiliary model on every client. Full payloads and times are reported in the supplementary material (Appendix~C, Table~A6).

\subsection{Knowledge Distillation Results}
\label{sec:kd_results}

We report the ModelNet40 distillation and combined stages on our primary testbed. The combined stage pairs federated teachers spanning the full quality range, from strong aggregators down to collapsed teachers, with seven distillation objectives (including a Basic~KD variant that keeps the cross-entropy term at full weight as a probe); this spread is what makes the recovery illusion measurable. These teachers are individual checkpoints, so their standalone accuracies differ from the multi-seed means in \Cref{tab:fl_results}: SCAFFOLD and FedDyn sit at $8.50\%$ and $16.67\%$ here against multi-seed means of $64.26\%$ and $66.00\%$, and the centralized reference reaches $92.44\%$, close to the $92.26\%$ multi-seed mean. The benchmark supports the full $13\times10$ cross-product, which \Cref{sec:realworld} reports at full multi-seed scale on the clinical task.

\begin{figure}[tb]
\centering
\begin{subfigure}[b]{0.52\linewidth}
\centering
\includegraphics[width=\linewidth]{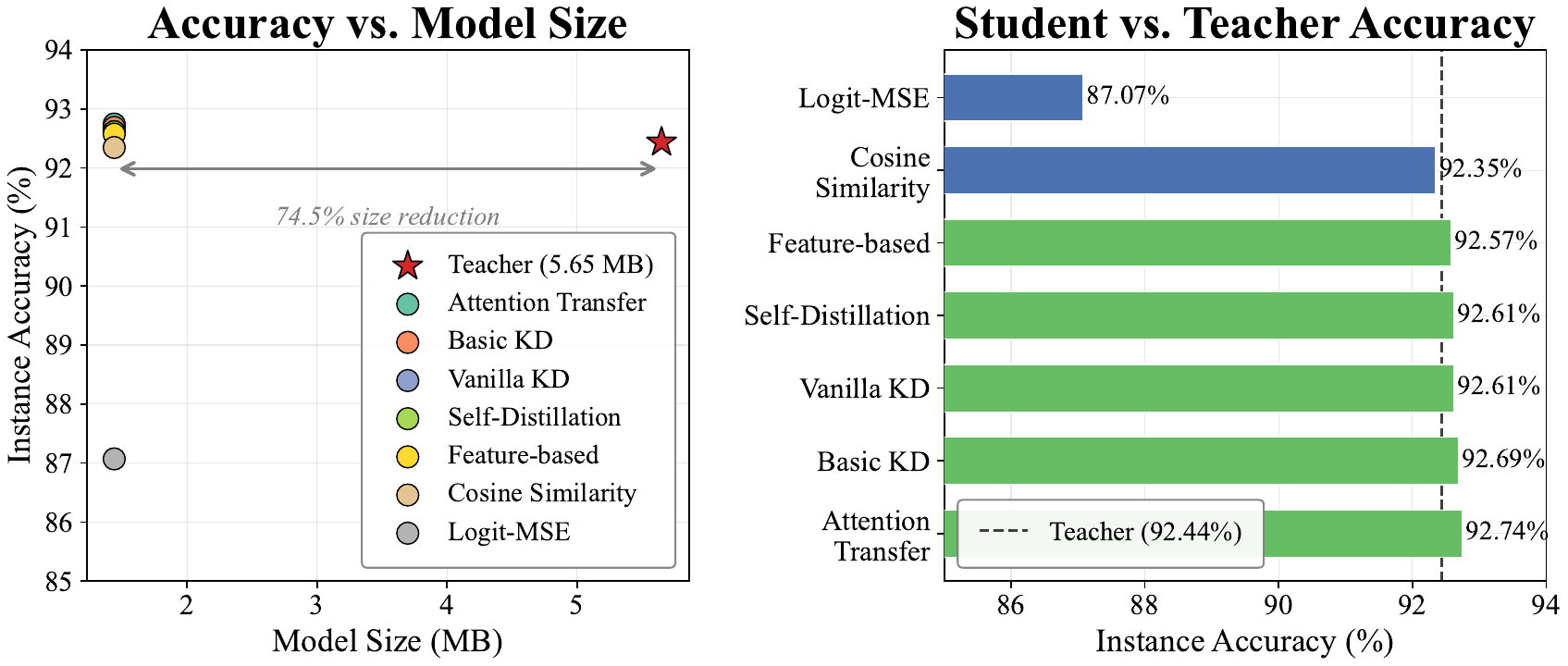}
\caption{Accuracy vs.\ model size}
\label{fig:efficiency_a}
\end{subfigure}
\hfill
\begin{subfigure}[b]{0.44\linewidth}
\centering
\includegraphics[width=\linewidth]{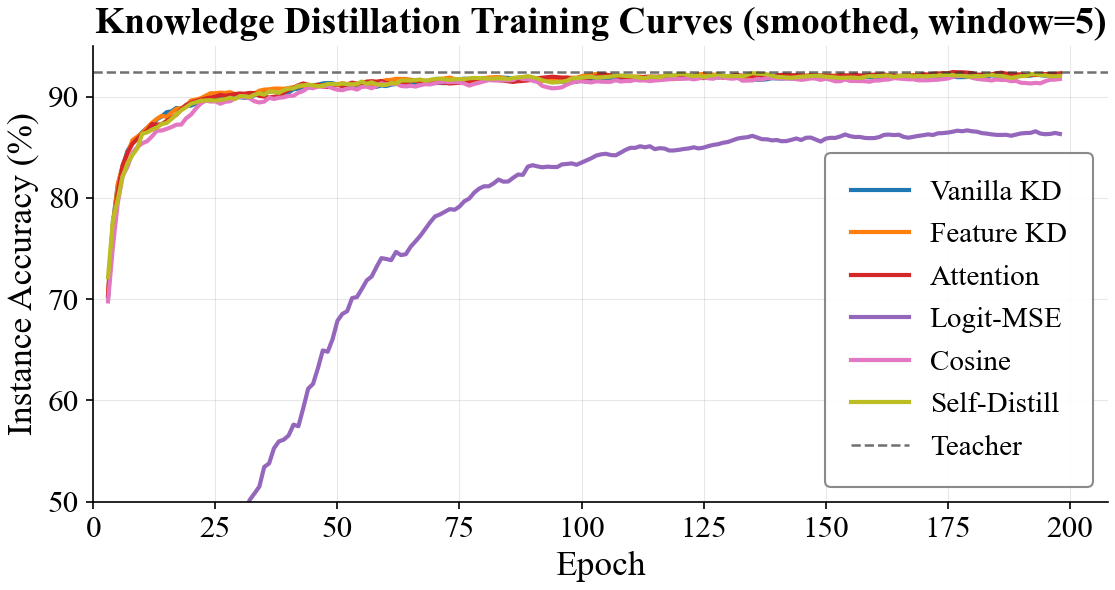}
\caption{Training-epoch convergence}
\label{fig:efficiency_b}
\end{subfigure}
\caption{\textbf{Efficiency and distillation performance on ModelNet40.}
\textbf{(a)} Instance accuracy versus model size (left) and each objective's student accuracy relative to the $92.44\%$ teacher (right): all SmallPointNet2 students ($1.44$\,MB) are $74.51\%$ smaller than the teacher ($5.65$\,MB), with most objectives within $0.3$ points of the teacher.
\textbf{(b)} Training-epoch accuracy for six of the seven objectives (Basic~KD overlaps Vanilla~KD and is omitted for clarity): Logit-MSE converges slower and plateaus below the teacher baseline, whereas the others asymptote at or above $92.35\%$ within $100$ epochs.}
\label{fig:efficiency}
\end{figure}

\Cref{tab:kd_results} reports instance accuracy for each distillation objective distilled from the centralized ModelNet40 teacher ($92.44\%$), and \Cref{fig:efficiency} relates accuracy to model size together with the student-versus-teacher training dynamics.

\begin{table}[tb]
\caption{KD results on ModelNet40. Teacher: PointNet++ SSG ($5.65$\,MB, $92.44\%$). Student: SmallPointNet2 ($1.44$\,MB, $74.51\%$ size reduction, approximately $50\%$ lower inference latency). Instance accuracy is the peak over all training epochs. All strategies are run from the same pre-trained teacher checkpoint. The seven shown are six of the ten benchmark objectives plus a full-weight Basic~KD probe; all ten are evaluated on the clinical task (\Cref{sec:realworld}).}
\label{tab:kd_results}
\centering
\small
\begin{tabular}{@{}lccc@{}}
\toprule
Strategy & Inst. Acc. (\%) $\uparrow$ & Class Acc. (\%) $\uparrow$ & vs.\ Teacher (\%) $\uparrow$ \\
\midrule
Teacher (PointNet++ SSG)~\cite{pointnet2} & 92.44 & 90.24 & 0.00 \\
\midrule
Attention Transfer~\cite{zagoruyko2016paying} & \textbf{92.74} & 90.05 & \textbf{$+$0.30} \\
Basic KD~\cite{gou2021knowledge}              & 92.69 & 89.30 & $+$0.25 \\
Vanilla KD~\cite{hinton2015distilling}        & 92.61 & 89.43 & $+$0.17 \\
Self-Distillation~\cite{furlanello2018born}   & 92.61 & 89.66 & $+$0.17 \\
Feature KD~\cite{romero2014fitnets}           & 92.57 & 89.52 & $+$0.13 \\
Cosine Similarity~\cite{luo2018cosine}        & 92.35 & \textbf{90.12} & $-$0.09 \\
Logit-MSE~\cite{ba2014deep}                   & 87.07 & 75.08 & $-$5.37 \\
\bottomrule
\end{tabular}
\end{table}

Distillation is an effective compressor for this architecture (\Cref{fig:efficiency}). Five of the seven objectives meet or surpass the $92.44\%$ teacher on the compact student: Attention Transfer ($92.74\%$), Basic~KD ($92.69\%$), Vanilla~KD ($92.61\%$), Self-Distillation ($92.61\%$), and Feature~KD ($92.57\%$); Cosine~Similarity ($92.35\%$) trails by $0.09$ points. The lone outlier is Logit-MSE ($87.07\%$): a mean-squared error on the output vectors is scale-sensitive relative to the cross-entropy term and unbalances the loss, slowing convergence (\Cref{fig:efficiency_b}). Compression is uniform across objectives ($74.51\%$ smaller, from $5.65$ to $1.44$\,MB, and about $50\%$ lower inference latency). This near-uniform success under a strong teacher is exactly what makes standalone KD a poor probe of the federated teacher: when the labels alone reach the ceiling, the objective is barely tested. The combined experiment exposes that.

\subsection{Combined Federated Learning and Knowledge Distillation Results}
\label{sec:combined_results}

\begin{figure}[tb]
\centering
\begin{subfigure}[b]{0.46\linewidth}
\centering
\includegraphics[width=\linewidth]{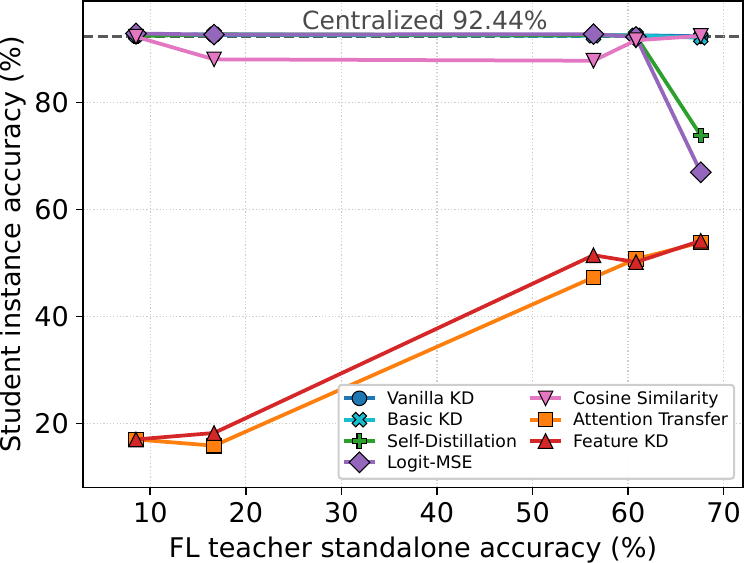}
\caption{Student vs.\ teacher accuracy}
\label{fig:combined_robustness_a}
\end{subfigure}
\hfill
\begin{subfigure}[b]{0.46\linewidth}
\centering
\includegraphics[width=\linewidth]{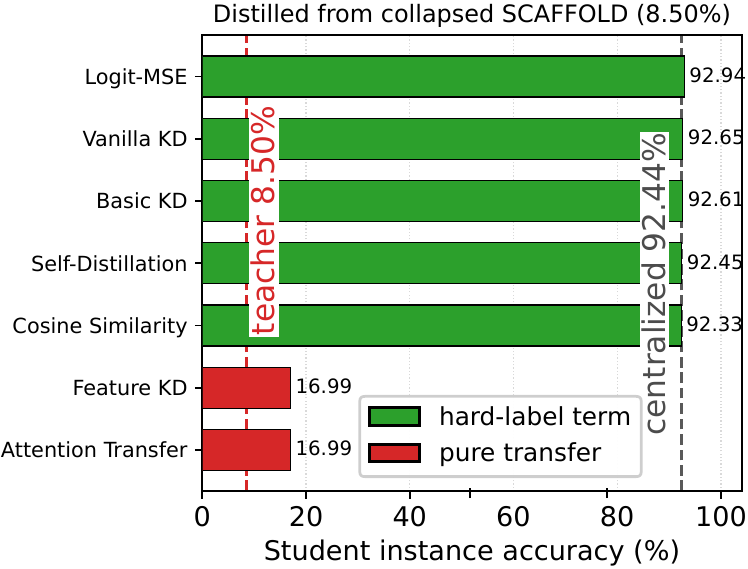}
\caption{Recovery from a collapsed teacher}
\label{fig:combined_robustness_b}
\end{subfigure}
\caption{\textbf{The recovery illusion on ModelNet40 (focused round).}
\textbf{(a)} Student instance accuracy versus FL-teacher standalone accuracy for the seven distillation objectives of \Cref{tab:combined_results}: the five objectives with a positive cross-entropy coefficient (Vanilla~KD, Basic~KD, Self-Distillation, Logit-MSE, Cosine~Similarity) stay near the $92\%$ centralized line for every teacher, even one collapsed to $8.50\%$, because the labeled proxy split supervises the student directly, whereas Attention Transfer and Feature~KD, whose cross-entropy coefficient is exactly $0$, fall in step with the teacher.
\textbf{(b)} Student accuracy distilled from the collapsed SCAFFOLD teacher ($8.50\%$ standalone) under each objective, ordered by accuracy: the five hard-label objectives recover to the $92.44\%$ centralized ceiling, a gap of up to $84.4$ points above the teacher, while Attention Transfer and Feature~KD stay near the teacher's own level. The full $13\times10$ multi-seed heatmaps appear in the supplementary material.}
\label{fig:combined_robustness}
\end{figure}

\Cref{tab:combined_results} shows instance accuracy for the two-stage combined pipeline across the five FL teachers and the seven KD objectives, and \Cref{fig:combined_robustness} plots each objective's sensitivity to teacher quality. These five teachers span the full standalone-accuracy range, from FedProx at $67.61\%$ down to SCAFFOLD at $8.50\%$ (FedAvg reaches $60.83\%$ with size-weighted and $56.38\%$ with uniform averaging, and FedDyn $16.67\%$); we also evaluated FedMedian at $31.16\%$ and the server-side FedAvgM and FedAdam at $4.05\%$, excluding them as teachers because their near-chance logits carry no usable distillation gradient. It shows a consistent dichotomy that turns on a single design choice, the weight of the hard-label cross-entropy term.

\begin{table}[tb]
\caption{Two-stage combined pipeline: instance accuracy (\%) on ModelNet40 (focused round, individual checkpoints). Each cell is the best instance accuracy of a SmallPointNet2 student trained with the given KD objective after federated pre-training of the teacher with the given FL strategy; ``Std.'' is the teacher's own standalone instance accuracy. Column headers: Vanilla~KD (VKD)~\cite{hinton2015distilling}, Basic~KD (BKD)~\cite{gou2021knowledge}, Self-Distillation (SD)~\cite{furlanello2018born}, Attention Transfer (AT)~\cite{zagoruyko2016paying}, Feature~KD (FKD)~\cite{romero2014fitnets}, Logit-MSE (MSE)~\cite{ba2014deep}, Cosine~Similarity (Cos)~\cite{luo2018cosine}. FedAvg (van.) uses uniform (unweighted) client averaging; the other FedAvg uses size-weighted averaging. \textbf{Bold} marks the best value in each KD-objective column. $\dagger$~Denotes runs stopped before $200$ epochs; values are the best accuracy recorded to that point. These five teachers were chosen to span the full standalone-accuracy range (``Std.''); the complete $13\times10$ multi-seed cross-products appear in the supplementary material (ModelNet40) and in \Cref{sec:realworld} (clinical task).}
\label{tab:combined_results}
\centering
\footnotesize
\setlength{\tabcolsep}{3.2pt}
\begin{tabular}{@{}lrrrrrrrr@{}}
\toprule
FL Teacher & Std. & VKD $\uparrow$ & BKD $\uparrow$ & SD $\uparrow$ & AT $\uparrow$ & FKD $\uparrow$ & MSE $\uparrow$ & Cos $\uparrow$ \\
\midrule
FedAvg~\cite{mcmahan2017communication} & $60.83$ & 92.61 & 92.57 & 92.45 & 50.74 & 50.16 & $92.29^\dagger$ & $91.67^\dagger$ \\
FedAvg (van.)~\cite{mcmahan2017communication} & $56.38$ & 92.57 & 92.53 & 92.65 & 47.31 & 51.46 & 92.82 & $87.84^\dagger$ \\
FedProx~\cite{li2020federated} & $67.61$ & 92.41          & 92.13          & $73.85^\dagger$ & \textbf{53.84} & \textbf{54.16} & $66.97^\dagger$ & \textbf{92.49} \\
FedDyn~\cite{acar2021federated} & $16.67$ & \textbf{92.65} & \textbf{92.77} & \textbf{92.78} & $15.82^\dagger$ & 18.20 & 92.73 & $88.10^\dagger$ \\
SCAFFOLD~\cite{karimireddy2020scaffold} & $\phantom{0}8.50$ & \textbf{92.65} & 92.61 & 92.45 & $16.99^\dagger$ & 16.99 & \textbf{92.94} & 92.33 \\
\bottomrule
\end{tabular}
\end{table}

\noindent\textbf{Objectives that keep a hard-label term mask the teacher.}
The five objectives with a positive cross-entropy coefficient (Vanilla~KD, Basic~KD, Self-Distillation, Logit-MSE, and Cosine~Similarity) recover near-centralized accuracy from every teacher: setting aside the runs that stopped before epoch~$200$ (marked~$\dagger$), every completed cell for these five lies between $92.13\%$ and $92.94\%$. The peak cell, SCAFFOLD\,+\,Logit-MSE at $92.94\%$, exceeds the $92.44\%$ centralized reference by $0.50$ points despite an FL teacher that converges to only $8.50\%$ on its own, a gap of $84.4$ points above the teacher it distilled from, which cannot reflect genuine teacher-to-student transfer. Seven of the ten highest-accuracy configurations pair a hard-label objective with a collapsed teacher (three from SCAFFOLD at $8.50\%$, four from FedDyn at $16.67\%$).

\noindent\textbf{Objectives with no hard-label term reveal the teacher.}
Feature- and attention-based objectives behave oppositely. For collapsed teachers (SCAFFOLD at $8.50\%$, FedDyn at $16.67\%$) they reach only $15.82\%$ to $18.20\%$, comparable to the teacher itself, and even with the FedProx teacher ($67.61\%$) they stay in the $53\%$ to $55\%$ range. At $\alpha{=}\beta{=}0.5$ their cross-entropy coefficient $(1-\alpha-\beta)$ is exactly $0$, removing the labeled-data fallback: both remaining terms depend entirely on teacher quality, so a collapsed teacher drives the student toward meaningless representations. These objectives are thus a label-free probe of teacher quality, showing genuine transfer unpropped by proxy labels.

\noindent\textbf{Interpretation.}
Both behaviors follow from the loss $\mathcal{L} = w_{\text{CE}}\,\mathcal{L}_{\text{CE}}(\hat{y}, y) + \mathcal{L}_{\text{transfer}}$ on the labeled centralized set. When $w_{\text{CE}}{>}0$ and the teacher has collapsed, its soft targets give little useful gradient and the well-conditioned cross-entropy term dominates, so the student learns the proxy labels directly; when $w_{\text{CE}}{=}0$, the teacher is the only signal, so a failed teacher yields a failed student. \emph{An FL-KD evaluation that uses hard labels therefore cannot measure the federated teacher: it reduces the pipeline to centralized training on the proxy labels, reusing the very labels whose privacy motivated federation.} A sound evaluation distills without hard labels, by aggregating client soft predictions on unlabeled data~\cite{lin2020ensemble} or omitting the cross-entropy term; the feature-based curves of \Cref{fig:combined_robustness} are exactly that measurement. Two controls expose the illusion: a \emph{no-cross-entropy} control ($\alpha{=}1$, $\mathcal{L}_{\text{CE}}$ removed) and an \emph{unlabeled-proxy} control. Both leave the teacher as the sole signal, so a sound student falls back toward its teacher's standalone accuracy (a SCAFFOLD-conditioned student toward $8.50\%$, not $92.94\%$); residual recovery signals cross-entropy-driven masking.

\subsection{Real-World Validation: Clinical Craniosynostosis}
\label{sec:realworld}

The ModelNet40 grid isolates the recovery illusion under controlled conditions; the clinical craniosynostosis dataset shows that it survives on real patient data, where the stakes are concrete. These head shapes sit in separate clinics, cannot be pooled freely~\cite{rieke2020future,sheller2020federated}, and must run on portable point-of-care scanners, so federation and compression are operational requirements rather than conveniences. We therefore run the complete cross-product of all thirteen FL teachers and ten KD objectives ($130$ teacher--objective pairs) at full multi-seed scale (three seeds). Standalone distillation compresses the $100\%$ centralized teacher into a $1.42$\,MB student ($74.7\%$ smaller, roughly $2\times$ faster) at near-ceiling accuracy across all ten objectives (supplementary material, Appendix~B, Table~A4); the unambiguous ceiling lets us read teacher quality directly.

The combined grid reproduces the ModelNet40 dichotomy, and the hard-label cross-entropy term again decides the outcome (\Cref{tab:combined_summary}; see also the supplementary material, Appendix~B, Figure~A4). The six objectives that keep a cross-entropy term recover the student from any teacher: their accuracy varies by at most $7.5$ points across the thirteen teachers and does not correlate with teacher quality ($r$ from $-0.53$ to $+0.37$). A teacher frozen at the $25\%$ chance level (FedAvgM) still yields a $97.5\%$ student under Vanilla~KD and $100\%$ under CRD. The four pure-transfer objectives, whose cross-entropy coefficient is $0$ at $\alpha{=}\beta{=}0.5$, instead track the teacher almost perfectly ($r{=}0.986$ to $0.990$): they span $57.5$ to $58.3$ points across teachers, fall to chance when the teacher does (FedAvgM with Attention Transfer or SP yields $25.0\%$), and recover only behind a strong teacher (FedProx with Feature~KD reaches $84.2\%$, tracking its $75.83\%$ teacher).

The clinical data thus turns the diagnosis into a deployment warning. A hard-label-distilled student here logs near-$100\%$ accuracy from a federated teacher no better than chance, while the cross-entropy term silently reuses the patient labels that federation was meant to protect. Only label-free distillation reports the federated teacher's true quality. Per-cell means, standard deviations, and per-seed values appear in the supplementary material (Appendix~C, Tables~A8 and~A9; Appendix~D, Table~A13).

\begin{table}[tb]
\caption{Two-stage combined pipeline on the clinical task, summarized by KD objective over all thirteen FL teachers (mean instance accuracy, \%, three seeds). ``CE wt.'' is the weight on the hard-label cross-entropy term. ``Chance teacher'' uses FedAvgM~\cite{hsu2019measuring}, a teacher at the $25\%$ chance level; ``best teacher'' uses FedProx~\cite{li2020federated} at $75.83\%$. ``Range'' is the spread of student accuracy across the thirteen teachers, and $r$ is the Pearson correlation between student accuracy and the teacher's standalone accuracy. The six objectives that keep a cross-entropy term recover the student regardless of teacher quality ($r\approx0$); the four with zero cross-entropy weight track the teacher ($r\approx0.99$). Per-cell values appear in the supplementary material (Appendix~C, Tables~A8 and~A9).}
\label{tab:combined_summary}
\centering
\small
\setlength{\tabcolsep}{4.5pt}
\begin{tabular}{@{}lccccc@{}}
\toprule
KD Objective & CE wt. & Chance teacher & Best teacher & Range & $r$ \\
\midrule
\multicolumn{6}{@{}l}{\emph{Objectives retaining a hard-label term}}\\
Vanilla KD~\cite{hinton2015distilling} & $0.5$ & $97.5$ & $99.6$ & $\phantom{0}2.5$ & $+0.37$ \\
Self-Distillation~\cite{furlanello2018born} & $0.5$ & $99.2$ & $99.6$ & $\phantom{0}1.2$ & $+0.19$ \\
Cosine Similarity~\cite{luo2018cosine} & $0.5$ & $99.6$ & $100.0$ & $\phantom{0}0.8$ & $-0.05$ \\
Contrastive (CRD)~\cite{tian2019contrastive} & $0.5$ & $100.0$ & $99.6$ & $\phantom{0}0.8$ & $-0.53$ \\
Logit-MSE~\cite{ba2014deep} & $0.5$ & $97.5$ & $99.2$ & $\phantom{0}3.8$ & $+0.10$ \\
Decoupled KD~\cite{zhao2022decoupled} & $1.0$ & $93.8$ & $96.7$ & $\phantom{0}7.5$ & $-0.10$ \\
\midrule
\multicolumn{6}{@{}l}{\emph{Pure teacher-transfer objectives ($1-\alpha-\beta=0$)}}\\
Feature KD~\cite{romero2014fitnets} & $0$ & $27.1$ & $84.2$ & $57.5$ & $+0.99$ \\
Attention Transfer~\cite{zagoruyko2016paying} & $0$ & $25.0$ & $83.3$ & $58.3$ & $+0.99$ \\
Relational KD~\cite{park2019relational} & $0$ & $27.5$ & $81.7$ & $57.5$ & $+0.99$ \\
Similarity-Preserving~\cite{tung2019similarity} & $0$ & $25.0$ & $82.9$ & $58.3$ & $+0.99$ \\
\bottomrule
\end{tabular}
\end{table}

% =====================================================================
\section{Discussion and Limitations}
\label{sec:discussion}

\noindent\textbf{Why does FL degrade severely for 3D architectures?}
Extreme label skew forces each client's layers to overfit local categories, producing gradient heterogeneity that standard aggregation cannot handle. Adaptive server optimizers suffer most, likely from ill-conditioned second-moment estimates under heterogeneous updates~\cite{reddi2020adaptive}. That this holds on both a $40$-class benchmark and a four-class clinical dataset shows it is not an artifact of label-space size.

\noindent\textbf{No universally robust algorithm.}
The dataset-dependent ranking (\Cref{tab:fl_results}) cautions against choosing an FL method from a single dataset: the leader on one dataset slips on the other, and even FedProx, the most consistent of the thirteen, never closes the gap to centralized training on both. That the contrastive, personalized, and adaptive server-side methods all leave this gap indicates these objectives do not by themselves overcome the extreme label skew studied here.

\noindent\textbf{The hard-label term governs the combined pipeline.}
An FL-KD pipeline's reported accuracy is governed by whether the distillation loss keeps a hard-label term, not by the federated teacher: with such a term the student reaches the ceiling from any teacher. A faithful assessment must therefore distill on unlabeled data or drop the cross-entropy term, and report teacher-tracking accuracy.

\noindent\textbf{Hierarchical versus transformer backbones.}
The degradation may also depend on the backbone. The fixed-radius neighborhood aggregation that may underlie PointNet++'s sensitivity to label skew is specific to hierarchical models. Transformer backbones such as PCT~\cite{guo2021pct} and Point Transformer V2~\cite{wu2022pointtransv2}, and kernel-point methods such as KPConv~\cite{thomas2019kpconv}, aggregate features through different mechanisms whose behavior under parameter averaging is not obvious in advance. Whether such architectures degrade less than hierarchical models under non-IID federation is an open question that our single-backbone study cannot settle.

\noindent\textbf{Limitations.}
Several boundaries remain. First, we evaluate two datasets, ModelNet40 and a statistical-shape-model craniosynostosis set; larger real-world scanned collections remain to be tested. Second, the ModelNet40 combined diagnosis uses teachers spanning the full quality range down to collapsed checkpoints, and the clinical dataset confirms it at full multi-seed scale over the complete $13\times10$ grid, whose $100\%$ ceiling saturates the standalone KD numbers and leaves the combined grid as the diagnostic.
Third, we study one non-IID regime, the extreme disjoint label-skew partition; softer partitions would narrow the gap, and a heterogeneity sweep is left to future work. Fourth, the seeds vary initialization and optimization but not the deterministic partition, so the reported deviations capture training-time variance. Fifth, the architecture is a fixed PointNet++ SSG teacher and compact student; whether the effects persist for transformer or kernel-point backbones is open. Finally, all runs fix five clients, twenty rounds, and homogeneous client architectures; heterogeneous client models would require aggregation across dissimilar parameter spaces.

% =====================================================================
\section{Conclusion}
\label{sec:conclusion}

We presented a multi-seed benchmark for FL and KD in 3D point cloud classification, spanning thirteen FL algorithms, ten KD objectives, and their $130$-pair cross-product. Standalone federated training degrades sharply under extreme label skew on both datasets, the method ranking is dataset-dependent, and the four server-side optimizers collapse to near the chance level, so no single algorithm is robust. Distillation compresses the strong centralized teacher into a student $74.51\%$ smaller and roughly twice as fast at near-ceiling accuracy. The combined teacher-by-objective grid then exposes the central caution: any objective that keeps a hard-label cross-entropy term recovers the student to the centralized ceiling whatever the federated teacher, so its reported accuracy measures the proxy labels, not the federated model; only label-free objectives track teacher quality. The clinical craniosynostosis data shows the effect is not specific to ModelNet40: on a privacy-critical task, hard-label distillation reports a near-$100\%$ student from a chance-level teacher, while reusing the very labels federation was meant to protect. We therefore recommend evaluating FL-KD pipelines with label-free distillation, so that the reported accuracy measures the federated teacher.

For reproducibility, the source code and the trained model checkpoints for every reported run are publicly available at \url{https://ezharjan.github.io/FLKD3DBenchmark}, and new backbones, tasks, datasets, algorithms, and student architectures integrate through one shared interface. 
We envision this benchmark and its label-free evaluation protocol as a robust foundation for advancing privacy-preserving, edge-deployable 3D point cloud systems, and we welcome community contributions to expand its scope.

\clearpage

% ---- Bibliography ----
\bibliographystyle{splncs04}
\bibliography{ref}

\clearpage
\title{Benchmarking Federated Learning \& Knowledge Distillation for Point Cloud Classification
\\[6pt]{\large Supplementary Material}}

\titlerunning{FL and KD for Point Cloud Classification: Supplementary Material}

% Author list (identical to the main paper).
\author{Aizierjiang Aiersilan}
\authorrunning{Aizierjiang Aiersilan}
\institute{University of Macau \\
    \email{ezharjan@outlook.com}}

\maketitle

% ---------------------------------------------------------------
% appendix-style numbering
% (Appendix A--D; Tables, Figures and Algorithms prefixed with ``A'').
\appendix
\renewcommand{\thesection}{\Alph{section}}
\renewcommand{\thetable}{A\arabic{table}}
\renewcommand{\thefigure}{A\arabic{figure}}
\renewcommand{\thealgorithm}{A\arabic{algorithm}}
\setcounter{section}{0}
\setcounter{table}{0}
\setcounter{figure}{0}
\setcounter{algorithm}{0}
\crefalias{section}{appendix}
\Crefname{appendix}{Appendix}{Appendices}
\crefname{appendix}{Appendix}{Appendices}
\Crefname{figure}{Figure}{Figures}
\crefname{figure}{Figure}{Figures}

% =====================================================================
\noindent This document is the supplementary material for the paper ``Benchmarking Federated Learning \& Knowledge Distillation for Point Cloud Classification.'' \Cref{app:impl} documents the implementation in enough detail to reproduce all $504$ runs: the shared optimization settings, the teacher and student architectures, the data augmentation, the deterministic non-independent and identically distributed (non-IID) label-skew partition, the per-method federated learning (FL) update rules, the ten knowledge distillation (KD) loss formulations, and the two diagnostic controls, each accompanied by supporting pseudocode or a parameter table. \Cref{app:clinical} presents the complete craniosynostosis study, covering both the standalone distillation results and the full $13\times10$ combined-pipeline grid. \Cref{app:tables} gives the complete per-configuration numbers behind the summary tables and, in its final subsection, the full multi-seed ModelNet40 combined-pipeline grid; \Cref{app:perseed} lists every individual per-seed accuracy, for both the standalone runs and the clinical and ModelNet40 combined grids, before averaging; and \Cref{app:moredata} reports a generalization study on two further point-cloud datasets (ModelNet10 and OmniObject3D). References to the main paper are given by explicit numbers (for example, ``Section~4.5 of the main paper''), whereas references prefixed with the letter~A (for example, \Cref{tab:combined_mean}) point to tables, figures, and algorithms within this supplement. Throughout, accuracies are test instance accuracy in percent, reported as the mean over the three seeds $\{7,42,123\}$ with standard deviation where shown.

% =====================================================================
\section*{Notation and Abbreviations}
\addcontentsline{toc}{section}{Notation and Abbreviations}

\noindent The abbreviations and symbols used in the main paper and in this supplement are collected here so that they can be looked up at a glance. Each abbreviation is also defined at its first use in the running text.

\smallskip
\noindent\textbf{Abbreviations.}
\nopagebreak\par\smallskip
{\small
\setlength{\tabcolsep}{6pt}
\renewcommand{\arraystretch}{1.18}
\begin{longtable}{@{}p{0.155\textwidth} p{0.80\textwidth}@{}}
\toprule
Abbreviation & Meaning \\
\midrule
\endfirsthead
\toprule
Abbreviation & Meaning \\
\midrule
\endhead
\bottomrule
\endfoot
FL & Federated learning \\
KD & Knowledge distillation \\
non-IID & Non-independent and identically distributed (the data-partition regime) \\
SSG & Single-scale grouping (the PointNet++ teacher and student backbone) \\
MLP & Multi-layer perceptron \\
CE & Cross-entropy (hard-label loss) \\
KL & Kullback--Leibler (divergence) \\
MSE & Mean squared error \\
AT & Attention Transfer (KD objective) \\
SP & Similarity-Preserving distillation (KD objective) \\
RKD & Relational Knowledge Distillation (KD objective) \\
CRD & Contrastive Representation Distillation (KD objective) \\
DKD & Decoupled Knowledge Distillation (KD objective) \\
TCKD / NCKD & Target-class / non-target-class KD (the two terms of DKD) \\
DGCNN & Dynamic Graph Convolutional Neural Network \\
PCT & Point Cloud Transformer \\
KPConv & Kernel Point Convolution \\
GPU & Graphics processing unit \\
MB & Megabyte (model-size unit) \\
\end{longtable}
}

\smallskip
\noindent\textbf{Symbols.}
\nopagebreak\par\smallskip
{\small
\setlength{\tabcolsep}{6pt}
\renewcommand{\arraystretch}{1.25}
\begin{longtable}{@{}p{0.205\textwidth} p{0.75\textwidth}@{}}
\toprule
Symbol & Meaning \\
\midrule
\endfirsthead
\toprule
Symbol & Meaning \\
\midrule
\endhead
\bottomrule
\endfoot
$K$ & Number of federated clients (here $K{=}5$) \\
$\mathcal{D}_k$ & Private data shard held by client $k$ \\
$\theta,\ \theta^{g}$ & Local and global (broadcast) model parameters \\
$\alpha$ & Soft-target (KD) weight ($\alpha{=}0.5$) \\
$\beta$ & Intermediate-representation weight ($\beta{=}0.5$) \\
$T$ & Distillation temperature ($T{=}2.0$) \\
$\mathcal{L}_{\text{CE}}$ & Hard-label cross-entropy loss on the true labels \\
$\mathcal{L}_{\text{transfer}}$ & Generic teacher-transfer loss term \\
$w_{\text{CE}}$ & Weight applied to the hard-label cross-entropy term \\
$D_{\text{KL}}$ & Kullback--Leibler divergence \\
$p_s^{T},\ p_t^{T}$ & Temperature-softened student / teacher class distributions \\
$z_s,\ z_t$ & Student / teacher output (log-probability) vectors \\
$f_s,\ f_t$ & Student / teacher penultimate feature descriptors \\
$\phi$ & Learnable linear projection mapping the student feature to the teacher width \\
$\hat{f}$ & $\ell_2$-normalized feature, $\hat{f}{=}f/\lVert f\rVert_2$ \\
$A_s,\ A_t$ & Attention descriptors used by Attention Transfer \\
$G_s,\ G_t$ & Row-normalized Gram (sample-similarity) matrices used by SP \\
$\psi_D,\ \psi_A$ & Pairwise-distance and triplet-angle potentials used by RKD \\
$\mathcal{L}_{\text{CRD}}$ & In-batch contrastive (InfoNCE-style) CRD loss \\
$\eta_s$ & Server-side learning rate (server-optimizer methods) \\
$\mu,\ \lambda,\ \tau$ & Method-specific regularization / contrastive-temperature hyperparameters \\
$\alpha_{\text{D}},\ \beta_{\text{D}}$ & DKD weights on the TCKD and NCKD terms ($1$ and $8$) \\
$r$ & Pearson correlation between student and standalone teacher accuracy \\
\end{longtable}
}

% The two Notation longtables above are reference lists, not numbered data
% tables; reset the counter so the numbered tables of the appendix start at A1.
\setcounter{table}{0}

\section{Implementation Details}
\label{app:impl}

We document the full experimental protocol so that every number reported in the paper can be reproduced. The description is stated at the level of the learning procedure rather than any particular software interface, and the three benchmark stages (federated training, distillation, and their two-stage combination) are summarized as Algorithms~\ref{alg:partition}, \ref{alg:fl}, and~\ref{alg:pipeline}.

\smallskip\noindent\textbf{Shared optimization protocol.}
Every model receives an input tensor of $1{,}024$ points with three spatial coordinates and no surface normals, that is, a tensor of shape $3\times1{,}024$. All training, whether centralized, federated, or distillation, uses the Adam optimizer ($\beta_1{=}0.9$, $\beta_2{=}0.999$, $\epsilon{=}10^{-8}$, weight decay $10^{-4}$) with learning rate $0.001$ and batch size $24$. The centralized teacher and every distilled student are trained for $200$ epochs under a step learning-rate schedule that multiplies the rate by $0.7$ every $20$ epochs. Federated training runs for $20$ communication rounds of $5$ local epochs each across five clients; the per-method federated hyperparameters are listed in \Cref{tab:fl_params}. For every run we retain the checkpoint with the highest test instance accuracy, and all reported accuracies use that best checkpoint.

\smallskip\noindent\textbf{Output convention.}
Both backbones emit log-probability vectors, that is, a log-softmax over the class logits, rather than raw logits, and all distillation objectives are defined consistently on these outputs. The soft-target Kullback--Leibler term and the hard-label cross-entropy term therefore follow their standard formulation, while the two logit-space objectives (Logit-MSE and Cosine) act on the log-probability output vectors. This single convention keeps the objectives directly comparable, because every loss receives inputs of the same form.

\smallskip\noindent\textbf{Network architectures.}
The teacher is a PointNet++ single-scale-grouping classifier with three hierarchical set-abstraction levels. The first two levels sample $512$ and $128$ centroids by ball-query grouping at radii $0.2$ and $0.4$ with $32$ and $64$ neighbors respectively, and the third level pools all remaining points into a single global feature. The per-level shared multi-layer-perceptron widths are $[64,64,128]$, $[128,128,256]$, and $[256,512,1024]$, so the teacher produces a $1{,}024$-dimensional global descriptor, which is passed through two fully connected layers of widths $512$ and $256$ (each with batch normalization and dropout $0.4$) before the linear classification head. The compact student preserves this three-level hierarchy but narrows every stage: it samples $256$ and $64$ centroids at the same radii $0.2$ and $0.4$ with $16$ and $32$ neighbors, then pools globally, with multi-layer-perceptron widths $[32,32,64]$, $[64,64,128]$, and $[128,256,512]$ and a classification head of widths $256$ and $128$. The student therefore yields a $512$-dimensional penultimate descriptor, which it exposes so that the feature-based distillation objectives have an intermediate signal to match.

\smallskip\noindent\textbf{Data augmentation.}
Each training batch is augmented on the fly by three operations applied in sequence. First, random per-cloud point dropout removes a fraction of points drawn uniformly in $[0,0.875]$, resetting each dropped point to the cloud's first point so that the point count is preserved. Second, random isotropic scaling multiplies the coordinates by a factor drawn uniformly in $[0.8,1.25]$. Third, random translation shifts the cloud by an offset drawn uniformly in $[-0.1,0.1]$ independently along each axis. Only the spatial coordinates are transformed, and test clouds are evaluated without any augmentation.

\smallskip\noindent\textbf{Non-IID label-skew partition.}
Stage~1 uses the deterministic label-skew partition of Algorithm~\ref{alg:partition}: the training set is sorted by class label and cut into five contiguous slices of equal size, one per client, so that clients hold disjoint segments of the label space. This is a deliberately severe, worst-case form of heterogeneity, and it is held fixed across seeds so that the reported variance reflects training-time randomness rather than partition luck. On the balanced four-class clinical data ($80$ training shapes per class) the rule yields the composition in \Cref{tab:partition}: two clients see a single class and the other three see two adjacent classes, so no client observes more than half of the label space and the central classes are split across clients. On ModelNet40 the same rule gives each client roughly $1{,}969$ shapes spanning about eight contiguous categories; because the ModelNet40 class sizes are not uniform, the category boundaries do not fall exactly on slice edges, yet each client still observes a disjoint block of the $40$ categories. Softer schemes such as a Dirichlet partition would reduce the heterogeneity.

\begin{algorithm}[t]
\caption{Deterministic label-skew partition used in Stage~1. Sorting by label and cutting into equal contiguous slices gives each client a disjoint block of the label space, the worst-case non-IID regime studied here. The procedure depends only on the dataset, so it is identical across seeds.}
\label{alg:partition}
\hrule\vspace{3pt}
\begin{algsteps}
\item \algkw{Input:} training set with class labels $y_1,\dots,y_N$; number of clients $K$
\item sort the sample indices by their class label into a single ordered sequence
\item cut the sorted sequence into $K$ contiguous slices of near-equal size
\item assign slice $k$ to client $k$ as its private shard $\mathcal{D}_k$ \algcom{disjoint label blocks}
\item \algkw{return} the client shards $\mathcal{D}_1,\dots,\mathcal{D}_K$
\end{algsteps}
\vspace{3pt}\hrule
\end{algorithm}

\begin{table}[t]
\caption{Client composition of the label-skew partition on the four-class clinical training set ($320$ shapes, $80$ per class, classes ordered as the control group and the coronal, metopic, and sagittal fusions). Each client holds $64$ shapes drawn from at most two adjacent classes.}
\label{tab:partition}
\centering
\small
\setlength{\tabcolsep}{8pt}
\begin{tabular}{@{}lccccc@{}}
\toprule
& Client 1 & Client 2 & Client 3 & Client 4 & Client 5 \\
\midrule
Control  & 64 & 16 & -- & -- & -- \\
Coronal  & -- & 48 & 32 & -- & -- \\
Metopic  & -- & -- & 32 & 48 & -- \\
Sagittal & -- & -- & -- & 16 & 64 \\
\midrule
Total    & 64 & 64 & 64 & 64 & 64 \\
\bottomrule
\end{tabular}
\end{table}

\smallskip\noindent\textbf{FL update rules.}
Every federated round follows the generic template of Algorithm~\ref{alg:fl}: the server broadcasts the current global model, each of the five clients optimizes it locally for five epochs on its disjoint label-skewed shard, and the server aggregates the returned models. The thirteen algorithms differ only in the local objective, the server aggregation rule, or both, and we group them by the mechanism they add. Client optimization uses the shared Adam configuration for all methods except SCAFFOLD, whose control-variate correction is applied through explicit gradient-descent steps at the same learning rate.

The \emph{drift-correction} family keeps the server average but constrains the local update. FedProx adds a proximal penalty $\tfrac{\mu}{2}\lVert\theta-\theta^{g}\rVert^2$ that pulls each client toward the broadcast global model $\theta^{g}$, with $\mu{=}0.01$. FedDyn adds a dynamic regularizer that combines a linear term in an accumulated per-client gradient with a quadratic proximal term of strength $\alpha{=}0.01$, so that the local stationary points align with the global one over rounds. SCAFFOLD maintains server and client control variates $c$ and $c_k$ and replaces the local gradient $g$ by the corrected direction $g+c-c_k$, refreshing $c_k$ from the option-II rule after each client pass. FedNova rescales each client's accumulated update by its number of local steps $\tau_k$ before averaging and then reapplies the effective step count, which removes the objective inconsistency that arises when clients take different numbers of steps because their shards differ in size.

The \emph{server-optimization} family keeps a plain local objective and instead treats the average client displacement $\Delta=\theta^{g}-\bar{\theta}$ as a pseudo-gradient at the server. FedAvgM applies momentum, $v\leftarrow\beta v+\Delta$ then $\theta\leftarrow\theta-\eta_s v$, with $\eta_s{=}1.0$ and $\beta{=}0.9$. FedAdam, FedYogi, and FedAdagrad maintain first- and second-moment estimates $m$ and $v$ of the pseudo-gradient and update $\theta\leftarrow\theta+\eta_s\,m/(\sqrt{v}+\epsilon)$; the three share $\eta_s{=}0.05$, $\beta_1{=}0.9$, $\epsilon{=}10^{-3}$, and differ only in the second-moment recursion (an exponential moving average for Adam, a sign-stabilized variant for Yogi, and a cumulative sum for Adagrad). The moderate $\eta_s{=}0.05$ already reflects a tuned, non-trivial step size for the three adaptive optimizers, so their collapse under label skew is not a mere step-size artifact.

The \emph{robust-aggregation} method, FedMedian, replaces the coordinate-wise mean with the coordinate-wise median across client models. The remaining methods address feature shift and personalization. FedBN averages all parameters except the batch-normalization statistics, which each client keeps local. MOON adds a model-contrastive term that pulls a client's representation toward that of the global model and away from that of its previous local model, with weight $\mu{=}1.0$ and temperature $\tau{=}0.5$. Ditto trains the shared model with a plain local objective and, in addition, a per-client personalized model regularized toward the shared model with strength $\lambda{=}0.1$; the shared model is what the server aggregates.

\begin{algorithm}[t]
\caption{One communication round of Stage~1 federated training. The template is shared by all thirteen algorithms; the local objective in line~5 and the server rule in line~8 are the method-specific components described above and parameterized in \Cref{tab:fl_params}.}
\label{alg:fl}
\hrule\vspace{3pt}
\begin{algsteps}
\item \algkw{Input:} global model $\theta$; clients $1,\dots,K$ with disjoint label-skewed shards $\mathcal{D}_1,\dots,\mathcal{D}_K$; local epochs $E$
\item \algkw{for} each client $k=1,\dots,K$ \algkw{in parallel do}
\item \algind broadcast the global model: $\theta_k \leftarrow \theta$
\item \algind \algkw{for} $E$ local epochs over $\mathcal{D}_k$ \algkw{do}
\item \algind\algind $\theta_k \leftarrow$ local optimization step under the method's objective \algcom{plain, proximal, dynamic, control-variate, or contrastive}
\item \algind \algkw{end for}
\item \algkw{end for}
\item $\theta' \leftarrow \textsc{Aggregate}(\theta_1,\dots,\theta_K)$ under the method's server rule \algcom{weighted mean, median, normalized average, or server optimizer}
\item evaluate $\theta'$ on the held-out test set and update the best checkpoint
\item \algkw{return} $\theta'$ as the new global model
\end{algsteps}
\vspace{3pt}\hrule
\end{algorithm}

\begin{table}[t]
\caption{Per-method hyperparameters of the thirteen FL algorithms, grouped by mechanism family. The notation $x(v)$ means hyperparameter $x$ is set to value $v$, and $\eta_s$ is the server-side learning rate. A dash marks methods that add no hyperparameter beyond the shared federated setting (five clients, five local epochs, twenty rounds, and the shared Adam configuration). Client optimization is the shared Adam configuration for every method except SCAFFOLD, whose corrected update uses gradient descent at the same learning rate.}
\label{tab:fl_params}
\centering
\small
\renewcommand{\arraystretch}{1.15}
\begin{tabular}{@{} l l @{}}
\toprule
Configuration & Key Parameters \\
\midrule
FedAvg~\cite{mcmahan2017communication} & -- \\
FedProx~\cite{li2020federated} & $\mu(0.01)$ \\
SCAFFOLD~\cite{karimireddy2020scaffold} & -- \\
FedNova~\cite{wang2020tackling} & -- \\
FedDyn~\cite{acar2021federated} & $\alpha(0.01)$ \\
MOON~\cite{li2021model} & $\mu(1.0)$, $\tau(0.5)$ \\
Ditto~\cite{li2021ditto} & $\lambda(0.1)$ \\
FedBN~\cite{li2021fedbn} & -- \\
FedMedian~\cite{yin2018byzantine} & -- \\
FedAvgM~\cite{hsu2019measuring} & $\eta_s(1.0)$, $\beta(0.9)$ \\
FedAdam~\cite{reddi2020adaptive} & $\eta_s(0.05)$, $\beta_1(0.9)$, $\beta_2(0.999)$, $\epsilon(10^{-3})$ \\
FedYogi~\cite{reddi2020adaptive} & $\eta_s(0.05)$, $\beta_1(0.9)$, $\beta_2(0.999)$, $\epsilon(10^{-3})$ \\
FedAdagrad~\cite{reddi2020adaptive} & $\eta_s(0.05)$, $\beta_1(0.9)$, $\epsilon(10^{-3})$ \\
\bottomrule
\end{tabular}
\end{table}

\smallskip\noindent\textbf{Distillation loss formulations.}
\Cref{tab:kd_strategies} gives the loss formulation of each of the ten distillation objectives. All objectives are defined on the log-probability outputs described above and share the intermediate-representation weight $\beta{=}0.5$ where one applies and the temperature $T{=}2.0$ in every soft-target term; all but DKD also share the soft-target weight $\alpha{=}0.5$. Each Kullback--Leibler term is scaled by $T^2$, the standard correction that keeps the magnitude of the soft-target gradient comparable to that of the hard-label gradient. In the notation of the table, $p_s^T$ and $p_t^T$ are the temperature-softened student and teacher distributions and $z_s$, $z_t$ are the output vectors. The intermediate features are the penultimate global descriptors; because the student descriptor is $512$-dimensional and the teacher descriptor is $1{,}024$-dimensional, a learnable linear projection $\phi$ maps the student feature to the teacher width, after which both are $\ell_2$-normalized (written $\hat{f}{=}f/\lVert f\rVert_2$), so Feature KD matches $\widehat{\phi(f_s)}$ to $\hat{f}_t$ under a mean-squared error. The attention descriptors $A_s$ and $A_t$ are the channel-wise $\ell_2$-normalized squared activations of these descriptors, that is, the attention-transfer construction adapted to a globally pooled feature rather than a spatial map. $G_s$ and $G_t$ are the row-normalized sample-similarity (Gram) matrices computed over the batch and used by SP, and $\psi_D$, $\psi_A$ are the pairwise-distance and triplet-angle relational potentials used by RKD. The contrastive term $\mathcal{L}_{\text{CRD}}$ is an in-batch normalized-temperature cross-entropy between student and teacher embeddings: each is mapped by a learnable head to a $128$-dimensional space and compared at contrastive temperature $0.07$, taking the teacher embedding of the same sample as the positive and the other in-batch teacher embeddings as negatives. Decoupled KD (DKD) splits the soft target into a target-class term (TCKD) and a non-target-class term (NCKD), weighted by $\alpha_{\text{D}}{=}1$ and $\beta_{\text{D}}{=}8$ and combined with a full-weight cross-entropy term, all at the shared temperature $T{=}2.0$.

\begin{table}[t]
\caption{The ten KD loss objectives, using $\alpha{=}0.5$, $\beta{=}0.5$, and $T{=}2.0$ except where noted (DKD). $\mathcal{L}_{\text{CE}}$ denotes cross-entropy on true labels; $\hat{f}{=}f/\|f\|_2$ denotes an $\ell_2$-normalized feature vector and $\phi$ a learnable linear projection. The feature, attention, SP, and RKD objectives weight $\mathcal{L}_{\text{CE}}$ by $(1-\alpha-\beta){=}0$ at $\alpha{=}\beta{=}0.5$, leaving no hard-label term, and every soft-target Kullback--Leibler term carries the $T^2$ scaling. For Self-Distillation, $p_{s_{\text{prev}}}^T$ is the previous student generation. For DKD, TCKD and NCKD are the target-class and non-target-class soft terms weighted by $\alpha_{\text{D}}{=}1$ and $\beta_{\text{D}}{=}8$ and combined with a full-weight $\mathcal{L}_{\text{CE}}$, all at the shared $T{=}2.0$.}
\label{tab:kd_strategies}
\centering
\small
\renewcommand{\arraystretch}{1.4}
\begin{tabular*}{\textwidth}{@{\extracolsep{\fill}}ll@{}}
\toprule
Strategy & Loss Formulation \\
\midrule
Vanilla KD~\cite{hinton2015distilling} & $(1-\alpha) \mathcal{L}_{\text{CE}} + \alpha T^2 D_{\text{KL}}(p_t^T \parallel p_s^T)$ \\
Self-Distill.~\cite{furlanello2018born} & $(1-\alpha) \mathcal{L}_{\text{CE}} + \alpha T^2 D_{\text{KL}}(p_{s_{\text{prev}}}^T \parallel p_s^T)$ \\
Logit-MSE~\cite{ba2014deep} & $(1-\alpha) \mathcal{L}_{\text{CE}} + \alpha \mathcal{L}_{\text{MSE}}(z_s, z_t)$ \\
Cosine~\cite{luo2018cosine} & $(1-\alpha) \mathcal{L}_{\text{CE}} + \alpha \big(1 - \cos(z_s, z_t)\big)$ \\
Feature KD~\cite{romero2014fitnets} & $(1-\alpha-\beta) \mathcal{L}_{\text{CE}} + \alpha T^2 D_{\text{KL}}(p_t^T \parallel p_s^T) + \beta \mathcal{L}_{\text{MSE}}\big(\widehat{\phi(f_s)}, \hat{f}_t\big)$ \\
Attention~\cite{zagoruyko2016paying} & $(1-\alpha-\beta) \mathcal{L}_{\text{CE}} + \alpha T^2 D_{\text{KL}}(p_t^T \parallel p_s^T) + \beta \mathcal{L}_{\text{MSE}}(A_s, A_t)$ \\
SP~\cite{tung2019similarity} & $(1-\alpha-\beta) \mathcal{L}_{\text{CE}} + \alpha T^2 D_{\text{KL}}(p_t^T \parallel p_s^T) + \beta \lVert G_s - G_t \rVert_F^2$ \\
RKD~\cite{park2019relational} & $(1-\alpha-\beta) \mathcal{L}_{\text{CE}} + \alpha T^2 D_{\text{KL}}(p_t^T \parallel p_s^T) + \beta (\psi_D + \psi_A)$ \\
CRD~\cite{tian2019contrastive} & $(1-\alpha) \mathcal{L}_{\text{CE}} + \alpha \mathcal{L}_{\text{CRD}}(f_s, f_t)$ \\
DKD~\cite{zhao2022decoupled} & $\mathcal{L}_{\text{CE}} + \alpha_{\text{D}}\,\text{TCKD} + \beta_{\text{D}}\,\text{NCKD}$ \\
\bottomrule
\end{tabular*}
\end{table}

\smallskip\noindent\textbf{Two-stage pipeline and diagnostic controls.}
The combined pipeline runs the two stages in sequence (Algorithm~\ref{alg:pipeline}). Stage~1 trains a federated teacher under the label-skew partition and keeps its best checkpoint. Stage~2 freezes that teacher and distills the compact student on the centralized training split, which acts as a labeled server-side proxy, using one of the distillation objectives. Because the proxy split is labeled, the hard-label cross-entropy term, where it is present, can drive the student directly from the labels rather than from the teacher, which is the mechanism behind the recovery illusion analyzed in Section~4.4 of the main paper. To separate genuine teacher-to-student transfer from this label-driven recovery, the benchmark implements two controls that leave the teacher as the only learning signal. The no-cross-entropy control removes the hard-label term (equivalently, it sets the soft-target weight to $\alpha{=}1$ for the logit objectives), distilling on the same proxy inputs without using their labels. The unlabeled-proxy control withholds the proxy labels entirely, so the student is supervised only by the teacher's outputs on the proxy inputs. Under either control a faithful pipeline should fall back toward the teacher's own standalone accuracy, and any accuracy well above it signals label-driven masking rather than transfer.

\begin{algorithm}[t]
\caption{Two-stage federated-then-distillation pipeline with the two diagnostic controls. The default mode distills on the labeled proxy split; the two controls remove the label signal so that the teacher is the only supervision and the measured accuracy reflects genuine transfer.}
\label{alg:pipeline}
\hrule\vspace{3pt}
\begin{algsteps}
\item \algkw{Input:} client shards $\mathcal{D}_1,\dots,\mathcal{D}_K$; proxy split $\mathcal{P}$; distillation objective $\mathcal{L}$; mode $\in\{$labeled, no-CE, unlabeled$\}$
\item \algkw{Stage 1.} run Algorithm~\ref{alg:fl} for $R$ rounds and keep the best teacher checkpoint $\theta_T$; then freeze $\theta_T$
\item \algkw{Stage 2.} initialize the compact student $\theta_S$
\item \algind set the hard-label weight $w_{\text{CE}}$ to the objective's default \algkw{if} mode is labeled, \algkw{else} $0$ \algcom{both controls drop $\mathcal{L}_{\text{CE}}$}
\item \algind \algkw{for} $200$ epochs over the proxy split $\mathcal{P}$ \algkw{do}
\item \algind\algind \algkw{for} each batch $x\subset\mathcal{P}$ \algkw{do}
\item \algind\algind\algind $y \leftarrow$ proxy labels of $x$ \algkw{if} mode is not unlabeled, \algkw{else} none
\item \algind\algind\algind read the frozen teacher's outputs $\theta_T(x)$
\item \algind\algind\algind update $\theta_S$ by minimizing $\mathcal{L}$ on $\theta_T(x)$ and, weighted by $w_{\text{CE}}$, on $y$
\item \algind\algind \algkw{end for}
\item \algind \algkw{end for}
\item evaluate $\theta_S$ on the test set and keep the best checkpoint
\item \algkw{return} the distilled student $\theta_S$
\end{algsteps}
\vspace{3pt}\hrule
\end{algorithm}

\smallskip\noindent\textbf{Evaluation protocol.}
For every run we record the best test instance and mean-class accuracy over training, together with the model size (parameters plus buffers), the inference latency of a held-out evaluation batch on a single GPU, and, for federated runs, the per-round communication payload and wall-clock time. Accuracy is computed in fp32 from a full confusion matrix, from which we read both the overall instance accuracy and the macro per-class accuracy (the mean of the per-class recalls). On the balanced clinical test set ($20$ shapes per class) the macro class accuracy coincides with the instance accuracy, which is why the two columns are equal there. The label-skew partition is deterministic given the dataset, so the only source of run-to-run variation is the random seed, which controls the model initialization, the mini-batch ordering, and the augmentation sampling.

\smallskip\noindent\textbf{Training environment and footprint.}
All runs use a single shared implementation and are trained on a single GPU per run in fp32. Reported model footprints are the combined size of the parameter and buffer tensors at fp32, which is why they grow slightly with the width of the output layer: the teacher occupies $5.65$\,MB on the $40$-class ModelNet40 and $5.62$\,MB on the four-class clinical data, and the compact student occupies $1.44$\,MB and $1.42$\,MB respectively, a reduction of $74.51\%$ and $74.7\%$. Inference latency is reported as the mean over $100$ timed forward passes on a single held-out evaluation batch, taken after $10$ warm-up passes with on-device timing, and is therefore subject to hardware-load variation. Communication cost is identical across strategies because each transmits the full model, and the per-round payloads and wall-clock times are listed in \Cref{tab:fl_compute}.

\smallskip\noindent\textbf{Benchmark scope.}
The benchmark comprises $504$ training runs. On ModelNet40 we run a centralized baseline and the thirteen FL algorithms ($14$ configurations). On the clinical dataset we run the centralized baseline, the thirteen FL algorithms, the ten KD objectives, and the full combined grid of thirteen FL teachers crossed with ten KD objectives ($1\,{+}\,13\,{+}\,10\,{+}\,130 = 154$ configurations). Each of the $168$ configurations is repeated over the three seeds $\{7, 42, 123\}$, giving $42$ runs on ModelNet40 and $462$ on the clinical dataset. The standalone distillation and combined-pipeline study on ModelNet40 reported in the main paper comes from a separate focused round that predates this standardized multi-seed protocol and is not counted among the $504$ runs; it is retained because it spans federated teachers across the full quality range, down to collapsed checkpoints, which is what makes the recovery illusion measurable. For completeness we also ran the full multi-seed ModelNet40 combined grid (\Cref{app:mn40grid}); like the additional-dataset runs of \Cref{app:moredata}, it likewise lies outside the $504$-run count.

\smallskip\noindent\textbf{Reproducibility and extensibility.}
The study is built to be reproduced and extended. Data access goes through a single dataset-agnostic interface that already covers six point-cloud sources (ModelNet40~\cite{wu20153d}, ModelNet10~\cite{wu20153d}, OmniObject3D~\cite{wu2023omniobject3d}, YCB~\cite{calli2015ycb}, GazeboSim~\cite{downs2022google}, and the clinical craniosynostosis dataset~\cite{schaufelberger2021statistical}) with stratified per-class splitting and automatic handling of under-populated classes, so a new dataset can be added without changing the training protocol. Beyond the reported configuration, the framework provides a heterogeneous compact DGCNN student (about $1.5$\,MB) as an alternative to the same-family PointNet++ student, a multi-teacher distillation objective in addition to the ten benchmarked here, a sweep over the distillation temperature $T\in\{1,2,4,8\}$, and Dirichlet label partitions that interpolate between the worst-case split studied here and milder heterogeneity. The two diagnostic controls of Algorithm~\ref{alg:pipeline} are first-class options, so the recommended label-free evaluation can be reproduced directly. Standalone part- and semantic-segmentation baselines are also included for completeness, though they fall outside the federated-distillation scope of this study. Finally, every training job checkpoints and resumes idempotently and the full grid is enumerated automatically, which is how the multi-seed benchmark of $504$ runs was executed at scale; this design lets the community add new datasets, federated algorithms, distillation objectives, and student architectures through the same shared interface.

\smallskip\noindent\textbf{Project page, model checkpoints, and community.}
The full benchmark, comprising the source code, configuration files, and step-by-step instructions for reproducing every one of the $504$ runs, is hosted at the project page, \url{https://ezharjan.github.io/FLKD3DBenchmark}. The complete set of trained model checkpoints, namely the saved network weights produced by every centralized, federated, distilled, and combined configuration in this study, is openly archived in the accompanying model repository at \url{https://huggingface.co/ezharjan/FLKD3DBenchmark}, so that any reported result can be reloaded and verified directly without retraining. Our broader aim is to offer the community a shared and openly extensible foundation for privacy-preserving, edge-deployable 3D point cloud learning, one in which new federated algorithms, distillation objectives, datasets, and student architectures can be contributed through a single common interface and compared under the same label-free evaluation protocol we advocate here. We regard this not as a finished artifact but as an ongoing endeavour: together we can build benchmarks that make 3D vision trustworthy to deploy wherever data cannot be centralized. Contributions, extensions, and critiques are warmly welcomed, and issues, pull requests, and new benchmark submissions are all encouraged.

% =====================================================================
\section{Clinical Craniosynostosis Study}
\label{app:clinical}

Craniosynostosis is the premature fusion of cranial sutures, which alters head shape in characteristic ways; statistical shape modeling has long been used to assess it from 3D surfaces~\cite{mendoza2014personalized}. Schaufelberger \etal~\cite{schaufelberger2021statistical} released a statistical shape model of patients together with model-generated instances for the principal suture-fusion types and a control group, a privacy-conscious source of patient-like head shapes that is naturally distributed across clinics and so motivates both federated training and compact models for point-of-care devices.

This appendix details the standalone clinical distillation behind the real-world validation of Section~4.5 of the main paper, together with the complete numerical tables. All clinical results use the standardized three-seed protocol over the full cross-product of thirteen FL teachers and ten KD objectives, and the $100\%$ centralized ceiling makes teacher quality unambiguous. The clinical federated convergence is plotted alongside ModelNet40 in \Cref{fig:fl_convergence}, and the standalone FL accuracies are listed in Table~1 of the main paper and \Cref{tab:fl_full}.

\subsection{Standalone Knowledge Distillation}

\begin{figure}[tb]
\centering
\includegraphics[width=\linewidth]{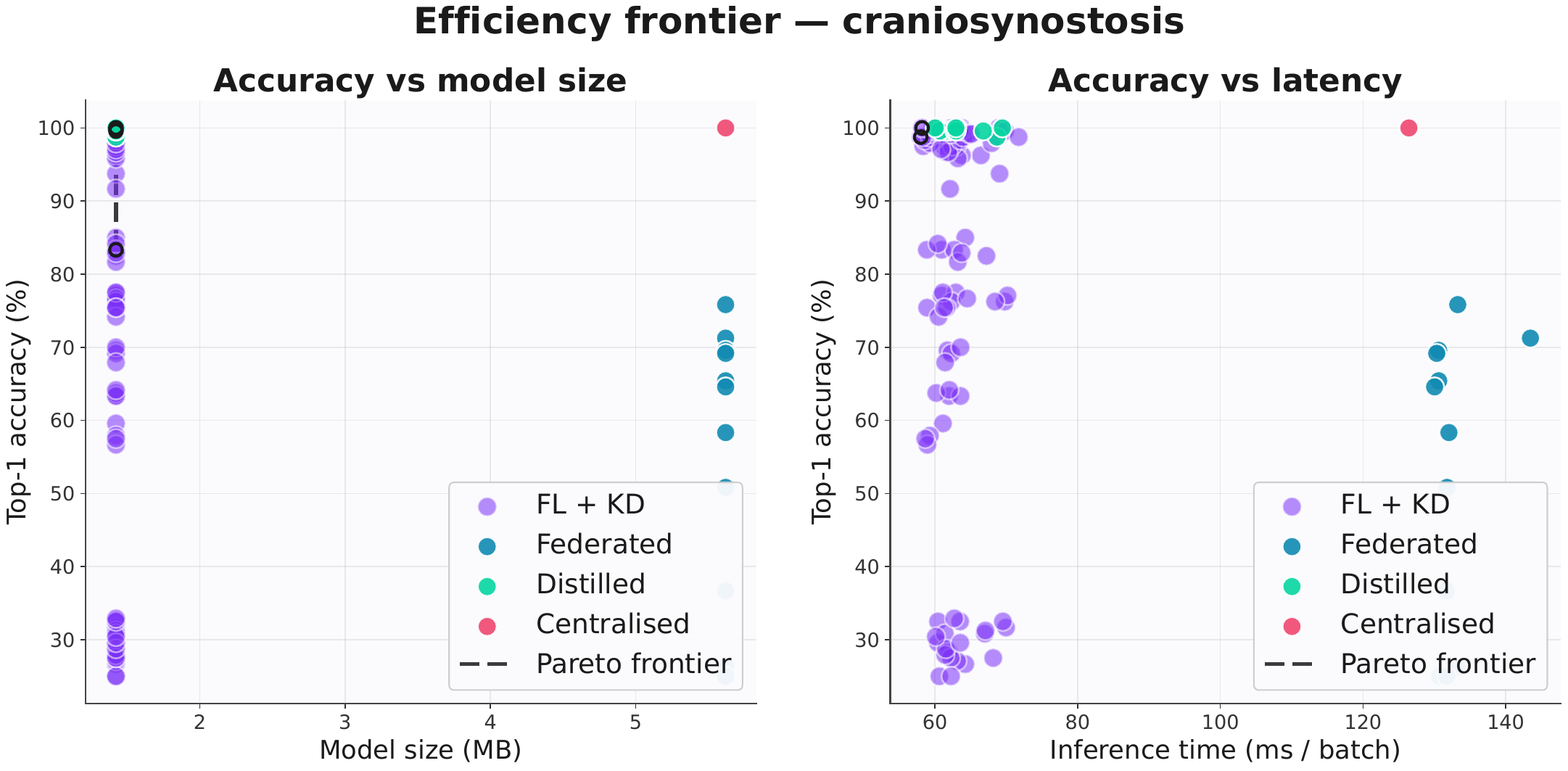}
\caption{\textbf{Efficiency of distillation on the clinical task.}
Test accuracy versus model size \textbf{(left)} and versus inference latency
\textbf{(right)}, colored by family. The compact KD students cluster at the top
left, near the teacher's accuracy at $1.42$\,MB and about $60$ to $69$\,ms per
evaluation batch, against the $5.62$\,MB, ${\approx}126$\,ms teacher. The
standalone FL models keep the teacher's size and latency yet span the full
accuracy range, and the combined students inherit the compact size while their
accuracy spreads according to objective (Section~4.4 of the main paper).}
\label{fig:efficiency_clinical}
\end{figure}

\begin{figure}[tb]
\centering
\includegraphics[width=0.99\linewidth]{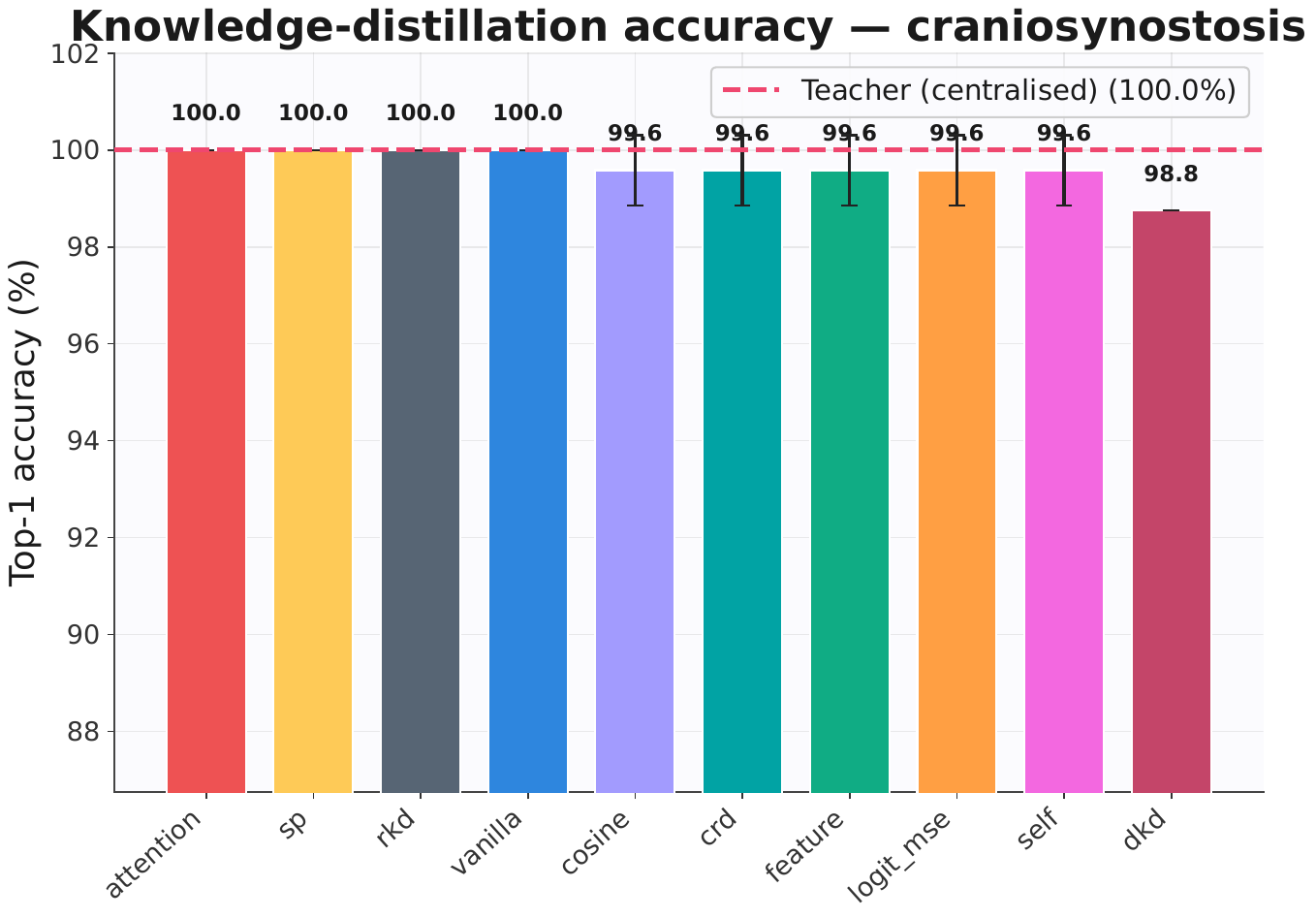}
\caption{\textbf{Standalone distillation accuracy on the clinical task.}
Test instance accuracy of the ten KD objectives distilling the centralized
($100\%$) teacher into the compact student (mean$\pm$std over three seeds); the
dashed line marks the teacher. Four objectives reach $100\%$, five reach
$99.58\%$, and Decoupled~KD trails at $98.75\%$, so every objective lands within
$1.25$ points of the teacher. This is the bar-chart view of \Cref{tab:kd_results_clinical}.}
\label{fig:kd_ranking_clinical}
\end{figure}

\begin{figure}[tb]
\centering
\includegraphics[width=\linewidth]{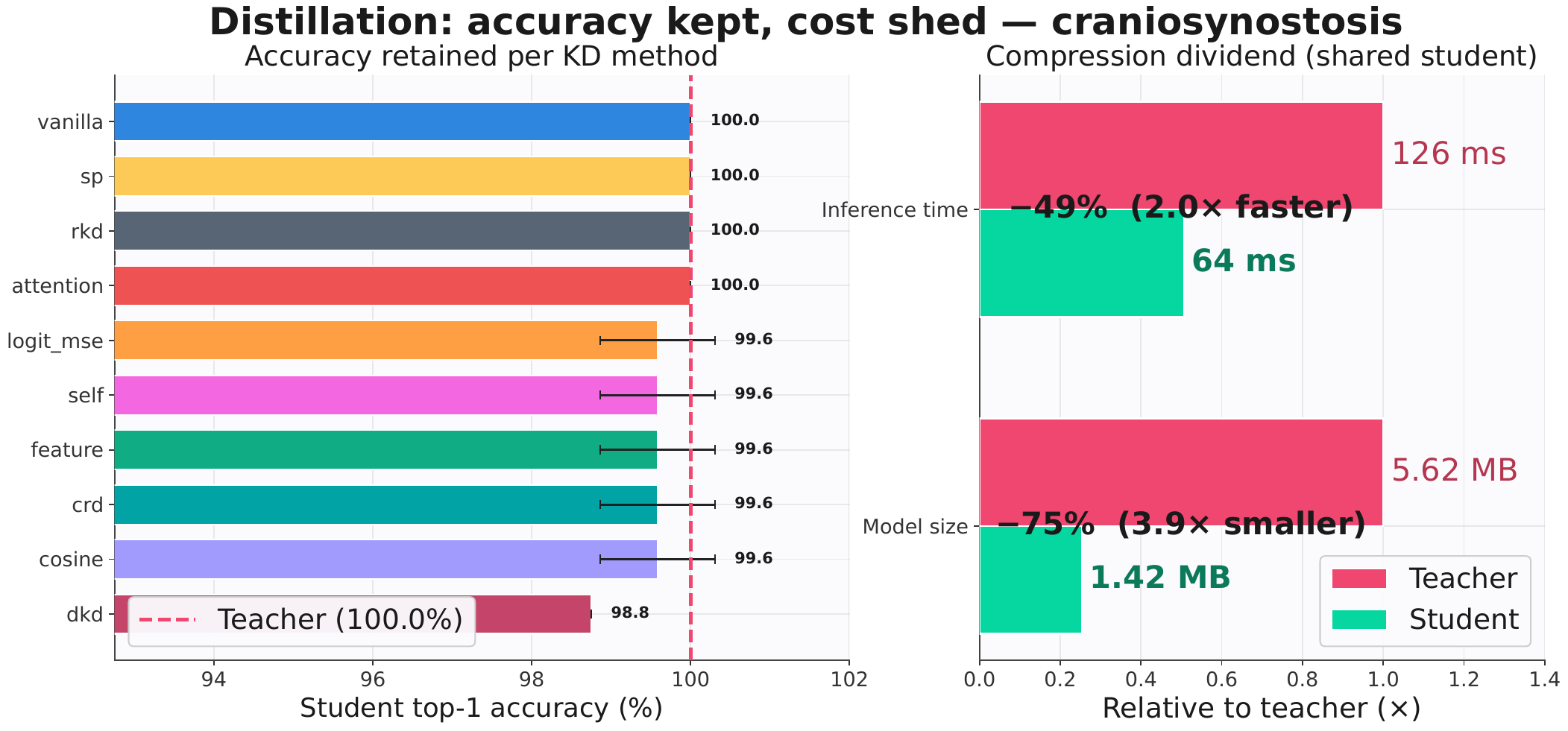}
\caption{\textbf{Distillation keeps accuracy while shedding cost (clinical task).}
\textbf{(left)} Accuracy retained by each objective, all within $1.25$ points of
the $100\%$ teacher. \textbf{(right)} The compression dividend of the shared
compact student: model size falls from $5.62$ to $1.42$\,MB (a $74.7\%$
reduction, $3.9\times$ smaller) and mean inference latency from about $126$ to
$64$\,ms ($-49\%$, $2.0\times$ faster). Latencies are wall-clock averages on a
single GPU and are subject to hardware-load variation.}
\label{fig:kd_compression_clinical}
\end{figure}

\Cref{tab:kd_results_clinical} reports accuracy for each of the ten distillation objectives on the clinical task; \Cref{fig:kd_ranking_clinical} shows the same ranking as a bar chart, \Cref{fig:efficiency_clinical} relates accuracy to model size and inference latency, and \Cref{fig:kd_compression_clinical} isolates the compression dividend. Per-seed values are in \Cref{tab:kd_full}.

\begin{table}[tb]
\caption{KD instance accuracy (\%, best per run, mean$\pm$std over three seeds) on the clinical task, for the compact student distilled from the PointNet++ SSG teacher. Methods are ordered by accuracy. All ten objectives transfer the teacher's accuracy into the compact student with at most a $1.25$-point loss. On the balanced test set the mean-class accuracy equals the instance accuracy.}
\label{tab:kd_results_clinical}
\centering
\small
\setlength{\tabcolsep}{8pt}
\begin{tabular}{@{}lc@{}}
\toprule
Strategy & Craniosynostosis $\uparrow$ \\
\midrule
Teacher (PointNet++ SSG)~\cite{pointnet2} & $100.00$\sd{0.00} \\
\midrule
Attention Transfer~\cite{zagoruyko2016paying} & $\mathbf{100.00}$\sd{0.00} \\
Relational KD~\cite{park2019relational} & $\mathbf{100.00}$\sd{0.00} \\
Similarity-Preserving~\cite{tung2019similarity} & $\mathbf{100.00}$\sd{0.00} \\
Vanilla KD~\cite{hinton2015distilling} & $\mathbf{100.00}$\sd{0.00} \\
Cosine Similarity~\cite{luo2018cosine} & $99.58$\sd{0.72} \\
Contrastive (CRD)~\cite{tian2019contrastive} & $99.58$\sd{0.72} \\
Feature KD~\cite{romero2014fitnets} & $99.58$\sd{0.72} \\
Logit-MSE~\cite{ba2014deep} & $99.58$\sd{0.72} \\
Self-Distillation~\cite{furlanello2018born} & $99.58$\sd{0.72} \\
Decoupled KD~\cite{zhao2022decoupled} & $98.75$\sd{0.00} \\
\bottomrule
\end{tabular}
\end{table}

When the teacher is the strong centralized model, distillation is a dependable compressor. All ten objectives land within $1.25$ points of the $100\%$ teacher on the compact student: Attention Transfer, Relational~KD, Similarity-Preserving distillation, and Vanilla~KD reach $100\%$ exactly, five further objectives reach $99.58\%$, and DKD trails marginally at $98.75\%$. The compression is uniform across objectives because they share the same teacher and student: the student is $74.7\%$ smaller (from $5.62$ to $1.42$\,MB) and roughly twice as fast at inference, requiring about $64$\,ms per held-out evaluation batch on average against the teacher's $126$\,ms, close to a halving of latency. These timings are wall-clock averages on a single GPU and are subject to hardware-load variation. The near-uniform success here is exactly what makes the standalone KD result a poor probe of the federated teacher in the combined pipeline: when the supervisory labels alone suffice to reach the ceiling, the distillation objective is barely tested.

\subsection{Combined Federated Learning and Knowledge Distillation Pipeline}

\begin{figure}[tb]
\centering
\includegraphics[width=0.99\linewidth]{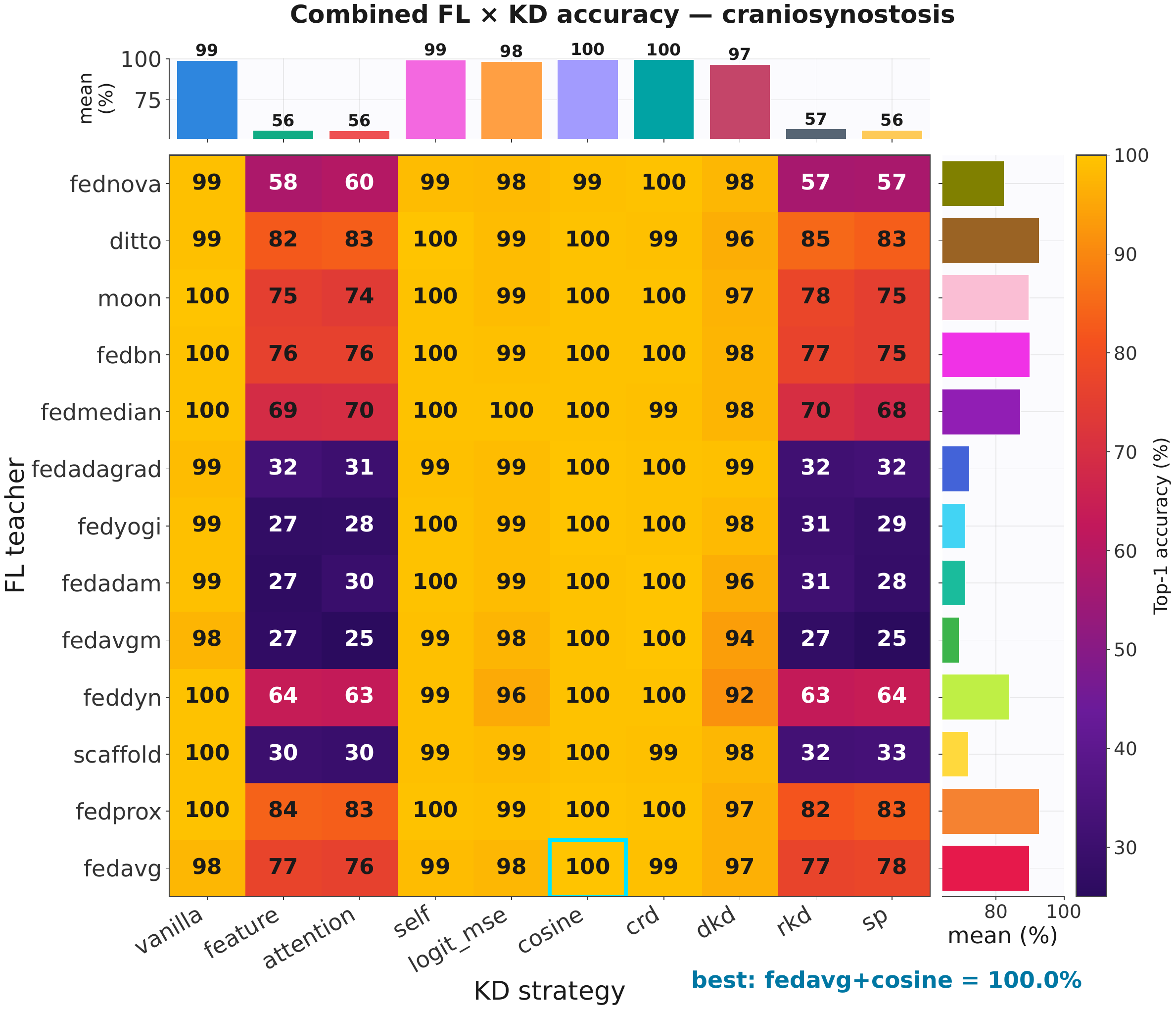}
\caption{\textbf{The recovery illusion and its control on the clinical task.}
Student accuracy for each FL teacher (rows) and KD objective (columns); the bars
above each column give its mean over the thirteen teachers. The six
columns that keep a hard-label cross-entropy term (vanilla, self, logit-MSE,
cosine, CRD, DKD) are uniformly bright: the student recovers to roughly $92\%$ to
$100\%$ even when the teacher is at the $25\%$ chance level. The four
feature-based columns (feature, attention, RKD, SP) instead darken in step with
the teacher, collapsing to near-chance accuracy for the failed server-side optimizers and
recovering only for the better teachers. Both axes follow the benchmark's
native order rather than being grouped, so the hard-label and feature-based
columns are interleaved; \Cref{tab:combined_mean} regroups them.}
\label{fig:combined_heatmap_clinical}
\end{figure}

This subsection reports the full evidence for the two-stage pipeline on the clinical task, in which each of the thirteen federated teachers is frozen after Stage~1 and then distilled into the compact student under each of the ten objectives, forming the complete $13\times10$ grid of $130$ teacher--objective pairs evaluated over three seeds. \Cref{fig:combined_heatmap_clinical} renders the resulting student accuracy as a teacher-by-objective heatmap, and \Cref{tab:combined_mean,tab:combined_std} give the per-cell means and standard deviations behind it, from which the per-objective summary of Table~4 of the main paper is computed; \Cref{tab:seed_combined} lists the individual per-seed values. As the analysis in Section~4.5 of the main paper establishes, the outcome is governed by a single design choice, the weight the objective places on the hard-label cross-entropy term, and the numbers tabulated here make that dependence explicit.

The heatmap separates the ten objectives into two visually distinct groups. The six objectives that keep a cross-entropy term (Vanilla~KD, Self-Distillation, Cosine, CRD, DKD, and Logit-MSE) form a uniformly bright band: the student recovers to between $91.7\%$ and $100\%$ for every teacher, including the server-side optimizers frozen at the $25\%$ four-class chance level. A teacher that learned nothing thus appears to yield a near-perfect student, because the labeled proxy split supplies a supervised signal that bypasses the teacher entirely. The four pure-transfer objectives (Feature~KD, Attention Transfer, RKD, and SP), whose cross-entropy coefficient $(1-\alpha-\beta)$ is exactly $0$ at $\alpha{=}\beta{=}0.5$, instead darken in step with the teacher: they reach the low-to-mid $80\%$ range behind the strongest teacher (FedProx, $75.83\%$ standalone) and fall to near the $25\%$ chance level behind the collapsed server-side optimizers. Because the heatmap lists both axes in the benchmark's native order rather than grouping them, the two kinds of column are interleaved rather than contiguous; \Cref{tab:combined_mean,tab:combined_std} instead group the six hard-label objectives apart from the four pure-transfer ones and order the teachers by descending standalone accuracy, so that the trend along each column can be read directly.

The per-cell numbers make the dichotomy quantitative. Reading down the six hard-label columns of \Cref{tab:combined_mean}, student accuracy stays near the ceiling as the teacher degrades from FedProx ($75.83\%$ standalone) to the chance-level optimizers: across the thirteen teachers it varies by at most $7.5$ points (decoupled KD) and by under $4$ points for the other five, and its correlation with teacher quality is negligible (Pearson $r$ between $-0.53$ and $+0.37$ in Table~4 of the main paper). The four pure-transfer columns behave oppositely, spanning $57.5$ to $58.3$ points from the best to the worst teacher and tracking standalone teacher accuracy almost perfectly ($r$ from $0.986$ to $0.990$). Two cells make the contrast concrete: behind FedAvgM, a teacher stuck at the $25\%$ chance level, Vanilla~KD still records a $97.5\%$ student and CRD a $100\%$ student, whereas Attention Transfer and SP report $25.0\%$, exactly the teacher's own level. The standard deviations in \Cref{tab:combined_std} reinforce this reading: the hard-label cells are tight across seeds (typically under $1.5$ points), whereas the pure-transfer cells reach standard deviations above $13$ points, because each seed retrains the federated teacher and the pure-transfer student inherits that variation instead of being anchored by the fixed proxy labels. \Cref{tab:seed_combined} exposes the same pattern run by run, confirming that the hard-label recovery holds on every individual seed while the pure-transfer accuracy rises and falls with the teacher checkpoint. The clinical grid therefore reproduces, on real patient-derived shapes and at full multi-seed scale, the recovery illusion isolated under controlled conditions: hard-label distillation reports a near-ceiling student from a federated teacher that never learned the task, while silently reusing the patient labels that federation is meant to protect, so that only the label-free objectives report the teacher's true quality.

\begin{figure}[tb]
\centering
\includegraphics[width=0.99\linewidth]{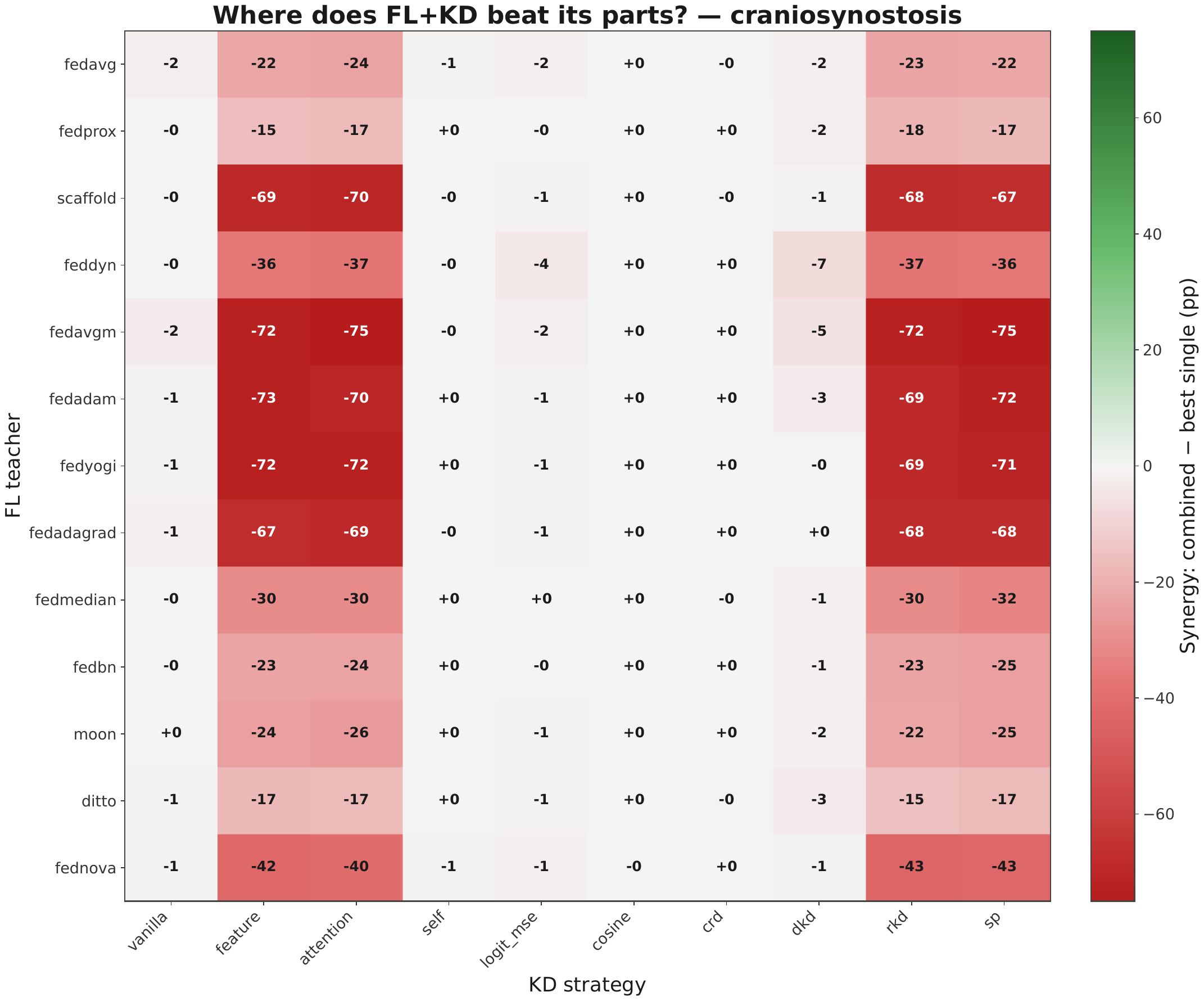}
\caption{\textbf{The combined pipeline shows no positive synergy (clinical task).}
For every FL teacher (rows) and KD objective (columns), the cell gives the combined student's instance accuracy minus the better of its two stand-alone components (the federated teacher alone and the student distilled from the centralized teacher), in percentage points (mean over three seeds). Values at or below zero mean the two-stage pipeline does not improve on simply distilling from the centralized teacher. The six hard-label objectives (vanilla, self, logit-MSE, cosine, CRD, DKD) stay close to zero, reproducing centralized distillation through the proxy labels, whereas the four pure-transfer objectives (feature, attention, RKD, SP) fall up to roughly $75$ points below their best component for the collapsed server-side optimizers.}
\label{fig:synergy_clinical}
\end{figure}

\begin{figure}[tb]
\centering
\includegraphics[width=0.99\linewidth]{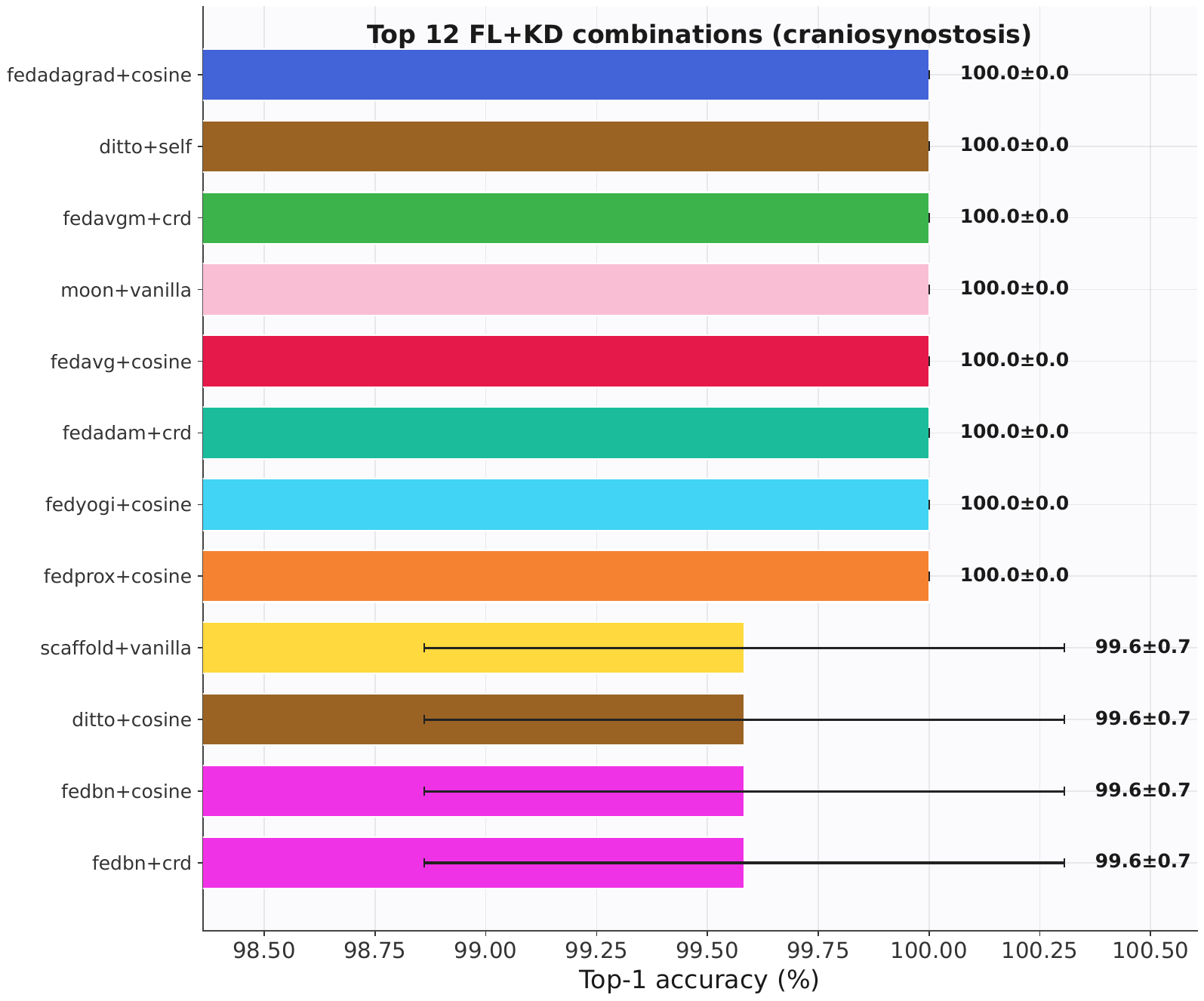}
\caption{\textbf{The leaderboard rewards collapsed teachers (clinical task).}
The twelve highest-accuracy FL$+$KD combinations on the clinical task, each as mean$\pm$std over three seeds. All exceed $99.5\%$, yet several pair a hard-label objective (cosine, CRD, vanilla, self) with a federated teacher frozen near the $25\%$ four-class chance level (FedAdagrad, FedAvgM, FedAdam, FedYogi), so the ranking reflects the proxy labels rather than the quality of the federated teacher.}
\label{fig:best_combos_clinical}
\end{figure}

\smallskip\noindent\textbf{No positive synergy, and a leaderboard that rewards collapse.}
Two further views sharpen the same conclusion. \Cref{fig:synergy_clinical} reports, for every teacher--objective cell, the combined student's accuracy minus the better of its two components in isolation (the federated teacher on its own and the student distilled directly from the centralized teacher). No configuration improves on its best single component by more than about half a point. The six hard-label objectives sit essentially at zero: the two-stage pipeline merely reproduces what centralized distillation already delivers through the proxy labels, adding no transfer of its own. The four pure-transfer objectives are sharply negative, falling as much as $75$ points below their best component for the collapsed server-side optimizers, because once the hard-label term is removed the failed teacher is the only remaining signal. The combined pipeline therefore never synthesizes a student better than its parts, and the apparent gains under hard-label distillation are an artifact of the labeled proxy rather than genuine synergy. \Cref{fig:best_combos_clinical} makes the same point from the top of the ranking: of the twelve highest-scoring teacher--objective combinations, four pair a hard-label objective with a server-side optimizer frozen near the $25\%$ four-class chance level (FedAdagrad$+$Cosine, FedAvgM$+$CRD, FedAdam$+$CRD, and FedYogi$+$Cosine), each reaching $100\%$. A leaderboard built on hard-label distillation would therefore crown federated teachers that never learned the task.

\clearpage
% =====================================================================
\section{Complete Numerical Results}
\label{app:tables}

This appendix collects the complete per-configuration numbers behind every summary in the main paper, so that each reported value can be traced to its source. It is organised in three movements: the standalone federated results, then the clinical combined pipeline, and finally the ModelNet40 counterpart.

\begin{figure}[t]
\centering
\begin{subfigure}{0.49\textwidth}\centering
\includegraphics[width=\linewidth]{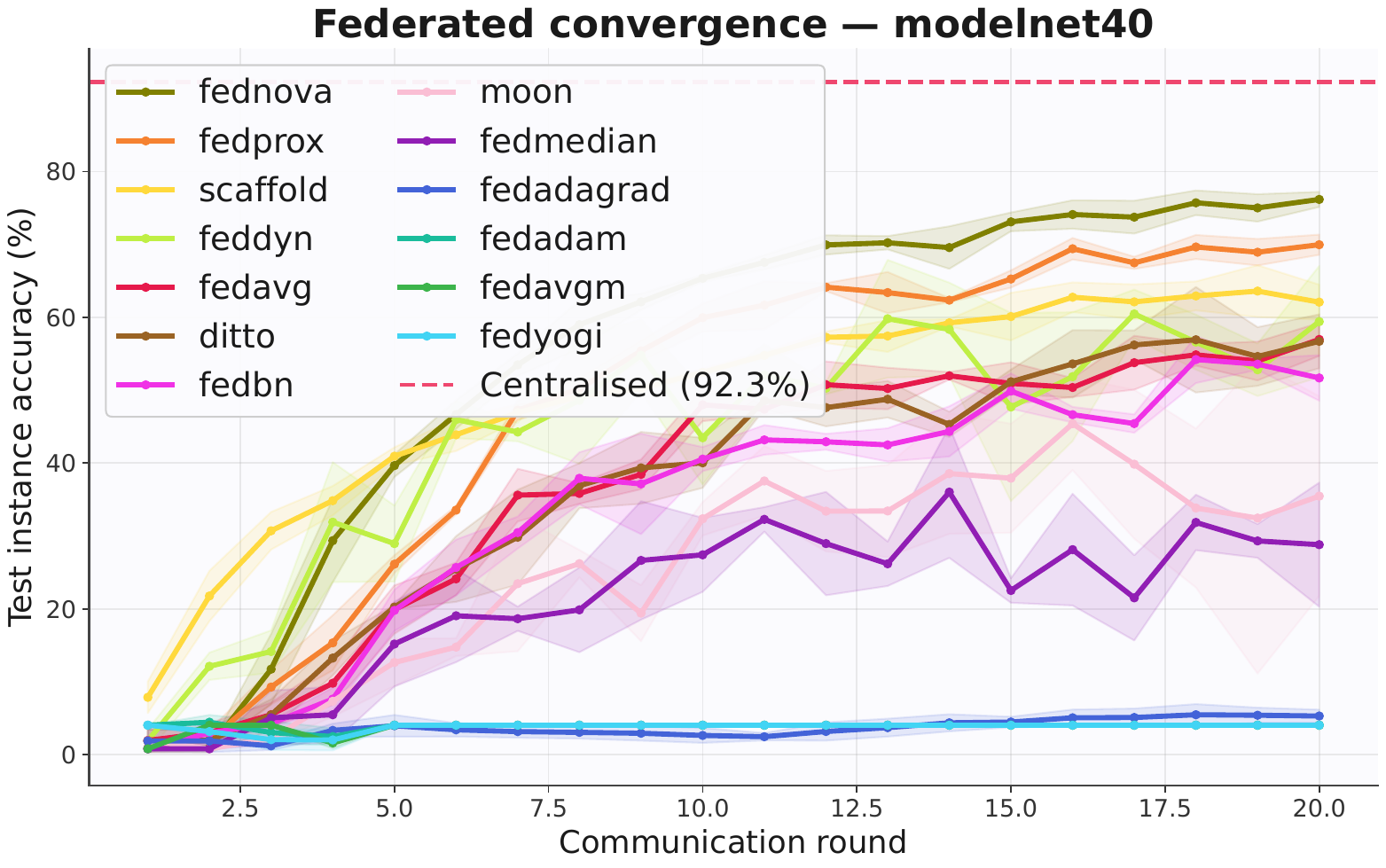}
\caption{ModelNet40}
\label{fig:fl_convergence_mn40}
\end{subfigure}\hfill
\begin{subfigure}{0.49\textwidth}\centering
\includegraphics[width=\linewidth]{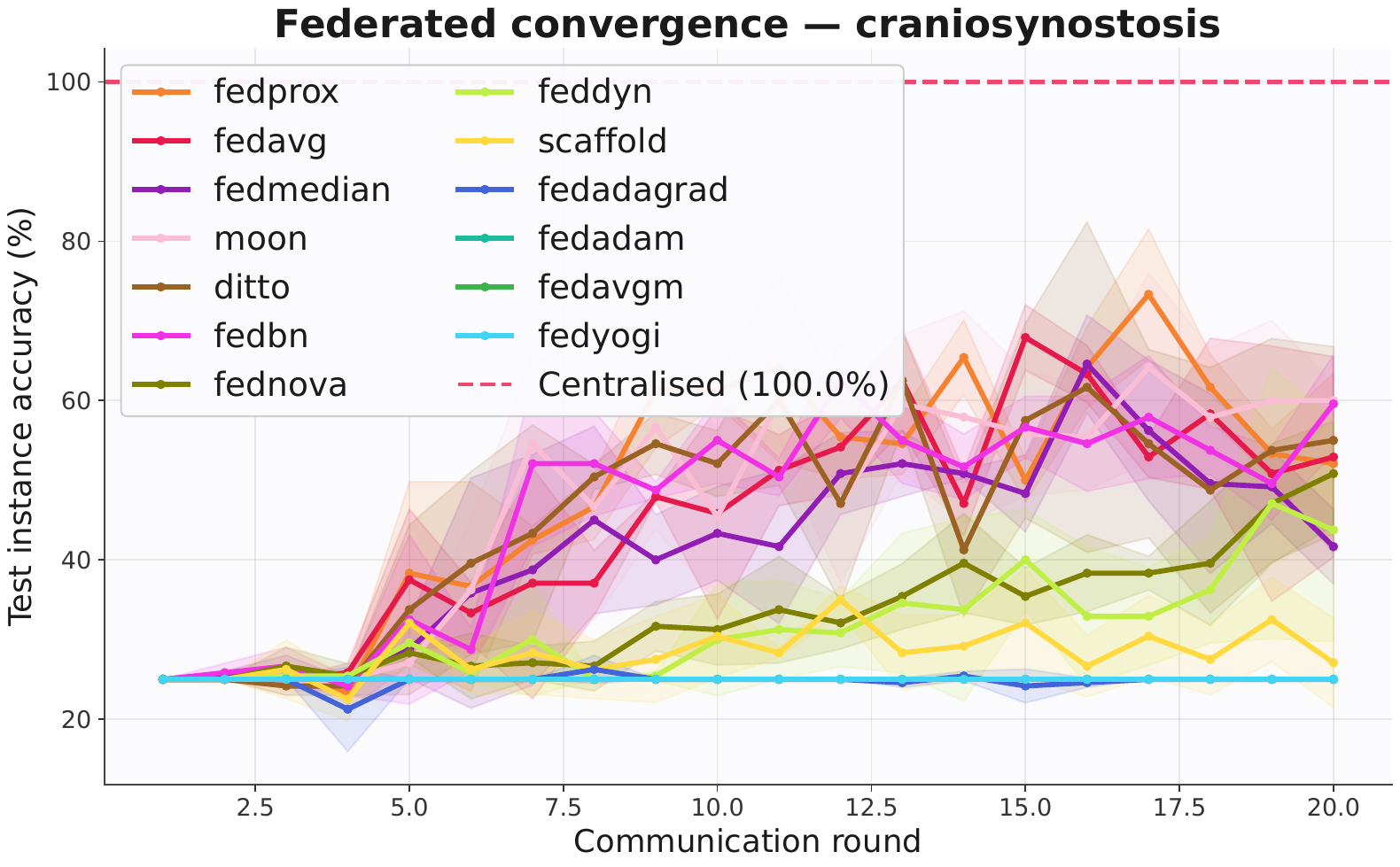}
\caption{Craniosynostosis}
\label{fig:fl_convergence_cranio}
\end{subfigure}
\caption{\textbf{FL convergence under the non-IID label-skew partition:} per-round test instance accuracy (mean over three seeds) for all thirteen algorithms; the dashed line marks the centralized reference. \textbf{(a)} On ModelNet40, FedNova and FedProx lead, a broad middle band of FedDyn, SCAFFOLD, Ditto, FedAvg, FedBN, and MOON follows, FedMedian trails near $39\%$, and the four server-side optimizers stay near the chance level. \textbf{(b)} On the clinical task, FedProx leads, a broad middle group follows, SCAFFOLD stalls near $37\%$, and the four server-side optimizers stay at the $25\%$ four-class chance level; the trajectories are noisier because the test set has only $80$ shapes. Each legend is ordered by final accuracy.}
\label{fig:fl_convergence}
\end{figure}

\begin{figure}[t]
\centering
\begin{subfigure}{0.49\textwidth}\centering
\includegraphics[width=\linewidth]{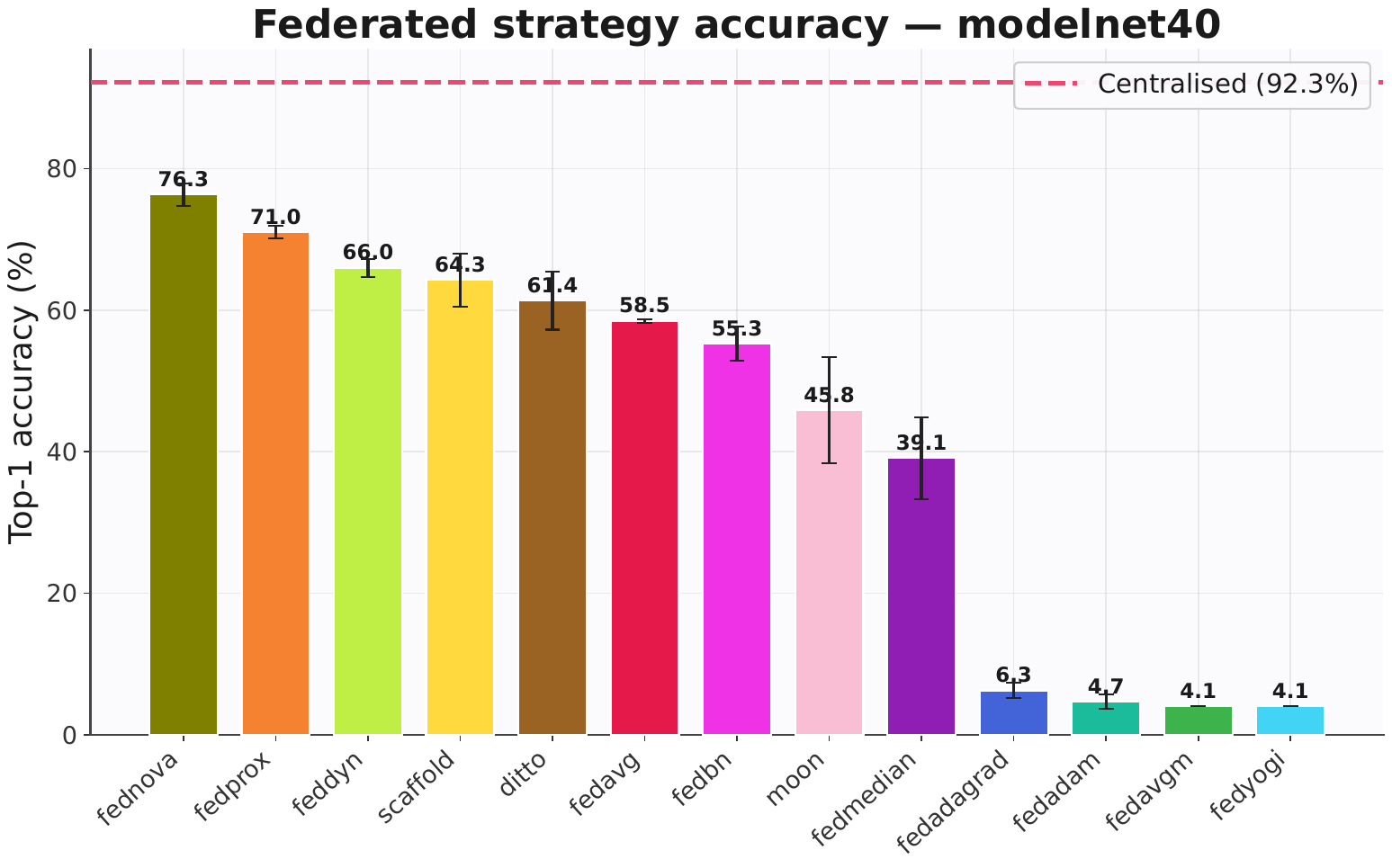}
\caption{ModelNet40}
\label{fig:fl_ranking_mn40}
\end{subfigure}\hfill
\begin{subfigure}{0.49\textwidth}\centering
\includegraphics[width=\linewidth]{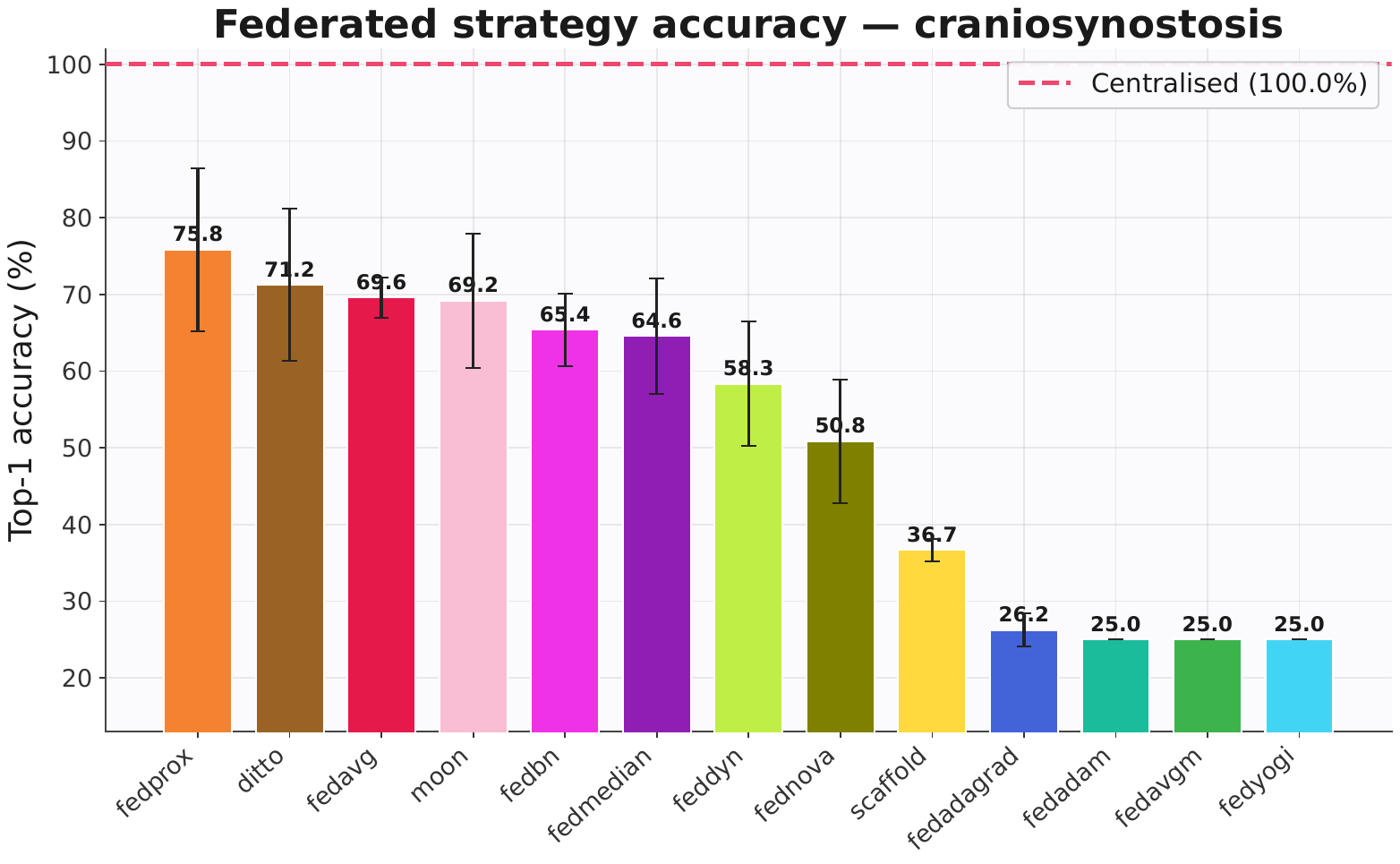}
\caption{Craniosynostosis}
\label{fig:fl_ranking_cranio}
\end{subfigure}
\caption{\textbf{Federated strategy accuracy ranked (best per run, mean$\pm$std over three seeds).}
Bar-chart view of \Cref{tab:fl_full}; the dashed line marks the centralized reference.
\textbf{(a)} On ModelNet40 the order is FedNova, FedProx, FedDyn, SCAFFOLD, Ditto, FedAvg, FedBN, MOON, FedMedian, then the four server-side optimizers near the chance level.
\textbf{(b)} On the clinical task the order changes to FedProx, Ditto, FedAvg, MOON, FedBN, FedMedian, FedDyn, FedNova, SCAFFOLD, then the four server-side optimizers at the $25\%$ chance level. The reordering between panels is the dataset-dependent ranking discussed in Section~4.2 of the main paper.}
\label{fig:fl_ranking}
\end{figure}

\begin{figure}[t]
\centering
\begin{subfigure}{0.49\textwidth}\centering
\includegraphics[width=\linewidth]{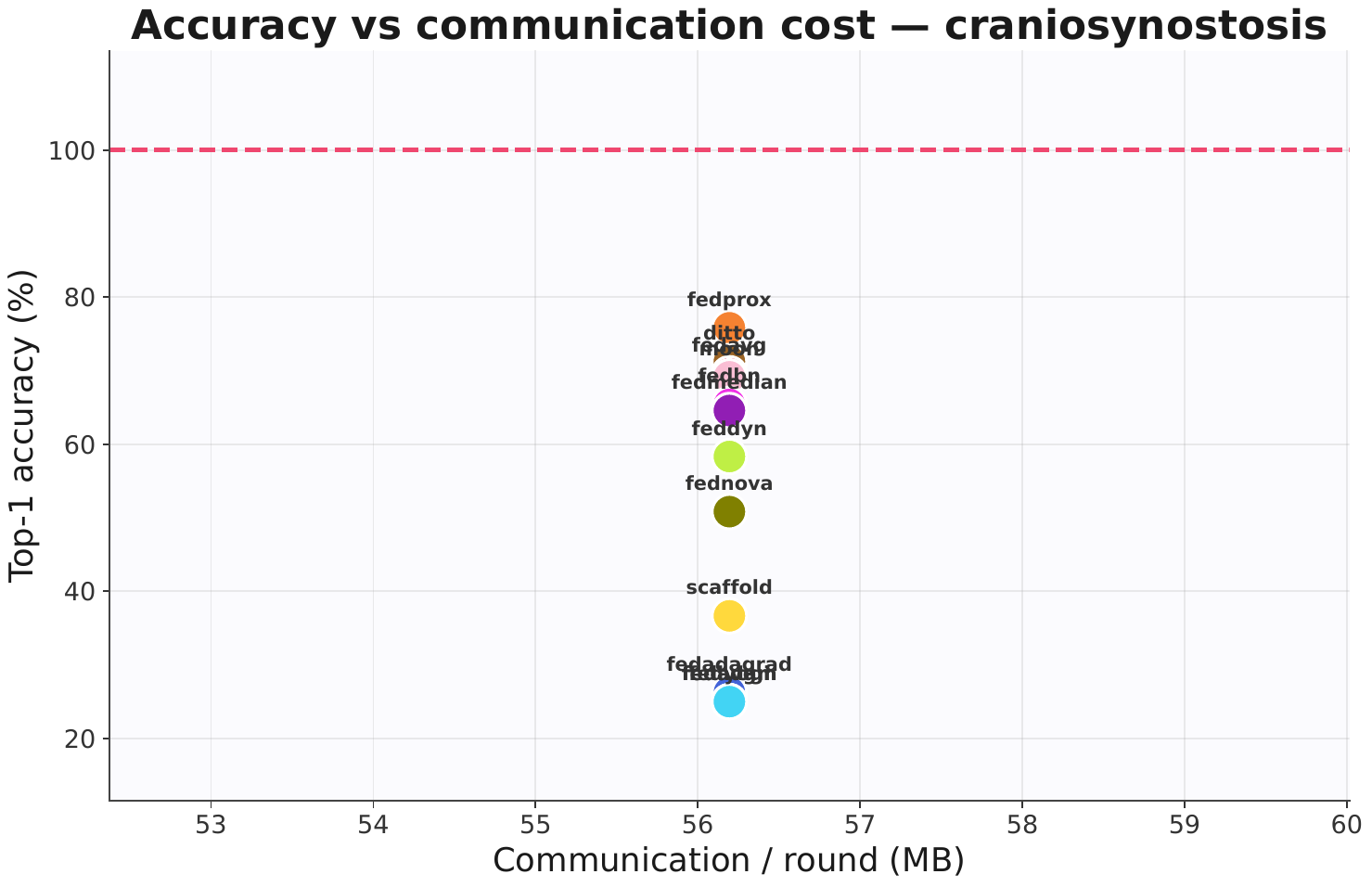}
\caption{Craniosynostosis}
\label{fig:fl_comm_cranio}
\end{subfigure}\hfill
\begin{subfigure}{0.49\textwidth}\centering
\includegraphics[width=\linewidth]{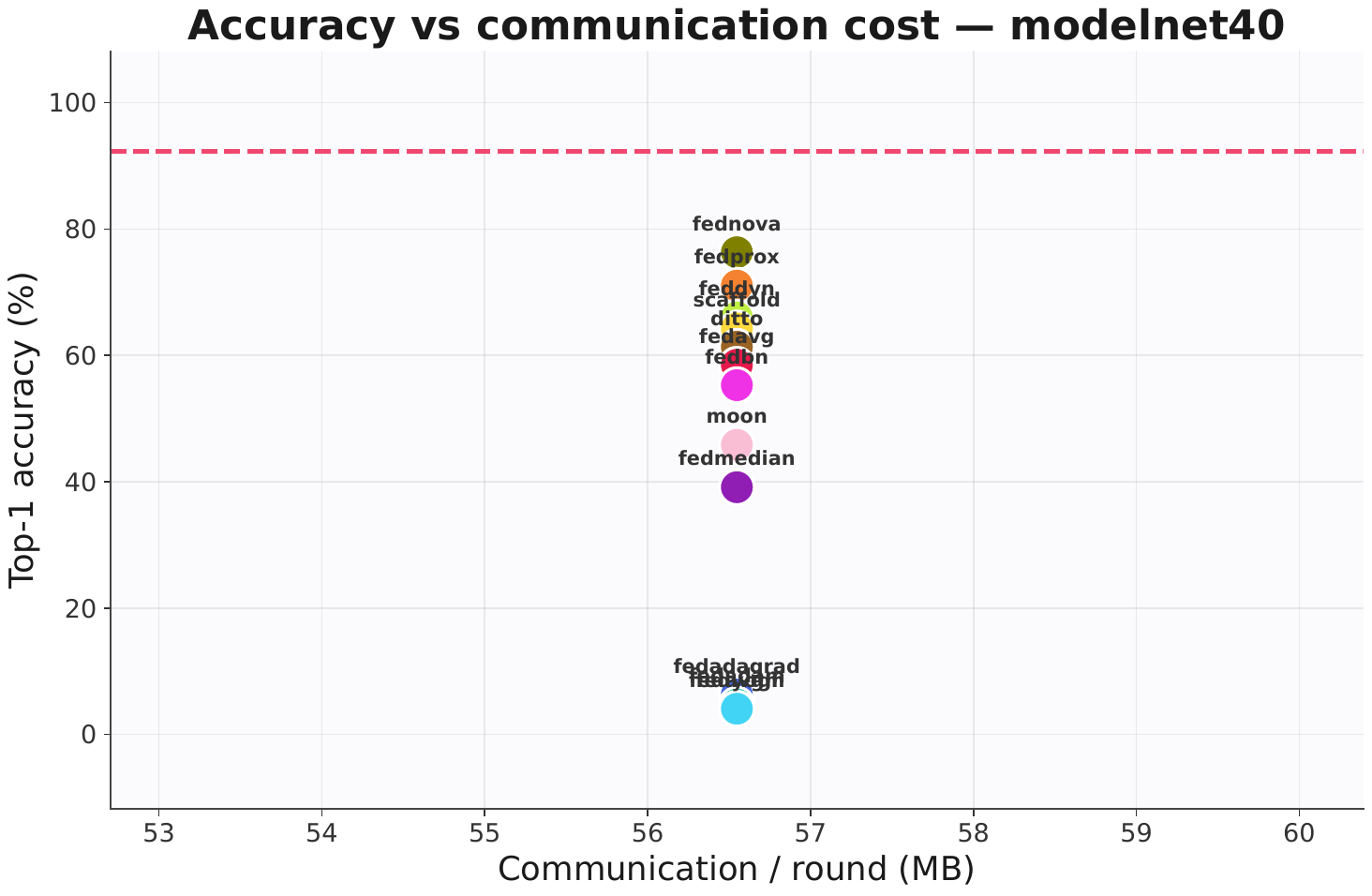}
\caption{ModelNet40}
\label{fig:fl_comm_mn40}
\end{subfigure}
\caption{\textbf{Accuracy versus per-round communication cost.}
Best-checkpoint test accuracy against the per-round communication payload (MB, summed over the five clients' uplink and downlink). Because every strategy transmits the full model, the per-round payload is identical across methods (about $56.2$\,MB on the clinical task and $56.5$\,MB on ModelNet40; \Cref{tab:fl_compute}), so all strategies share a single vertical line and accuracy varies only along it: higher accuracy is not bought with more communication. The dashed line marks the centralized reference. \textbf{(a)} On the clinical task the strategies spread from FedProx at the top down to the four server-side optimizers at the $25\%$ chance level. \textbf{(b)} On ModelNet40 they spread from FedNova down to the four server-side optimizers near the floor.}
\label{fig:fl_comm_pareto}
\end{figure}

The standalone federated results open the appendix. \Cref{tab:fl_full} extends the comparison of Table~1 of the main paper with mean-class accuracy on both datasets and confirms that the instance-accuracy ranking survives under the class-balanced metric. Two figures then make those same endpoints legible at a glance. \Cref{fig:fl_convergence} traces the per-round accuracy of all thirteen algorithms, so that the drift-correction methods are seen climbing steadily across the twenty rounds while the four server-side optimizers never leave the chance level, the clinical curves looking noisier only because that test set holds eighty shapes. \Cref{fig:fl_ranking} then redraws the final accuracies as ranked bar charts, whose reordering between the two datasets is precisely the dataset-dependence discussed in Section~4.2 of the main paper. The cost behind these accuracies is reported beside them: \Cref{tab:fl_compute} gives the per-round communication payload, identical across strategies because each transmits the full model, together with the wall-clock time that sets the lightweight aggregators apart from the contrastive and personalized methods (MOON and Ditto) that train an auxiliary model on every client, and \Cref{fig:fl_comm_pareto} turns that observation into a picture in which every method stacks on a single vertical line, since higher accuracy is never bought with more communication.

The clinical combined pipeline follows. \Cref{tab:kd_full} unpacks each standalone-KD mean into its three per-seed values and shows that the sub-point spread in \Cref{tab:kd_results_clinical} is seed noise, after which \Cref{tab:combined_mean,tab:combined_std} lay out the full $13\times10$ teacher-by-objective matrix, mean and per-cell standard deviation, that underlies the summary of Table~4 of the main paper. Read from top to bottom, the six hard-label columns hold near the ceiling as the federated teacher degrades while the four pure-transfer columns fall in step with it, which is the recovery illusion of Figure~3 of the main paper made numerical. The ModelNet40 material then arrives as \Cref{app:mn40grid}, the full multi-seed combined grid, whose per-cell means and standard deviations appear in \Cref{tab:combined_mean_mn40,tab:combined_std_mn40}, reproducing the same split on the larger label space. Unless noted otherwise, every value is a mean over the three seeds $\{7,42,123\}$, with the standard deviation given wherever it is shown.

\begin{table}[t]
\caption{Complete standalone FL results on both datasets (5 clients, non-IID label-skew, 20 rounds, 5 local epochs per round), as mean$\pm$std over three seeds. Methods are ordered by ModelNet40 instance accuracy; Inst.\ and Class denote instance and mean-class accuracy (\%). On the balanced clinical test set the two coincide.}
\label{tab:fl_full}
\centering
\footnotesize
\setlength{\tabcolsep}{6pt}
\begin{tabular}{@{}l cc cc@{}}
\toprule
& \multicolumn{2}{c}{ModelNet40} & \multicolumn{2}{c}{Craniosynostosis} \\
\cmidrule(lr){2-3}\cmidrule(lr){4-5}
Method & Inst. $\uparrow$ & Class $\uparrow$ & Inst. $\uparrow$ & Class $\uparrow$ \\
\midrule
Centralized & $92.26$\sd{0.04} & $89.58$\sd{0.10} & $100.00$\sd{0.00} & $100.00$\sd{0.00} \\
\midrule
FedNova~\cite{wang2020tackling} & $\mathbf{76.32}$\sd{1.55} & $\mathbf{71.28}$\sd{0.96} & $50.83$\sd{8.04} & $50.83$\sd{8.04} \\
FedProx~\cite{li2020federated} & $71.04$\sd{0.87} & $66.83$\sd{0.82} & $\mathbf{75.83}$\sd{10.63} & $\mathbf{75.83}$\sd{10.63} \\
FedDyn~\cite{acar2021federated} & $66.00$\sd{1.28} & $60.14$\sd{2.24} & $58.33$\sd{8.13} & $58.33$\sd{8.13} \\
SCAFFOLD~\cite{karimireddy2020scaffold} & $64.26$\sd{3.75} & $54.48$\sd{3.62} & $36.67$\sd{1.44} & $36.67$\sd{1.44} \\
Ditto~\cite{li2021ditto} & $61.36$\sd{4.11} & $55.00$\sd{5.42} & $71.25$\sd{9.92} & $71.25$\sd{9.92} \\
FedAvg~\cite{mcmahan2017communication} & $58.51$\sd{0.25} & $52.85$\sd{0.91} & $69.58$\sd{2.60} & $69.58$\sd{2.60} \\
FedBN~\cite{li2021fedbn} & $55.28$\sd{2.37} & $48.08$\sd{2.53} & $65.42$\sd{4.73} & $65.42$\sd{4.73} \\
MOON~\cite{li2021model} & $45.84$\sd{7.51} & $38.49$\sd{10.06} & $69.17$\sd{8.78} & $69.17$\sd{8.78} \\
FedMedian~\cite{yin2018byzantine} & $39.13$\sd{5.79} & $37.69$\sd{5.40} & $64.58$\sd{7.53} & $64.58$\sd{7.53} \\
FedAdagrad~\cite{reddi2020adaptive} & $\phantom{0}6.28$\sd{1.09} & $\phantom{0}4.19$\sd{1.13} & $26.25$\sd{2.17} & $26.25$\sd{2.17} \\
FedAdam~\cite{reddi2020adaptive} & $\phantom{0}4.71$\sd{0.98} & $\phantom{0}3.27$\sd{0.70} & $25.00$\sd{0.00} & $25.00$\sd{0.00} \\
FedAvgM~\cite{hsu2019measuring} & $\phantom{0}4.05$\sd{0.00} & $\phantom{0}2.50$\sd{0.00} & $25.00$\sd{0.00} & $25.00$\sd{0.00} \\
FedYogi~\cite{reddi2020adaptive} & $\phantom{0}4.05$\sd{0.00} & $\phantom{0}2.50$\sd{0.00} & $25.00$\sd{0.00} & $25.00$\sd{0.00} \\
\bottomrule
\end{tabular}
\end{table}

\begin{table}[t]
\caption{Per-round communication and wall-clock cost of federated training. Communication is the full model payload aggregated over the five clients (uplink plus downlink) and is therefore identical across strategies; wall-clock time is the mean seconds per communication round on a single GPU. The clinical dataset is far smaller, so its rounds are seconds rather than minutes; strategies that train an auxiliary model on every client (MOON, Ditto) are markedly slower per round on ModelNet40.}
\label{tab:fl_compute}
\centering
\small
\setlength{\tabcolsep}{4.5pt}
\begin{tabular}{@{}lcccc@{}}
\toprule
& \multicolumn{2}{c}{ModelNet40} & \multicolumn{2}{c}{Craniosynostosis} \\
\cmidrule(lr){2-3}\cmidrule(lr){4-5}
Method & Comm.\ (MB) $\downarrow$ & Time (s) $\downarrow$ & Comm.\ (MB) $\downarrow$ & Time (s) $\downarrow$ \\
\midrule
FedNova~\cite{wang2020tackling} & 56.5 & 558 & 56.2 & 13.0 \\
FedProx~\cite{li2020federated} & 56.5 & 502 & 56.2 & 13.6 \\
FedDyn~\cite{acar2021federated} & 56.5 & 506 & 56.2 & 14.3 \\
SCAFFOLD~\cite{karimireddy2020scaffold} & 56.5 & 500 & 56.2 & 13.2 \\
Ditto~\cite{li2021ditto} & 56.5 & 1081 & 56.2 & 25.7 \\
FedAvg~\cite{mcmahan2017communication} & 56.5 & 497 & 56.2 & 12.9 \\
FedBN~\cite{li2021fedbn} & 56.5 & 545 & 56.2 & 12.6 \\
MOON~\cite{li2021model} & 56.5 & 940 & 56.2 & 32.1 \\
FedMedian~\cite{yin2018byzantine} & 56.5 & 566 & 56.2 & \textbf{12.5} \\
FedAdagrad~\cite{reddi2020adaptive} & 56.5 & 565 & 56.2 & 12.6 \\
FedAdam~\cite{reddi2020adaptive} & 56.5 & \textbf{495} & 56.2 & 12.6 \\
FedAvgM~\cite{hsu2019measuring} & 56.5 & 504 & 56.2 & 12.7 \\
FedYogi~\cite{reddi2020adaptive} & 56.5 & 563 & 56.2 & 12.6 \\
\bottomrule
\end{tabular}
\end{table}

\begin{table}[t]
\caption{Per-seed standalone KD test instance accuracy (\%) on the clinical task, for the compact student distilled from the centralized teacher. Methods are ordered by mean accuracy. Column headers are the random seeds.}
\label{tab:kd_full}
\centering
\small
\setlength{\tabcolsep}{10pt}
\begin{tabular}{@{}lccc@{}}
\toprule
Strategy & $7$~$\uparrow$ & $42$~$\uparrow$ & $123$~$\uparrow$ \\
\midrule
Teacher~\cite{pointnet2} & 100.00 & 100.00 & 100.00 \\
\midrule
Attention Transfer~\cite{zagoruyko2016paying} & 100.00 & 100.00 & 100.00 \\
Relational KD~\cite{park2019relational} & 100.00 & 100.00 & 100.00 \\
Similarity-Preserving~\cite{tung2019similarity} & 100.00 & 100.00 & 100.00 \\
Vanilla KD~\cite{hinton2015distilling} & 100.00 & 100.00 & 100.00 \\
Cosine Similarity~\cite{luo2018cosine} & 100.00 & 98.75 & 100.00 \\
Contrastive (CRD)~\cite{tian2019contrastive} & 100.00 & 98.75 & 100.00 \\
Feature KD~\cite{romero2014fitnets} & 100.00 & 98.75 & 100.00 \\
Logit-MSE~\cite{ba2014deep} & 98.75 & 100.00 & 100.00 \\
Self-Distillation~\cite{furlanello2018born} & 100.00 & 98.75 & 100.00 \\
Decoupled KD~\cite{zhao2022decoupled} & 98.75 & 98.75 & 98.75 \\
\bottomrule
\end{tabular}
\end{table}

\begin{table}[t]
\caption{Complete combined-pipeline results on the clinical task: mean student instance accuracy (\%) over three seeds for every FL teacher (rows, with standalone accuracy in parentheses, ordered by it) and KD objective (columns). Columns: Van.\ (Vanilla KD~\cite{hinton2015distilling}), Self (Self-Distillation~\cite{furlanello2018born}), Cos.\ (Cosine~\cite{luo2018cosine}), CRD (Contrastive~\cite{tian2019contrastive}), DKD (Decoupled KD~\cite{zhao2022decoupled}), LMSE (Logit-MSE~\cite{ba2014deep}), Feat.\ (Feature KD~\cite{romero2014fitnets}), Att.\ (Attention~\cite{zagoruyko2016paying}), RKD (Relational KD~\cite{park2019relational}), SP (Similarity-Preserving~\cite{tung2019similarity}). The first six columns keep a hard-label term; the last four do not. Standard deviations are in \Cref{tab:combined_std}.}
\label{tab:combined_mean}
\centering
\footnotesize
\setlength{\tabcolsep}{3.2pt}
\resizebox{\textwidth}{!}{%
\begin{tabular}{@{}l cccccc cccc@{}}
\toprule
& \multicolumn{6}{c}{Hard-label objectives} & \multicolumn{4}{c}{Pure transfer objectives} \\
\cmidrule(lr){2-7}\cmidrule(lr){8-11}
Teacher (std.\%) & Van. $\uparrow$ & Self $\uparrow$ & Cos. $\uparrow$ & CRD $\uparrow$ & DKD $\uparrow$ & LMSE $\uparrow$ & Feat. $\uparrow$ & Att. $\uparrow$ & RKD $\uparrow$ & SP $\uparrow$ \\
\midrule
FedProx~\cite{li2020federated} (75.8) & 99.6 & 99.6 & \textbf{100.0} & 99.6 & 96.7 & 99.2 & \textbf{84.2} & \textbf{83.3} & 81.7 & \textbf{82.9} \\
Ditto~\cite{li2021ditto} (71.2) & 99.2 & \textbf{100.0} & 99.6 & 99.2 & 96.2 & 98.8 & 82.5 & \textbf{83.3} & \textbf{85.0} & 83.3 \\
FedAvg~\cite{mcmahan2017communication} (69.6) & 97.9 & 98.8 & \textbf{100.0} & 99.2 & 96.7 & 97.9 & 77.1 & 76.2 & 77.1 & 77.5 \\
MOON~\cite{li2021model} (69.2) & \textbf{100.0} & 99.6 & 99.6 & 99.6 & 97.1 & 98.8 & 75.4 & 74.2 & 77.5 & 75.4 \\
FedBN~\cite{li2021fedbn} (65.4) & 99.6 & 99.6 & 99.6 & 99.6 & 97.5 & 99.2 & 76.2 & 76.2 & 76.7 & 75.4 \\
FedMedian~\cite{yin2018byzantine} (64.6) & 99.6 & 99.6 & 99.6 & 99.2 & 97.5 & \textbf{99.6} & 69.2 & 69.6 & 70.0 & 67.9 \\
FedDyn~\cite{acar2021federated} (58.3) & 99.6 & 99.2 & 99.6 & 99.6 & 91.7 & 95.8 & 63.8 & 63.3 & 63.3 & 64.2 \\
FedNova~\cite{wang2020tackling} (50.8) & 99.2 & 98.8 & 99.2 & 99.6 & 97.9 & 98.3 & 57.9 & 59.6 & 56.7 & 57.5 \\
SCAFFOLD~\cite{karimireddy2020scaffold} (36.7) & 99.6 & 99.2 & 99.6 & 99.2 & 97.9 & 98.8 & 30.4 & 29.6 & 32.5 & 32.9 \\
FedAdagrad~\cite{reddi2020adaptive} (26.2) & 98.8 & 99.2 & \textbf{100.0} & 99.6 & \textbf{99.2} & 98.8 & 32.5 & 30.8 & 31.7 & 32.5 \\
FedAdam~\cite{reddi2020adaptive} (25.0) & 99.2 & 99.6 & 99.6 & \textbf{100.0} & 96.2 & 98.8 & 26.7 & 29.6 & 31.2 & 28.3 \\
FedAvgM~\cite{hsu2019measuring} (25.0) & 97.5 & 99.2 & 99.6 & \textbf{100.0} & 93.8 & 97.5 & 27.1 & 25.0 & 27.5 & 25.0 \\
FedYogi~\cite{reddi2020adaptive} (25.0) & 99.2 & 99.6 & \textbf{100.0} & 99.6 & 98.3 & 98.8 & 27.5 & 27.9 & 30.8 & 28.8 \\
\bottomrule
\end{tabular}}
\end{table}

\begin{table}[t]
\caption{Standard deviation (\%, three seeds) of the combined-pipeline student instance accuracy reported in \Cref{tab:combined_mean}. Column abbreviations as in \Cref{tab:combined_mean}. The pure-transfer objectives show the larger spread, reflecting their dependence on the variable federated teacher.}
\label{tab:combined_std}
\centering
\footnotesize
\setlength{\tabcolsep}{3.2pt}
\resizebox{\textwidth}{!}{%
\begin{tabular}{@{}l cccccc cccc@{}}
\toprule
& \multicolumn{6}{c}{Hard-label objectives} & \multicolumn{4}{c}{Pure transfer objectives} \\
\cmidrule(lr){2-7}\cmidrule(lr){8-11}
Teacher & Van. & Self & Cos. & CRD & DKD & LMSE & Feat. & Att. & RKD & SP \\
\midrule
FedProx~\cite{li2020federated} & 0.72 & 0.72 & 0.00 & 0.72 & 0.72 & 0.72 & 5.64 & 6.41 & 7.53 & 9.04 \\
Ditto~\cite{li2021ditto} & 0.72 & 0.00 & 0.72 & 0.72 & 1.25 & 1.25 & 3.31 & 0.72 & 2.50 & 1.91 \\
FedAvg~\cite{mcmahan2017communication} & 0.72 & 0.00 & 0.00 & 0.72 & 1.44 & 1.44 & 6.88 & 4.51 & 5.91 & 6.61 \\
MOON~\cite{li2021model} & 0.00 & 0.72 & 0.72 & 0.72 & 1.44 & 0.00 & 9.21 & 13.25 & 8.20 & 8.32 \\
FedBN~\cite{li2021fedbn} & 0.72 & 0.72 & 0.72 & 0.72 & 1.25 & 1.44 & 7.50 & 5.73 & 6.41 & 5.64 \\
FedMedian~\cite{yin2018byzantine} & 0.72 & 0.72 & 0.72 & 0.72 & 1.25 & 0.72 & 4.39 & 5.20 & 3.31 & 6.88 \\
FedDyn~\cite{acar2021federated} & 0.72 & 0.72 & 0.72 & 0.72 & 2.60 & 1.91 & 13.17 & 13.13 & 12.58 & 13.25 \\
FedNova~\cite{wang2020tackling} & 1.44 & 0.00 & 0.72 & 0.72 & 1.91 & 0.72 & 8.78 & 10.10 & 8.32 & 7.81 \\
SCAFFOLD~\cite{karimireddy2020scaffold} & 0.72 & 0.72 & 0.72 & 0.72 & 0.72 & 1.25 & 3.15 & 3.15 & 0.00 & 1.91 \\
FedAdagrad~\cite{reddi2020adaptive} & 1.25 & 1.44 & 0.00 & 0.72 & 1.44 & 0.00 & 0.00 & 1.91 & 1.91 & 1.25 \\
FedAdam~\cite{reddi2020adaptive} & 0.72 & 0.72 & 0.72 & 0.00 & 1.25 & 1.25 & 1.91 & 1.91 & 2.17 & 3.82 \\
FedAvgM~\cite{hsu2019measuring} & 1.25 & 1.44 & 0.72 & 0.00 & 2.50 & 0.00 & 2.60 & 0.00 & 2.17 & 0.00 \\
FedYogi~\cite{reddi2020adaptive} & 0.72 & 0.72 & 0.00 & 0.72 & 1.44 & 1.25 & 2.17 & 1.44 & 4.02 & 2.17 \\
\bottomrule
\end{tabular}}
\end{table}

\subsection{ModelNet40 Multi-Seed Combined Grid}
\label{app:mn40grid}

The controlled combined-pipeline diagnosis of Section~4.4 of the main paper uses a focused round of hand-picked ModelNet40 teachers spanning the full quality range down to collapsed checkpoints. For completeness we also ran the \emph{full} multi-seed combined grid on ModelNet40, crossing all thirteen standardized FL teachers with all ten KD objectives over the three seeds $\{7,42,123\}$ ($390$ runs). These reuse the standardized protocol but lie outside the $504$-run count of the main benchmark, exactly as the additional-dataset runs of \Cref{app:moredata} do; we report them because they reproduce, at multi-seed scale, the same dichotomy that the clinical grid shows.

\begin{figure}[tb]
\centering
\includegraphics[width=0.99\linewidth]{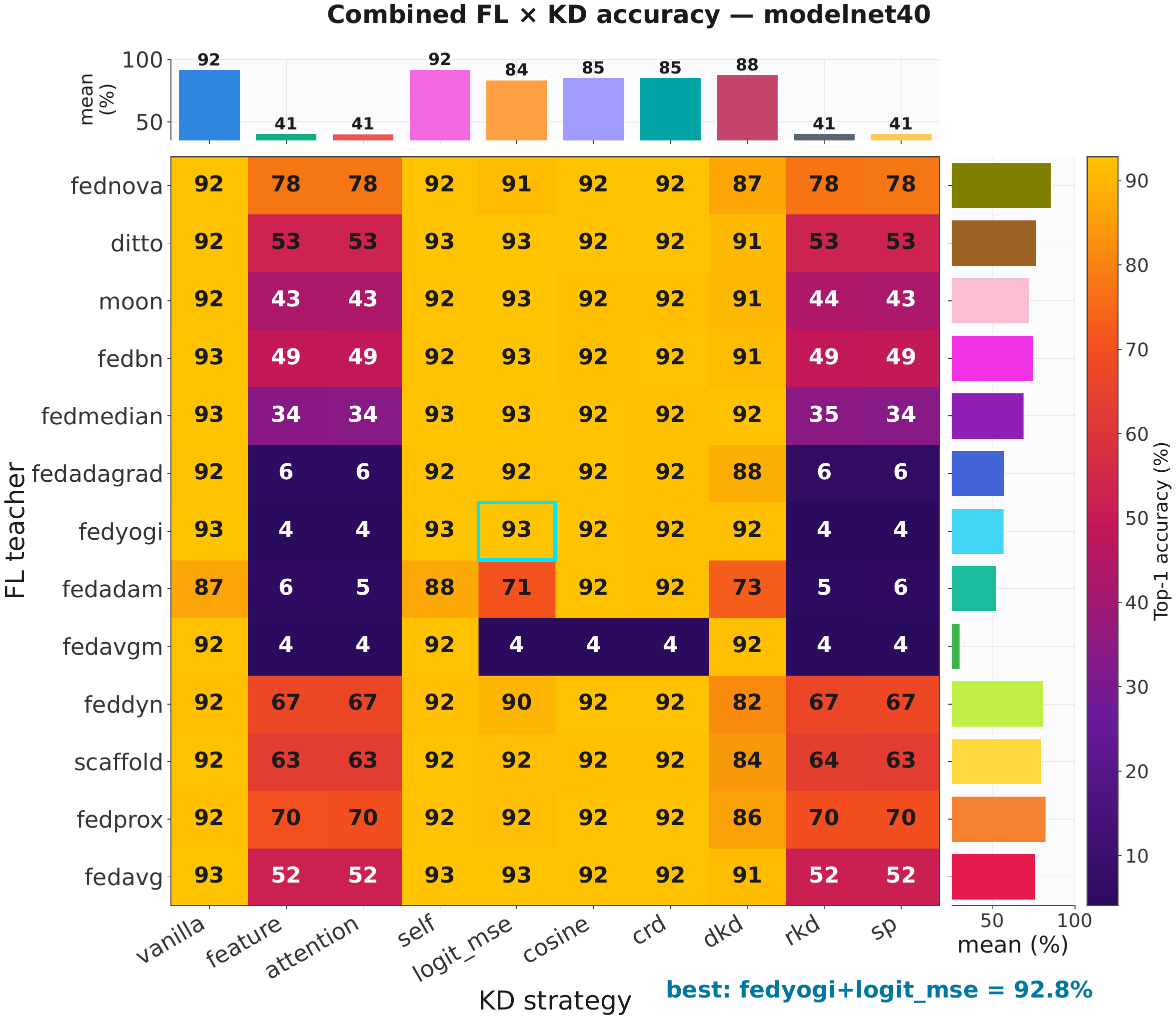}
\caption{\textbf{Multi-seed combined-pipeline grid on ModelNet40.}
Student instance accuracy for each FL teacher (rows) and KD objective (columns), mean over three seeds, with per-column means in the bars above. The hard-label columns (vanilla, self, logit-MSE, cosine, CRD, DKD) stay high while the pure-transfer columns (feature, attention, RKD, SP) track the teacher; the best cell pairs the chance-level FedYogi teacher with Logit-MSE at $92.8\%$. Both axes follow the benchmark's native order.}
\label{fig:combined_heatmap_mn40}
\end{figure}

\begin{figure}[tb]
\centering
\includegraphics[width=0.99\linewidth]{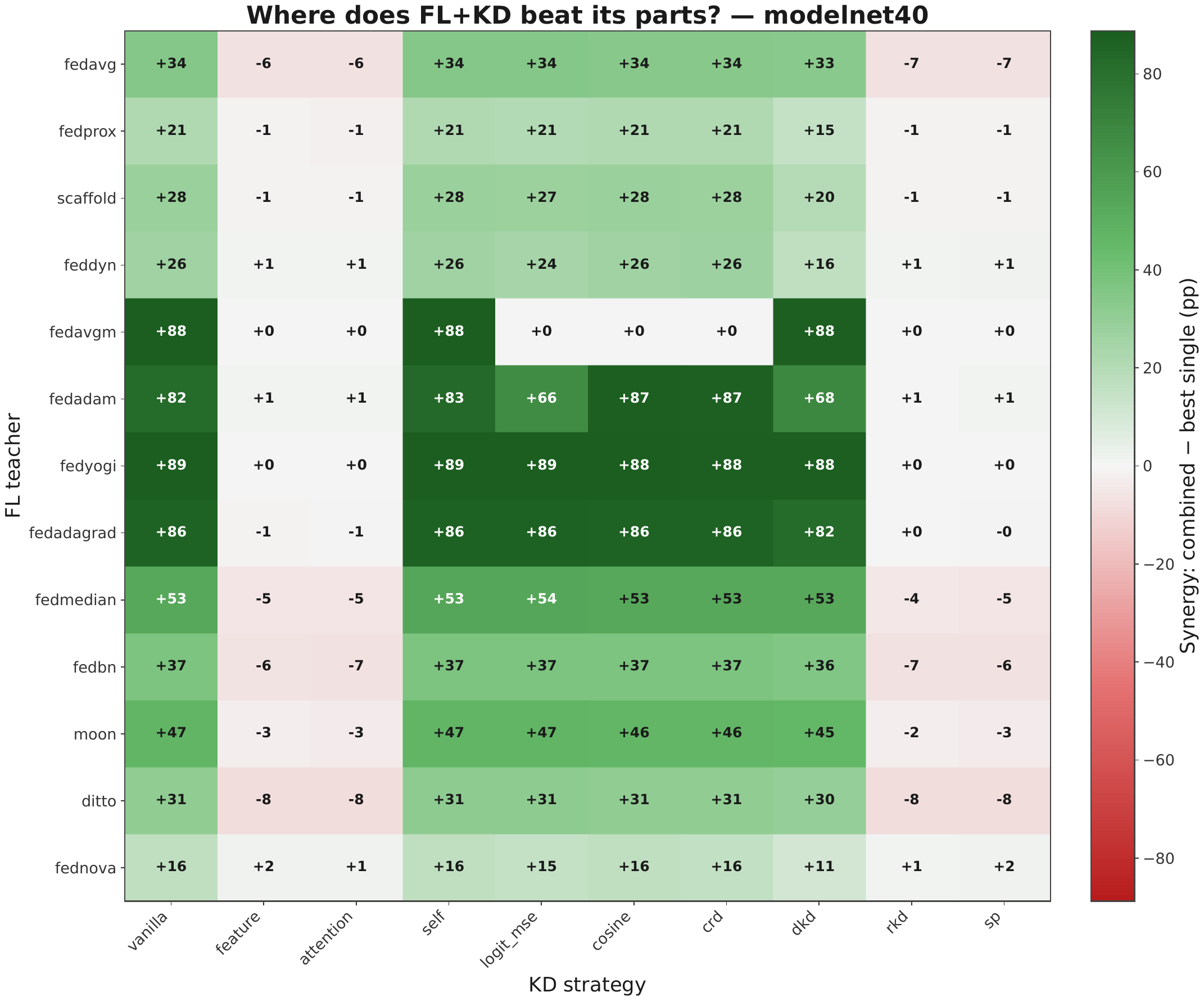}
\caption{\textbf{No positive synergy on ModelNet40.}
For every FL teacher (rows) and KD objective (columns), the combined student's instance accuracy minus the better of its two stand-alone components (the federated teacher alone and the student distilled from the centralized teacher), in percentage points (mean over three seeds). The six hard-label objectives stay near zero, while the four pure-transfer objectives fall well below their best component for the collapsed teachers, mirroring the clinical result of \Cref{fig:synergy_clinical}.}
\label{fig:synergy_mn40}
\end{figure}

\begin{figure}[tb]
\centering
\includegraphics[width=0.99\linewidth]{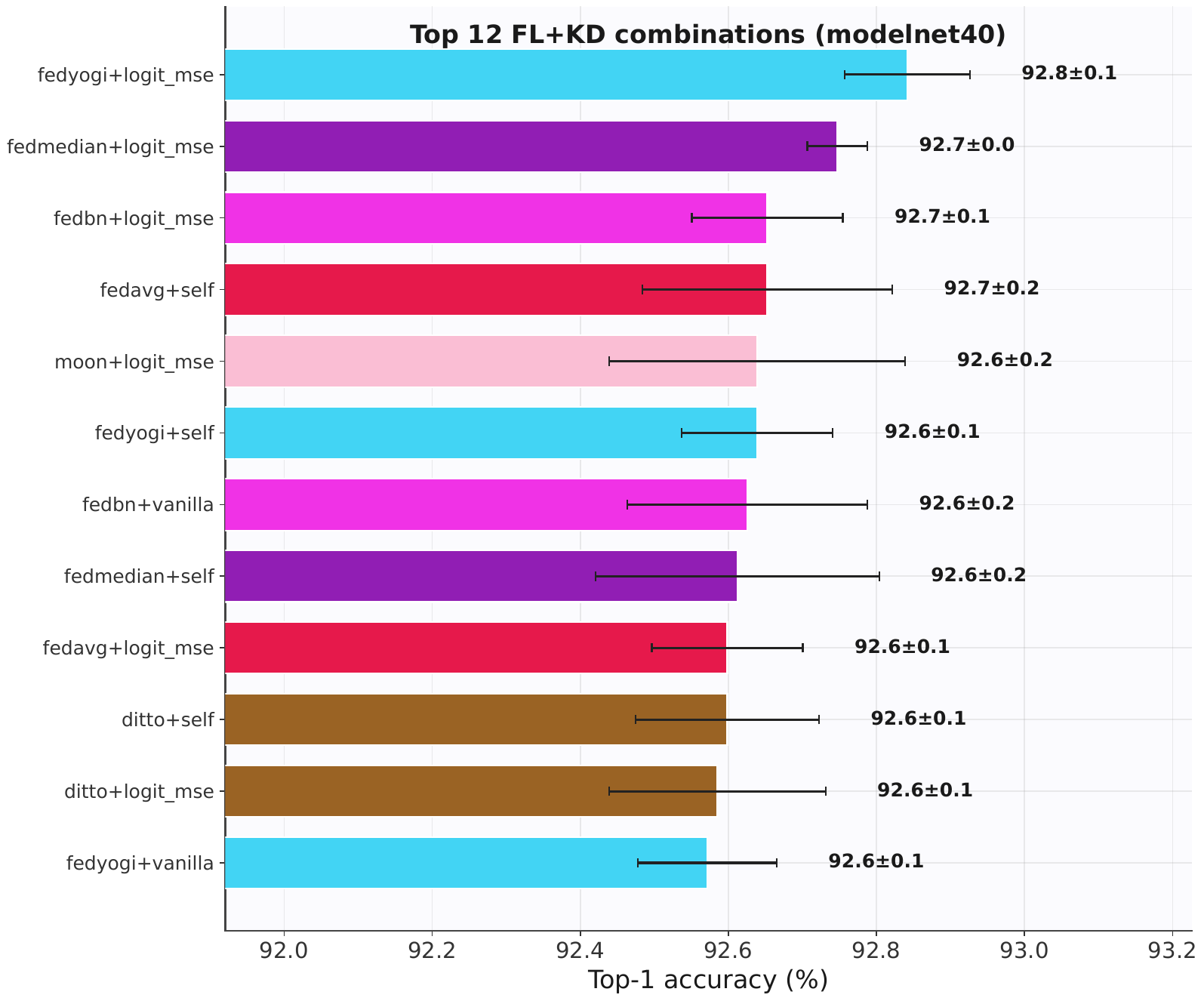}
\caption{\textbf{The ModelNet40 leaderboard also rewards collapsed teachers.}
The twelve highest-accuracy FL$+$KD combinations on ModelNet40, each as mean$\pm$std over three seeds. Several pair a hard-label objective with a federated teacher frozen near the $4\%$ chance level (for example FedYogi$+$Logit-MSE at $92.8\%$), so the ranking reflects the proxy labels rather than the quality of the federated teacher.}
\label{fig:best_combos_mn40}
\end{figure}

\begin{figure}[tb]
\centering
\includegraphics[width=\linewidth]{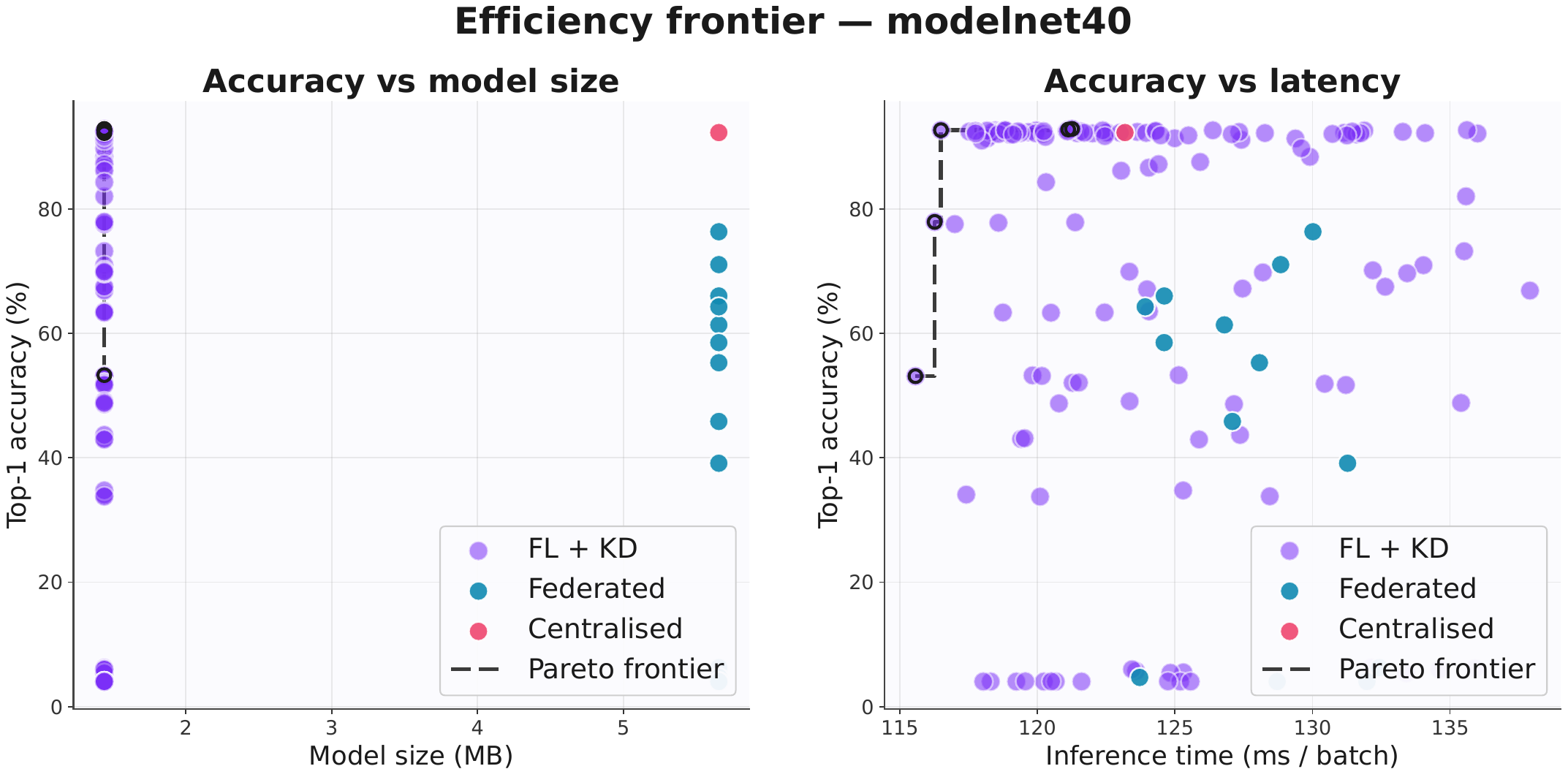}
\caption{\textbf{Accuracy versus model size and latency on ModelNet40.}
Test accuracy against model size \textbf{(left)} and inference latency \textbf{(right)}, colored by family. The compact students cluster at the upper left near the teacher's accuracy at $1.44$\,MB and at lower latency, the standalone federated models keep the teacher's $5.65$\,MB footprint while spanning the full accuracy range, and the combined students inherit the compact size. As on the clinical task, these wall-clock latencies are averages on a single GPU and are subject to hardware-load variation, so the latency margin in this multi-seed run is narrower than the roughly twofold speed-up measured for the same compact student under the lighter-load focused round (Section~4.3 of the main paper). This is the ModelNet40 counterpart of \Cref{fig:efficiency_clinical}.}
\label{fig:efficiency_mn40}
\end{figure}

\Cref{fig:combined_heatmap_mn40} renders the grid as a teacher-by-objective heatmap, and \Cref{tab:combined_mean_mn40,tab:combined_std_mn40} give the per-cell means and standard deviations behind it (teachers ordered by descending standalone accuracy and the objectives grouped into the six that keep a hard-label term and the four pure-transfer ones, as in the clinical \Cref{tab:combined_mean}). The six hard-label columns stay high across almost every teacher (column means about $84$ to $92\%$): the two plain Kullback--Leibler objectives (Vanilla~KD and Self-Distillation) recover the student to about $92\%$ from essentially every teacher, including the server-side optimizers frozen near the $4\%$ chance level, with the only material dip (FedAdam, to $86.6\%$ and $87.5\%$) traceable to a single divergent seed (\Cref{tab:combined_std_mn40}); Decoupled~KD likewise recovers from every teacher, though at a lower and more variable level (column mean about $88\%$); and the three objectives that instead match the teacher directly in logit or feature space (Logit-MSE, Cosine, CRD) recover from almost all teachers but can collapse on the single most degenerate teacher (FedAvgM), reflecting the greater sensitivity of direct logit- and feature-matching to a fully degenerate teacher on the $40$-class task. The four pure-transfer columns instead track teacher quality (column means ${\approx}41\%$), falling to the chance level when the teacher collapses. The strongest single cell pairs a FedYogi teacher frozen at the $4.05\%$ chance level with Logit-MSE at $92.8\%$, an $88.8$-point gap that again reflects the proxy labels rather than the teacher. \Cref{fig:synergy_mn40,fig:best_combos_mn40} confirm the reading from two further angles, exactly as on the clinical task: the combined pipeline shows no positive synergy, and the top of the leaderboard is populated by collapsed teachers paired with hard-label objectives. \Cref{fig:efficiency_mn40} shows the corresponding accuracy--size--latency trade-off.

\begin{table}[t]
\caption{Complete multi-seed combined-pipeline results on ModelNet40: mean student instance accuracy (\%) over three seeds for every FL teacher (rows, with standalone accuracy in parentheses, ordered by it) and KD objective (columns). Columns: Van.\ (Vanilla KD~\cite{hinton2015distilling}), Self (Self-Distillation~\cite{furlanello2018born}), Cos.\ (Cosine~\cite{luo2018cosine}), CRD (Contrastive~\cite{tian2019contrastive}), DKD (Decoupled KD~\cite{zhao2022decoupled}), LMSE (Logit-MSE~\cite{ba2014deep}), Feat.\ (Feature KD~\cite{romero2014fitnets}), Att.\ (Attention~\cite{zagoruyko2016paying}), RKD (Relational KD~\cite{park2019relational}), SP (Similarity-Preserving~\cite{tung2019similarity}). The first six columns keep a hard-label term; the last four do not. \textbf{Bold} marks the best value in each column. This is the multi-seed counterpart of the focused-round grid in Table~3 of the main paper; standard deviations are in \Cref{tab:combined_std_mn40}.}
\label{tab:combined_mean_mn40}
\centering
\footnotesize
\setlength{\tabcolsep}{3.2pt}
\resizebox{\textwidth}{!}{%
\begin{tabular}{@{}l cccccc cccc@{}}
\toprule
& \multicolumn{6}{c}{Hard-label objectives} & \multicolumn{4}{c}{Pure transfer objectives} \\
\cmidrule(lr){2-7}\cmidrule(lr){8-11}
Teacher (std.\%) & Van. $\uparrow$ & Self $\uparrow$ & Cos. $\uparrow$ & CRD $\uparrow$ & DKD $\uparrow$ & LMSE $\uparrow$ & Feat. $\uparrow$ & Att. $\uparrow$ & RKD $\uparrow$ & SP $\uparrow$ \\
\midrule
FedNova~\cite{wang2020tackling} (76.3) & 92.4 & 92.4 & 92.3 & 92.2 & 87.2 & 91.3 & \textbf{77.8} & \textbf{77.8} & \textbf{77.6} & \textbf{77.9} \\
FedProx~\cite{li2020federated} (71.0) & 92.0 & 92.2 & 92.3 & 92.2 & 86.1 & 91.7 & 70.1 & 69.7 & 69.8 & 69.9 \\
FedDyn~\cite{acar2021federated} (66.0) & 91.8 & 91.8 & 92.2 & \textbf{92.4} & 82.0 & 89.7 & 67.2 & 67.1 & 66.9 & 67.5 \\
SCAFFOLD~\cite{karimireddy2020scaffold} (64.3) & 92.5 & 92.5 & 92.1 & 92.0 & 84.3 & 91.6 & 63.3 & 63.3 & 63.5 & 63.4 \\
Ditto~\cite{li2021ditto} (61.4) & 92.4 & 92.6 & 92.2 & 92.2 & 91.1 & 92.6 & 53.2 & 53.3 & 53.2 & 53.1 \\
FedAvg~\cite{mcmahan2017communication} (58.5) & 92.5 & \textbf{92.7} & \textbf{92.4} & 92.3 & 91.3 & 92.6 & 52.1 & 52.1 & 51.9 & 51.7 \\
FedBN~\cite{li2021fedbn} (55.3) & \textbf{92.6} & 92.5 & 92.1 & 92.2 & 91.3 & 92.7 & 49.1 & 48.7 & 48.6 & 48.8 \\
MOON~\cite{li2021model} (45.8) & 92.5 & 92.4 & 92.0 & 92.1 & 91.0 & 92.6 & 43.1 & 43.0 & 43.7 & 42.9 \\
FedMedian~\cite{yin2018byzantine} (39.1) & 92.5 & 92.6 & 92.3 & 92.1 & \textbf{92.2} & 92.7 & 34.1 & 33.8 & 34.8 & 33.8 \\
FedAdagrad~\cite{reddi2020adaptive} (6.3) & 92.3 & 92.4 & 92.2 & 92.0 & 88.4 & 92.4 & 5.5 & 5.8 & 6.3 & 6.2 \\
FedAdam~\cite{reddi2020adaptive} (4.7) & 86.6 & 87.5 & 92.2 & 92.0 & 73.2 & 70.9 & 6.0 & 5.5 & 5.4 & 6.0 \\
FedAvgM~\cite{hsu2019measuring} (4.1) & 92.2 & 92.3 & 4.1 & 4.1 & 91.9 & 4.1 & 4.1 & 4.1 & 4.1 & 4.1 \\
FedYogi~\cite{reddi2020adaptive} (4.1) & \textbf{92.6} & 92.6 & 92.2 & 91.9 & 91.8 & \textbf{92.8} & 4.1 & 4.1 & 4.1 & 4.1 \\
\bottomrule
\end{tabular}}
\end{table}

\begin{table}[t]
\caption{Standard deviation (\%, three seeds) of the ModelNet40 combined-pipeline student instance accuracy reported in \Cref{tab:combined_mean_mn40}. Column abbreviations as in \Cref{tab:combined_mean_mn40}. The pure-transfer objectives show the larger spread, reflecting their dependence on the variable federated teacher; the adaptive server-side teachers (notably FedAdam) additionally inflate a few hard-label cells, where a single seed failed to converge.}
\label{tab:combined_std_mn40}
\centering
\footnotesize
\setlength{\tabcolsep}{3.2pt}
\resizebox{\textwidth}{!}{%
\begin{tabular}{@{}l cccccc cccc@{}}
\toprule
& \multicolumn{6}{c}{Hard-label objectives} & \multicolumn{4}{c}{Pure transfer objectives} \\
\cmidrule(lr){2-7}\cmidrule(lr){8-11}
Teacher & Van. & Self & Cos. & CRD & DKD & LMSE & Feat. & Att. & RKD & SP \\
\midrule
FedNova~\cite{wang2020tackling} & 0.08 & 0.14 & 0.17 & 0.06 & 0.55 & 0.13 & 0.79 & 0.62 & 0.44 & 0.61 \\
FedProx~\cite{li2020federated} & 0.27 & 0.15 & 0.14 & 0.13 & 0.55 & 0.08 & 0.70 & 0.79 & 0.58 & 0.35 \\
FedDyn~\cite{acar2021federated} & 0.37 & 0.26 & 0.25 & 0.17 & 2.10 & 1.15 & 2.12 & 2.36 & 3.10 & 2.50 \\
SCAFFOLD~\cite{karimireddy2020scaffold} & 0.11 & 0.04 & 0.14 & 0.13 & 2.09 & 0.63 & 4.55 & 4.60 & 4.29 & 4.40 \\
Ditto~\cite{li2021ditto} & 0.15 & 0.12 & 0.31 & 0.10 & 0.47 & 0.15 & 3.44 & 3.58 & 3.64 & 3.32 \\
FedAvg~\cite{mcmahan2017communication} & 0.21 & 0.17 & 0.05 & 0.05 & 0.32 & 0.10 & 1.72 & 1.18 & 1.36 & 1.62 \\
FedBN~\cite{li2021fedbn} & 0.16 & 0.10 & 0.06 & 0.16 & 0.66 & 0.10 & 1.20 & 1.10 & 1.20 & 1.15 \\
MOON~\cite{li2021model} & 0.06 & 0.06 & 0.22 & 0.14 & 0.33 & 0.20 & 7.21 & 7.51 & 6.77 & 7.09 \\
FedMedian~\cite{yin2018byzantine} & 0.19 & 0.19 & 0.02 & 0.16 & 0.10 & 0.04 & 5.32 & 6.00 & 5.31 & 5.70 \\
FedAdagrad~\cite{reddi2020adaptive} & 0.47 & 0.42 & 0.15 & 0.19 & 6.00 & 0.41 & 1.67 & 1.31 & 1.39 & 1.32 \\
FedAdam~\cite{reddi2020adaptive} & 9.88 & 8.86 & 0.05 & 0.08 & 29.01 & 37.48 & 3.05 & 2.06 & 1.37 & 2.96 \\
FedAvgM~\cite{hsu2019measuring} & 0.02 & 0.17 & 0.00 & 0.00 & 0.23 & 0.00 & 0.00 & 0.00 & 0.00 & 0.00 \\
FedYogi~\cite{reddi2020adaptive} & 0.09 & 0.10 & 0.29 & 0.21 & 0.23 & 0.08 & 0.02 & 0.05 & 0.05 & 0.00 \\
\bottomrule
\end{tabular}}
\end{table}

\clearpage
% =====================================================================
\section{Per-Seed Results}
\label{app:perseed}

For full transparency we report the individual test instance accuracy of every run, before averaging. \Cref{tab:seed_fl} gives the per-seed standalone FL results on both datasets and \Cref{tab:kd_full} the per-seed standalone KD results, while \Cref{tab:seed_combined,tab:seed_combined_mn40} list the per-seed accuracy of all $130$ combined configurations on the clinical task and on ModelNet40 respectively, complementing the mean and standard-deviation matrices in \Cref{tab:combined_mean,tab:combined_std} (clinical) and \Cref{tab:combined_mean_mn40,tab:combined_std_mn40} (ModelNet40). The per-seed spread is small for the converged methods; the larger spread of the feature-based combined runs reflects their dependence on the variable federated teacher rather than seed noise, as is also presented in the analysis in Section~4.4 of the main paper.

\begin{table}[t]
\caption{Per-seed standalone FL test instance accuracy (\%) on both datasets. Methods are ordered by mean ModelNet40 accuracy. Column headers are the random seeds.}
\label{tab:seed_fl}
\centering
\footnotesize
\setlength{\tabcolsep}{6pt}
\begin{tabular}{@{}l ccc ccc@{}}
\toprule
& \multicolumn{3}{c}{ModelNet40} & \multicolumn{3}{c}{Craniosynostosis} \\
\cmidrule(lr){2-4}\cmidrule(lr){5-7}
Method & $7$~$\uparrow$ & $42$~$\uparrow$ & $123$~$\uparrow$ & $7$~$\uparrow$ & $42$~$\uparrow$ & $123$~$\uparrow$ \\
\midrule
Centralized & 92.26 & 92.30 & 92.22 & 100.00 & 100.00 & 100.00 \\
\midrule
FedNova~\cite{wang2020tackling} & \textbf{75.16} & \textbf{75.73} & \textbf{78.08} & 45.00 & 47.50 & 60.00 \\
FedProx~\cite{li2020federated} & 71.68 & 71.39 & 70.06 & \textbf{80.00} & \textbf{83.75} & 63.75 \\
FedDyn~\cite{acar2021federated} & 66.94 & 64.55 & 66.53 & 66.25 & 58.75 & 50.00 \\
SCAFFOLD~\cite{karimireddy2020scaffold} & 60.41 & 67.91 & 64.47 & 37.50 & 35.00 & 37.50 \\
Ditto~\cite{li2021ditto} & 59.64 & 66.05 & 58.39 & 60.00 & 75.00 & \textbf{78.75} \\
FedAvg~\cite{mcmahan2017communication} & 58.79 & 58.31 & 58.43 & 67.50 & 72.50 & 68.75 \\
FedBN~\cite{li2021fedbn} & 57.90 & 53.28 & 54.66 & 68.75 & 67.50 & 60.00 \\
MOON~\cite{li2021model} & 42.91 & 40.24 & 54.38 & 60.00 & 77.50 & 70.00 \\
FedMedian~\cite{yin2018byzantine} & 44.04 & 40.60 & 32.74 & 63.75 & 72.50 & 57.50 \\
FedAdagrad~\cite{reddi2020adaptive} & $\phantom{0}$7.54 & $\phantom{0}$5.75 & $\phantom{0}$5.55 & 25.00 & 25.00 & 28.75 \\
FedAdam~\cite{reddi2020adaptive} & $\phantom{0}$4.05 & $\phantom{0}$4.25 & $\phantom{0}$5.83 & 25.00 & 25.00 & 25.00 \\
FedAvgM~\cite{hsu2019measuring} & $\phantom{0}$4.05 & $\phantom{0}$4.05 & $\phantom{0}$4.05 & 25.00 & 25.00 & 25.00 \\
FedYogi~\cite{reddi2020adaptive} & $\phantom{0}$4.05 & $\phantom{0}$4.05 & $\phantom{0}$4.05 & 25.00 & 25.00 & 25.00 \\
\bottomrule
\end{tabular}
\end{table}

\clearpage
% =====================================================================
{\footnotesize
\setlength{\tabcolsep}{8pt}
\renewcommand{\arraystretch}{1.05}
\begin{longtable}{@{}ll ccc@{}}
\caption{Per-seed combined-pipeline test instance accuracy (\%) on the clinical task, for every FL teacher (with its standalone accuracy) crossed with every KD objective. Column headers are the random seeds. The six hard-label objectives recover the student regardless of teacher quality, whereas the four pure-transfer objectives (Feature KD, Attention Transfer, Relational KD, Similarity-Preserving) track the teacher and fall toward the $25\%$ chance level when it collapses. For reference, the ten KD objectives are Vanilla KD~\cite{hinton2015distilling}, Self-Distillation~\cite{furlanello2018born}, Cosine Similarity~\cite{luo2018cosine}, Contrastive Representation Distillation (CRD)~\cite{tian2019contrastive}, Decoupled KD~\cite{zhao2022decoupled}, Logit-MSE~\cite{ba2014deep}, Feature KD~\cite{romero2014fitnets}, Attention Transfer~\cite{zagoruyko2016paying}, Relational KD~\cite{park2019relational}, and Similarity-Preserving distillation~\cite{tung2019similarity}.}
\label{tab:seed_combined}\\
\toprule
FL Teacher (standalone) & KD Objective & $7$~$\uparrow$ & $42$~$\uparrow$ & $123$~$\uparrow$ \\
\midrule
\endfirsthead
\multicolumn{5}{@{}l}{\itshape Table~\ref{tab:seed_combined}, continued} \\
\toprule
FL Teacher (standalone) & KD Objective & $7$~$\uparrow$ & $42$~$\uparrow$ & $123$~$\uparrow$ \\
\midrule
\endhead
\midrule
\multicolumn{5}{r@{}}{\itshape continued on next page} \\
\endfoot
\bottomrule
\endlastfoot
FedProx~\cite{li2020federated} (75.8\%) & Vanilla KD & 100.00 & 98.75 & 100.00 \\
 & Self-Distillation & 100.00 & 98.75 & 100.00 \\
 & Cosine Similarity & 100.00 & 100.00 & 100.00 \\
 & Contrastive (CRD) & 100.00 & 98.75 & 100.00 \\
 & Decoupled KD & 97.50 & 96.25 & 96.25 \\
 & Logit-MSE & 98.75 & 98.75 & 100.00 \\
 & Feature KD & 83.75 & 90.00 & 78.75 \\
 & Attention Transfer & 85.00 & 88.75 & 76.25 \\
 & Relational KD & 82.50 & 88.75 & 73.75 \\
 & Similarity-Preserving & 87.50 & 88.75 & 72.50 \\
\addlinespace
Ditto~\cite{li2021ditto} (71.2\%) & Vanilla KD & 100.00 & 98.75 & 98.75 \\
 & Self-Distillation & 100.00 & 100.00 & 100.00 \\
 & Cosine Similarity & 100.00 & 98.75 & 100.00 \\
 & Contrastive (CRD) & 100.00 & 98.75 & 98.75 \\
 & Decoupled KD & 95.00 & 97.50 & 96.25 \\
 & Logit-MSE & 100.00 & 97.50 & 98.75 \\
 & Feature KD & 80.00 & 81.25 & 86.25 \\
 & Attention Transfer & 82.50 & 83.75 & 83.75 \\
 & Relational KD & 82.50 & 85.00 & 87.50 \\
 & Similarity-Preserving & 81.25 & 83.75 & 85.00 \\
\addlinespace
FedAvg~\cite{mcmahan2017communication} (69.6\%) & Vanilla KD & 98.75 & 97.50 & 97.50 \\
 & Self-Distillation & 98.75 & 98.75 & 98.75 \\
 & Cosine Similarity & 100.00 & 100.00 & 100.00 \\
 & Contrastive (CRD) & 100.00 & 98.75 & 98.75 \\
 & Decoupled KD & 97.50 & 97.50 & 95.00 \\
 & Logit-MSE & 98.75 & 96.25 & 98.75 \\
 & Feature KD & 73.75 & 85.00 & 72.50 \\
 & Attention Transfer & 75.00 & 81.25 & 72.50 \\
 & Relational KD & 75.00 & 83.75 & 72.50 \\
 & Similarity-Preserving & 75.00 & 85.00 & 72.50 \\
\addlinespace
MOON~\cite{li2021model} (69.2\%) & Vanilla KD & 100.00 & 100.00 & 100.00 \\
 & Self-Distillation & 98.75 & 100.00 & 100.00 \\
 & Cosine Similarity & 100.00 & 98.75 & 100.00 \\
 & Contrastive (CRD) & 100.00 & 98.75 & 100.00 \\
 & Decoupled KD & 98.75 & 96.25 & 96.25 \\
 & Logit-MSE & 98.75 & 98.75 & 98.75 \\
 & Feature KD & 65.00 & 78.75 & 82.50 \\
 & Attention Transfer & 60.00 & 76.25 & 86.25 \\
 & Relational KD & 68.75 & 78.75 & 85.00 \\
 & Similarity-Preserving & 66.25 & 77.50 & 82.50 \\
\addlinespace
FedBN~\cite{li2021fedbn} (65.4\%) & Vanilla KD & 100.00 & 98.75 & 100.00 \\
 & Self-Distillation & 100.00 & 98.75 & 100.00 \\
 & Cosine Similarity & 100.00 & 98.75 & 100.00 \\
 & Contrastive (CRD) & 100.00 & 98.75 & 100.00 \\
 & Decoupled KD & 98.75 & 96.25 & 97.50 \\
 & Logit-MSE & 100.00 & 97.50 & 100.00 \\
 & Feature KD & 76.25 & 83.75 & 68.75 \\
 & Attention Transfer & 75.00 & 82.50 & 71.25 \\
 & Relational KD & 75.00 & 83.75 & 71.25 \\
 & Similarity-Preserving & 75.00 & 81.25 & 70.00 \\
\addlinespace
FedMedian~\cite{yin2018byzantine} (64.6\%) & Vanilla KD & 100.00 & 100.00 & 98.75 \\
 & Self-Distillation & 98.75 & 100.00 & 100.00 \\
 & Cosine Similarity & 100.00 & 98.75 & 100.00 \\
 & Contrastive (CRD) & 100.00 & 98.75 & 98.75 \\
 & Decoupled KD & 98.75 & 97.50 & 96.25 \\
 & Logit-MSE & 100.00 & 100.00 & 98.75 \\
 & Feature KD & 73.75 & 68.75 & 65.00 \\
 & Attention Transfer & 73.75 & 71.25 & 63.75 \\
 & Relational KD & 73.75 & 68.75 & 67.50 \\
 & Similarity-Preserving & 75.00 & 67.50 & 61.25 \\
\addlinespace
FedDyn~\cite{acar2021federated} (58.3\%) & Vanilla KD & 100.00 & 98.75 & 100.00 \\
 & Self-Distillation & 100.00 & 98.75 & 98.75 \\
 & Cosine Similarity & 100.00 & 98.75 & 100.00 \\
 & Contrastive (CRD) & 100.00 & 98.75 & 100.00 \\
 & Decoupled KD & 93.75 & 88.75 & 92.50 \\
 & Logit-MSE & 96.25 & 93.75 & 97.50 \\
 & Feature KD & 76.25 & 65.00 & 50.00 \\
 & Attention Transfer & 76.25 & 63.75 & 50.00 \\
 & Relational KD & 75.00 & 65.00 & 50.00 \\
 & Similarity-Preserving & 76.25 & 66.25 & 50.00 \\
\addlinespace
FedNova~\cite{wang2020tackling} (50.8\%) & Vanilla KD & 100.00 & 97.50 & 100.00 \\
 & Self-Distillation & 98.75 & 98.75 & 98.75 \\
 & Cosine Similarity & 98.75 & 98.75 & 100.00 \\
 & Contrastive (CRD) & 100.00 & 98.75 & 100.00 \\
 & Decoupled KD & 100.00 & 97.50 & 96.25 \\
 & Logit-MSE & 98.75 & 97.50 & 98.75 \\
 & Feature KD & 48.75 & 58.75 & 66.25 \\
 & Attention Transfer & 48.75 & 61.25 & 68.75 \\
 & Relational KD & 47.50 & 58.75 & 63.75 \\
 & Similarity-Preserving & 48.75 & 60.00 & 63.75 \\
\addlinespace
SCAFFOLD~\cite{karimireddy2020scaffold} (36.7\%) & Vanilla KD & 100.00 & 98.75 & 100.00 \\
 & Self-Distillation & 98.75 & 98.75 & 100.00 \\
 & Cosine Similarity & 100.00 & 98.75 & 100.00 \\
 & Contrastive (CRD) & 98.75 & 98.75 & 100.00 \\
 & Decoupled KD & 97.50 & 97.50 & 98.75 \\
 & Logit-MSE & 98.75 & 97.50 & 100.00 \\
 & Feature KD & 27.50 & 33.75 & 30.00 \\
 & Attention Transfer & 26.25 & 32.50 & 30.00 \\
 & Relational KD & 32.50 & 32.50 & 32.50 \\
 & Similarity-Preserving & 32.50 & 35.00 & 31.25 \\
\addlinespace
FedAdagrad~\cite{reddi2020adaptive} (26.2\%) & Vanilla KD & 98.75 & 97.50 & 100.00 \\
 & Self-Distillation & 100.00 & 97.50 & 100.00 \\
 & Cosine Similarity & 100.00 & 100.00 & 100.00 \\
 & Contrastive (CRD) & 100.00 & 98.75 & 100.00 \\
 & Decoupled KD & 100.00 & 100.00 & 97.50 \\
 & Logit-MSE & 98.75 & 98.75 & 98.75 \\
 & Feature KD & 32.50 & 32.50 & 32.50 \\
 & Attention Transfer & 32.50 & 31.25 & 28.75 \\
 & Relational KD & 33.75 & 30.00 & 31.25 \\
 & Similarity-Preserving & 32.50 & 31.25 & 33.75 \\
\addlinespace
FedAdam~\cite{reddi2020adaptive} (25.0\%) & Vanilla KD & 100.00 & 98.75 & 98.75 \\
 & Self-Distillation & 100.00 & 98.75 & 100.00 \\
 & Cosine Similarity & 100.00 & 98.75 & 100.00 \\
 & Contrastive (CRD) & 100.00 & 100.00 & 100.00 \\
 & Decoupled KD & 95.00 & 97.50 & 96.25 \\
 & Logit-MSE & 100.00 & 97.50 & 98.75 \\
 & Feature KD & 25.00 & 26.25 & 28.75 \\
 & Attention Transfer & 27.50 & 30.00 & 31.25 \\
 & Relational KD & 32.50 & 32.50 & 28.75 \\
 & Similarity-Preserving & 27.50 & 25.00 & 32.50 \\
\addlinespace
FedAvgM~\cite{hsu2019measuring} (25.0\%) & Vanilla KD & 98.75 & 96.25 & 97.50 \\
 & Self-Distillation & 100.00 & 97.50 & 100.00 \\
 & Cosine Similarity & 100.00 & 98.75 & 100.00 \\
 & Contrastive (CRD) & 100.00 & 100.00 & 100.00 \\
 & Decoupled KD & 96.25 & 93.75 & 91.25 \\
 & Logit-MSE & 97.50 & 97.50 & 97.50 \\
 & Feature KD & 30.00 & 25.00 & 26.25 \\
 & Attention Transfer & 25.00 & 25.00 & 25.00 \\
 & Relational KD & 30.00 & 26.25 & 26.25 \\
 & Similarity-Preserving & 25.00 & 25.00 & 25.00 \\
\addlinespace
FedYogi~\cite{reddi2020adaptive} (25.0\%) & Vanilla KD & 98.75 & 98.75 & 100.00 \\
 & Self-Distillation & 100.00 & 98.75 & 100.00 \\
 & Cosine Similarity & 100.00 & 100.00 & 100.00 \\
 & Contrastive (CRD) & 100.00 & 98.75 & 100.00 \\
 & Decoupled KD & 97.50 & 97.50 & 100.00 \\
 & Logit-MSE & 98.75 & 97.50 & 100.00 \\
 & Feature KD & 26.25 & 30.00 & 26.25 \\
 & Attention Transfer & 28.75 & 28.75 & 26.25 \\
 & Relational KD & 26.25 & 33.75 & 32.50 \\
 & Similarity-Preserving & 31.25 & 27.50 & 27.50 \\
\end{longtable}
\par}

\clearpage
% =====================================================================
{\footnotesize
\setlength{\tabcolsep}{8pt}
\renewcommand{\arraystretch}{1.05}
\begin{longtable}{@{}ll ccc@{}}
\caption{Per-seed combined-pipeline test instance accuracy (\%) on ModelNet40, for every FL teacher (with its standalone accuracy) crossed with every KD objective. Column headers are the random seeds. As on the clinical task, the six hard-label objectives recover the student regardless of teacher quality, whereas the four pure-transfer objectives (Feature KD, Attention Transfer, Relational KD, Similarity-Preserving) track the teacher and fall toward the chance level (about $4\%$) when it collapses; the per-seed values also expose the single seeds in which a few hard-label cells under the near-degenerate FedAdam teacher failed to converge. This is the multi-seed counterpart of the focused-round grid in Table~3 of the main paper and of the per-cell summary in \Cref{tab:combined_mean_mn40,tab:combined_std_mn40}. For reference, the ten KD objectives are Vanilla KD~\cite{hinton2015distilling}, Self-Distillation~\cite{furlanello2018born}, Cosine Similarity~\cite{luo2018cosine}, Contrastive Representation Distillation (CRD)~\cite{tian2019contrastive}, Decoupled KD~\cite{zhao2022decoupled}, Logit-MSE~\cite{ba2014deep}, Feature KD~\cite{romero2014fitnets}, Attention Transfer~\cite{zagoruyko2016paying}, Relational KD~\cite{park2019relational}, and Similarity-Preserving distillation~\cite{tung2019similarity}.}
\label{tab:seed_combined_mn40}\\
\toprule
FL Teacher (standalone) & KD Objective & $7$~$\uparrow$ & $42$~$\uparrow$ & $123$~$\uparrow$ \\
\midrule
\endfirsthead
\multicolumn{5}{@{}l}{\itshape Table~\ref{tab:seed_combined_mn40}, continued} \\
\toprule
FL Teacher (standalone) & KD Objective & $7$~$\uparrow$ & $42$~$\uparrow$ & $123$~$\uparrow$ \\
\midrule
\endhead
\midrule
\multicolumn{5}{r@{}}{\itshape continued on next page} \\
\endfoot
\bottomrule
\endlastfoot
FedNova~\cite{wang2020tackling} (76.3\%) & Vanilla KD & 92.46 & 92.38 & 92.30 \\
 & Self-Distillation & 92.30 & 92.42 & 92.59 \\
 & Cosine Similarity & 92.14 & 92.46 & 92.38 \\
 & Contrastive (CRD) & 92.26 & 92.14 & 92.22 \\
 & Decoupled KD & 87.12 & 86.67 & 87.76 \\
 & Logit-MSE & 91.29 & 91.25 & 91.49 \\
 & Feature KD & 77.07 & 77.80 & 78.65 \\
 & Attention Transfer & 77.07 & 78.24 & 78.00 \\
 & Relational KD & 77.07 & 77.67 & 77.92 \\
 & Similarity-Preserving & 77.59 & 77.51 & 78.61 \\
\addlinespace
FedProx~\cite{li2020federated} (71.0\%) & Vanilla KD & 92.22 & 91.73 & 92.18 \\
 & Self-Distillation & 92.30 & 92.02 & 92.22 \\
 & Cosine Similarity & 92.14 & 92.30 & 92.42 \\
 & Contrastive (CRD) & 92.14 & 92.38 & 92.18 \\
 & Decoupled KD & 86.75 & 85.70 & 85.94 \\
 & Logit-MSE & 91.77 & 91.69 & 91.61 \\
 & Feature KD & 70.26 & 70.75 & 69.37 \\
 & Attention Transfer & 69.37 & 70.54 & 69.04 \\
 & Relational KD & 70.02 & 70.22 & 69.12 \\
 & Similarity-Preserving & 69.81 & 70.30 & 69.61 \\
\addlinespace
FedDyn~\cite{acar2021federated} (66.0\%) & Vanilla KD & 92.22 & 91.53 & 91.65 \\
 & Self-Distillation & 92.10 & 91.69 & 91.61 \\
 & Cosine Similarity & 92.38 & 91.90 & 92.26 \\
 & Contrastive (CRD) & 92.54 & 92.22 & 92.46 \\
 & Decoupled KD & 84.40 & 80.47 & 81.16 \\
 & Logit-MSE & 91.05 & 89.26 & 88.90 \\
 & Feature KD & 67.46 & 64.95 & 69.17 \\
 & Attention Transfer & 67.26 & 64.63 & 69.33 \\
 & Relational KD & 67.02 & 63.70 & 69.89 \\
 & Similarity-Preserving & 67.34 & 65.07 & 70.06 \\
\addlinespace
SCAFFOLD~\cite{karimireddy2020scaffold} (64.3\%) & Vanilla KD & 92.42 & 92.38 & 92.59 \\
 & Self-Distillation & 92.42 & 92.46 & 92.50 \\
 & Cosine Similarity & 92.06 & 92.06 & 92.30 \\
 & Contrastive (CRD) & 91.86 & 91.90 & 92.10 \\
 & Decoupled KD & 85.33 & 85.66 & 81.89 \\
 & Logit-MSE & 92.18 & 91.61 & 90.92 \\
 & Feature KD & 59.04 & 68.11 & 62.88 \\
 & Attention Transfer & 59.12 & 68.23 & 62.60 \\
 & Relational KD & 59.68 & 68.15 & 62.76 \\
 & Similarity-Preserving & 59.36 & 68.07 & 62.64 \\
\addlinespace
Ditto~\cite{li2021ditto} (61.4\%) & Vanilla KD & 92.26 & 92.54 & 92.30 \\
 & Self-Distillation & 92.71 & 92.63 & 92.46 \\
 & Cosine Similarity & 91.90 & 92.50 & 92.10 \\
 & Contrastive (CRD) & 92.26 & 92.14 & 92.06 \\
 & Decoupled KD & 90.68 & 91.61 & 91.00 \\
 & Logit-MSE & 92.46 & 92.75 & 92.54 \\
 & Feature KD & 50.12 & 56.93 & 52.67 \\
 & Attention Transfer & 49.92 & 57.05 & 52.88 \\
 & Relational KD & 49.84 & 57.05 & 52.59 \\
 & Similarity-Preserving & 50.00 & 56.60 & 52.76 \\
\addlinespace
FedAvg~\cite{mcmahan2017communication} (58.5\%) & Vanilla KD & 92.79 & 92.38 & 92.46 \\
 & Self-Distillation & 92.79 & 92.46 & 92.71 \\
 & Cosine Similarity & 92.38 & 92.38 & 92.46 \\
 & Contrastive (CRD) & 92.30 & 92.38 & 92.30 \\
 & Decoupled KD & 91.69 & 91.25 & 91.09 \\
 & Logit-MSE & 92.50 & 92.59 & 92.71 \\
 & Feature KD & 50.69 & 51.58 & 54.01 \\
 & Attention Transfer & 51.50 & 51.26 & 53.40 \\
 & Relational KD & 50.89 & 51.34 & 53.44 \\
 & Similarity-Preserving & 50.24 & 51.42 & 53.44 \\
\addlinespace
FedBN~\cite{li2021fedbn} (55.3\%) & Vanilla KD & 92.63 & 92.79 & 92.46 \\
 & Self-Distillation & 92.34 & 92.46 & 92.54 \\
 & Cosine Similarity & 92.10 & 92.18 & 92.06 \\
 & Contrastive (CRD) & 92.06 & 92.18 & 92.38 \\
 & Decoupled KD & 92.06 & 90.92 & 90.92 \\
 & Logit-MSE & 92.67 & 92.75 & 92.54 \\
 & Feature KD & 50.41 & 48.06 & 48.78 \\
 & Attention Transfer & 49.92 & 47.73 & 48.58 \\
 & Relational KD & 49.96 & 47.65 & 48.26 \\
 & Similarity-Preserving & 50.08 & 47.81 & 48.58 \\
\addlinespace
MOON~\cite{li2021model} (45.8\%) & Vanilla KD & 92.50 & 92.42 & 92.54 \\
 & Self-Distillation & 92.38 & 92.42 & 92.50 \\
 & Cosine Similarity & 92.30 & 91.94 & 91.90 \\
 & Contrastive (CRD) & 91.98 & 91.98 & 92.22 \\
 & Decoupled KD & 90.68 & 91.33 & 90.92 \\
 & Logit-MSE & 92.87 & 92.50 & 92.54 \\
 & Feature KD & 38.90 & 39.06 & 51.46 \\
 & Attention Transfer & 38.49 & 38.90 & 51.70 \\
 & Relational KD & 39.10 & 40.48 & 51.46 \\
 & Similarity-Preserving & 38.57 & 39.14 & 51.13 \\
\addlinespace
FedMedian~\cite{yin2018byzantine} (39.1\%) & Vanilla KD & 92.42 & 92.42 & 92.75 \\
 & Self-Distillation & 92.46 & 92.54 & 92.83 \\
 & Cosine Similarity & 92.30 & 92.26 & 92.26 \\
 & Contrastive (CRD) & 91.94 & 92.22 & 92.22 \\
 & Decoupled KD & 92.06 & 92.14 & 92.26 \\
 & Logit-MSE & 92.71 & 92.75 & 92.79 \\
 & Feature KD & 39.71 & 33.43 & 29.13 \\
 & Attention Transfer & 40.11 & 32.98 & 28.20 \\
 & Relational KD & 40.56 & 33.55 & 30.15 \\
 & Similarity-Preserving & 39.79 & 33.23 & 28.44 \\
\addlinespace
FedAdagrad~\cite{reddi2020adaptive} (6.3\%) & Vanilla KD & 92.42 & 92.63 & 91.73 \\
 & Self-Distillation & 92.46 & 92.71 & 91.90 \\
 & Cosine Similarity & 92.18 & 92.38 & 92.10 \\
 & Contrastive (CRD) & 92.18 & 91.94 & 91.82 \\
 & Decoupled KD & 91.37 & 92.26 & 81.44 \\
 & Logit-MSE & 92.46 & 92.75 & 91.94 \\
 & Feature KD & 7.33 & 4.05 & 5.19 \\
 & Attention Transfer & 7.25 & 4.86 & 5.15 \\
 & Relational KD & 7.86 & 5.75 & 5.23 \\
 & Similarity-Preserving & 7.66 & 5.67 & 5.15 \\
\addlinespace
FedAdam~\cite{reddi2020adaptive} (4.7\%) & Vanilla KD & 92.26 & 75.20 & 92.38 \\
 & Self-Distillation & 92.75 & 77.31 & 92.54 \\
 & Cosine Similarity & 92.18 & 92.10 & 92.18 \\
 & Contrastive (CRD) & 92.06 & 91.90 & 92.02 \\
 & Decoupled KD & 91.53 & 39.75 & 88.29 \\
 & Logit-MSE & 92.91 & 27.67 & 92.26 \\
 & Feature KD & 4.05 & 4.46 & 9.52 \\
 & Attention Transfer & 4.09 & 4.46 & 7.82 \\
 & Relational KD & 4.70 & 4.50 & 6.97 \\
 & Similarity-Preserving & 4.09 & 4.50 & 9.40 \\
\addlinespace
FedAvgM~\cite{hsu2019measuring} (4.0\%) & Vanilla KD & 92.22 & 92.26 & 92.26 \\
 & Self-Distillation & 92.18 & 92.26 & 92.50 \\
 & Cosine Similarity & 4.05 & 4.05 & 4.05 \\
 & Contrastive (CRD) & 4.05 & 4.05 & 4.05 \\
 & Decoupled KD & 92.14 & 91.69 & 91.98 \\
 & Logit-MSE & 4.05 & 4.05 & 4.05 \\
 & Feature KD & 4.05 & 4.05 & 4.05 \\
 & Attention Transfer & 4.05 & 4.05 & 4.05 \\
 & Relational KD & 4.05 & 4.05 & 4.05 \\
 & Similarity-Preserving & 4.05 & 4.05 & 4.05 \\
\addlinespace
FedYogi~\cite{reddi2020adaptive} (4.0\%) & Vanilla KD & 92.46 & 92.63 & 92.63 \\
 & Self-Distillation & 92.54 & 92.75 & 92.63 \\
 & Cosine Similarity & 92.38 & 92.42 & 91.90 \\
 & Contrastive (CRD) & 91.98 & 92.10 & 91.69 \\
 & Decoupled KD & 91.61 & 92.06 & 91.77 \\
 & Logit-MSE & 92.75 & 92.87 & 92.91 \\
 & Feature KD & 4.09 & 4.05 & 4.05 \\
 & Attention Transfer & 4.13 & 4.05 & 4.05 \\
 & Relational KD & 4.05 & 4.05 & 4.13 \\
 & Similarity-Preserving & 4.05 & 4.05 & 4.05 \\
\end{longtable}
\par}

\clearpage
% =====================================================================
\section{Generalization to Additional Datasets}
\label{app:moredata}

The dataset-agnostic loader described in \Cref{app:impl} lets the same standalone protocols run on point-cloud sources beyond the two datasets of the main multi-seed benchmark. To show that the framework and its conclusions are not specific to ModelNet40 and the clinical task, we report standalone federated-learning and knowledge-distillation results on two further datasets: ModelNet10~\cite{wu20153d}, a ten-class subset of the ModelNet shapes, and OmniObject3D~\cite{wu2023omniobject3d}, a large-vocabulary collection of real-scanned objects. These auxiliary runs reuse the identical training configuration, the same compact student, and the same deterministic non-IID label-skew partition; they lie outside the $504$-run count of the main paper and are provided only to illustrate the framework's reach. None of these results appears elsewhere in this supplement. Across both datasets the same two patterns recur that the main paper documents on ModelNet40 and the clinical task: most of the server-side federated optimizers collapse toward the chance level while the drift-correction methods learn, and distillation transfers a strong centralized teacher into the compact student at a small accuracy cost.

\begin{figure}[tb]
\centering
\begin{subfigure}{0.49\textwidth}\centering
\includegraphics[width=\linewidth]{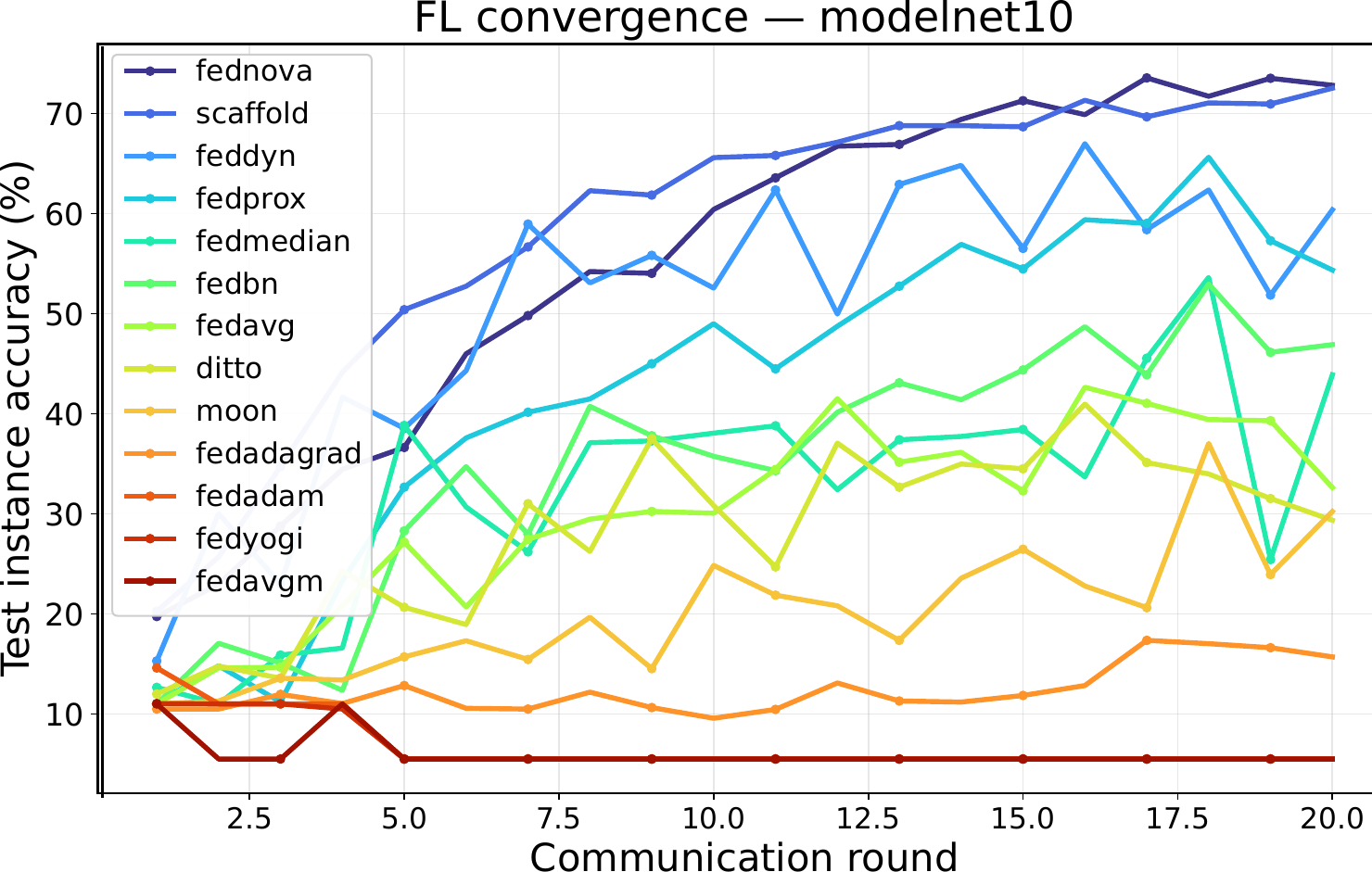}
\caption{ModelNet10}
\label{fig:fl_conv_mn10}
\end{subfigure}\hfill
\begin{subfigure}{0.49\textwidth}\centering
\includegraphics[width=\linewidth]{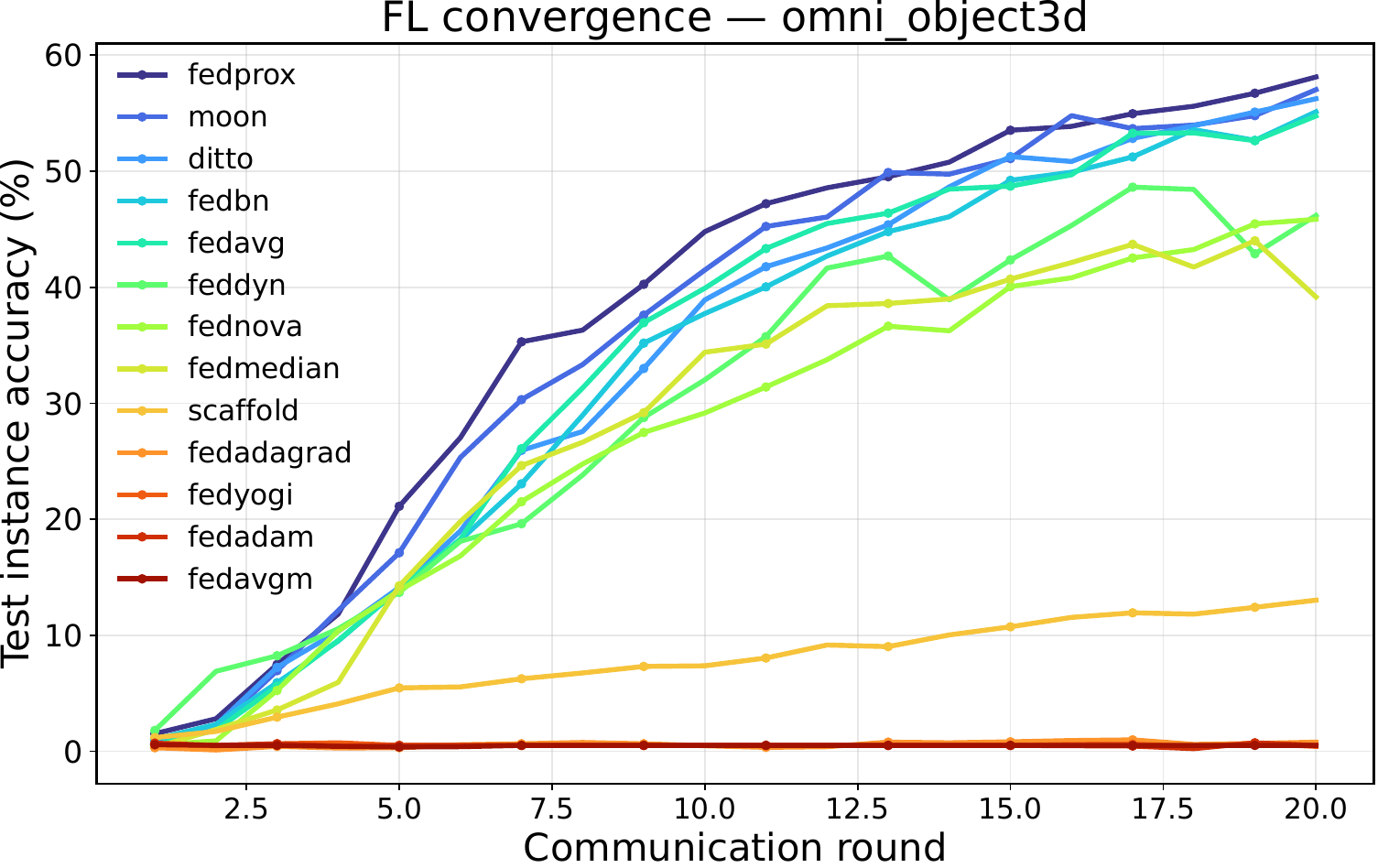}
\caption{OmniObject3D}
\label{fig:fl_conv_omni}
\end{subfigure}
\caption{\textbf{Federated convergence on two additional datasets.}
Per-round test instance accuracy of all thirteen FL algorithms under the non-IID label-skew partition. \textbf{(a)} On ModelNet10, FedNova leads near $74\%$, a broad middle band follows, and the four server-side optimizers trail at the bottom, with FedAvgM and FedYogi at the $10\%$ chance level. \textbf{(b)} On the large-vocabulary OmniObject3D, FedProx leads near $58\%$ and all four server-side optimizers collapse to near the chance level. The trends reproduce those of \Cref{fig:fl_convergence} on the two main datasets.}
\label{fig:fl_convergence_extra}
\end{figure}

\begin{figure}[tb]
\centering
\begin{subfigure}{0.49\textwidth}\centering
\includegraphics[width=\linewidth]{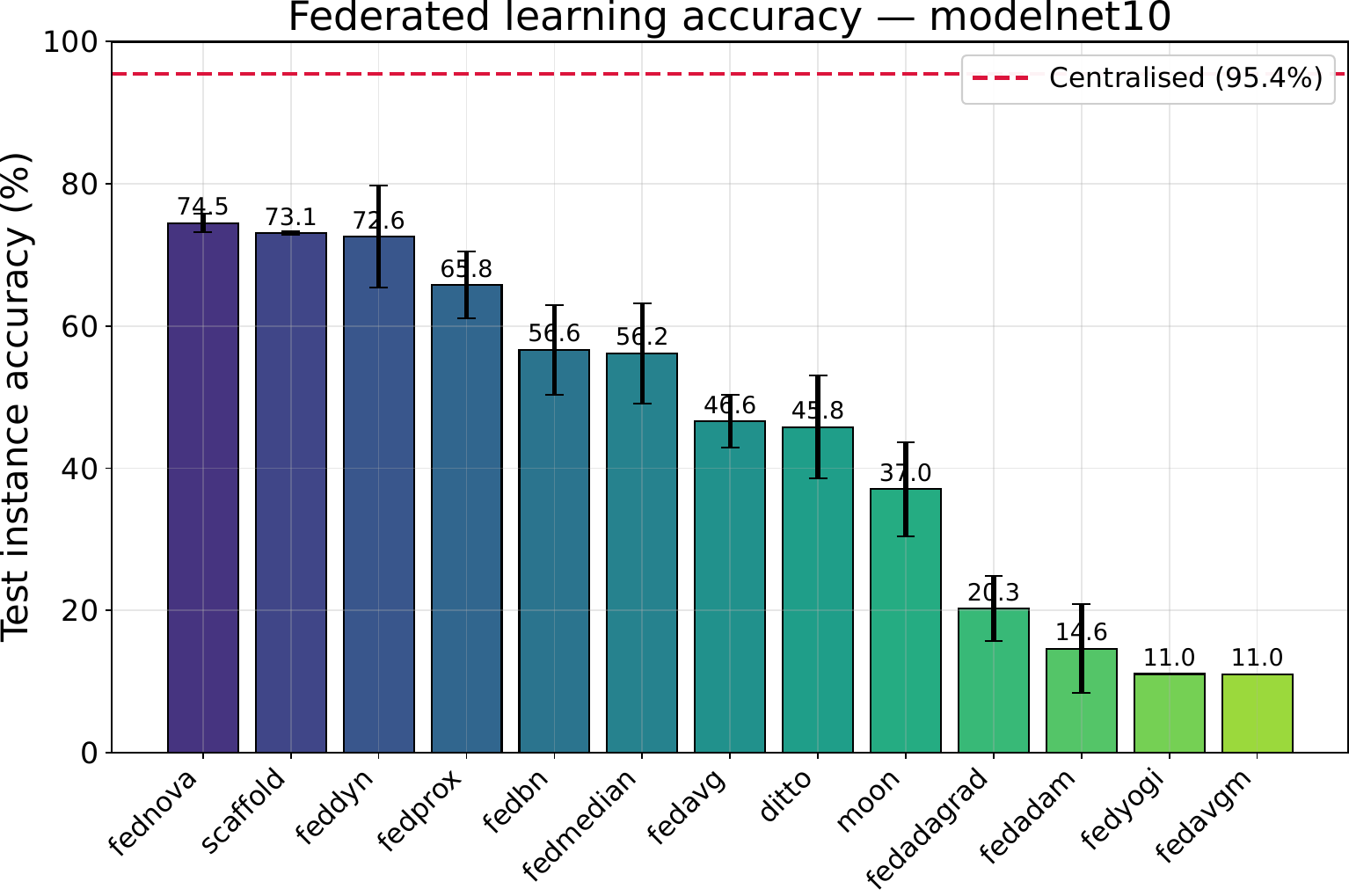}
\caption{ModelNet10}
\label{fig:fl_acc_mn10}
\end{subfigure}\hfill
\begin{subfigure}{0.49\textwidth}\centering
\includegraphics[width=\linewidth]{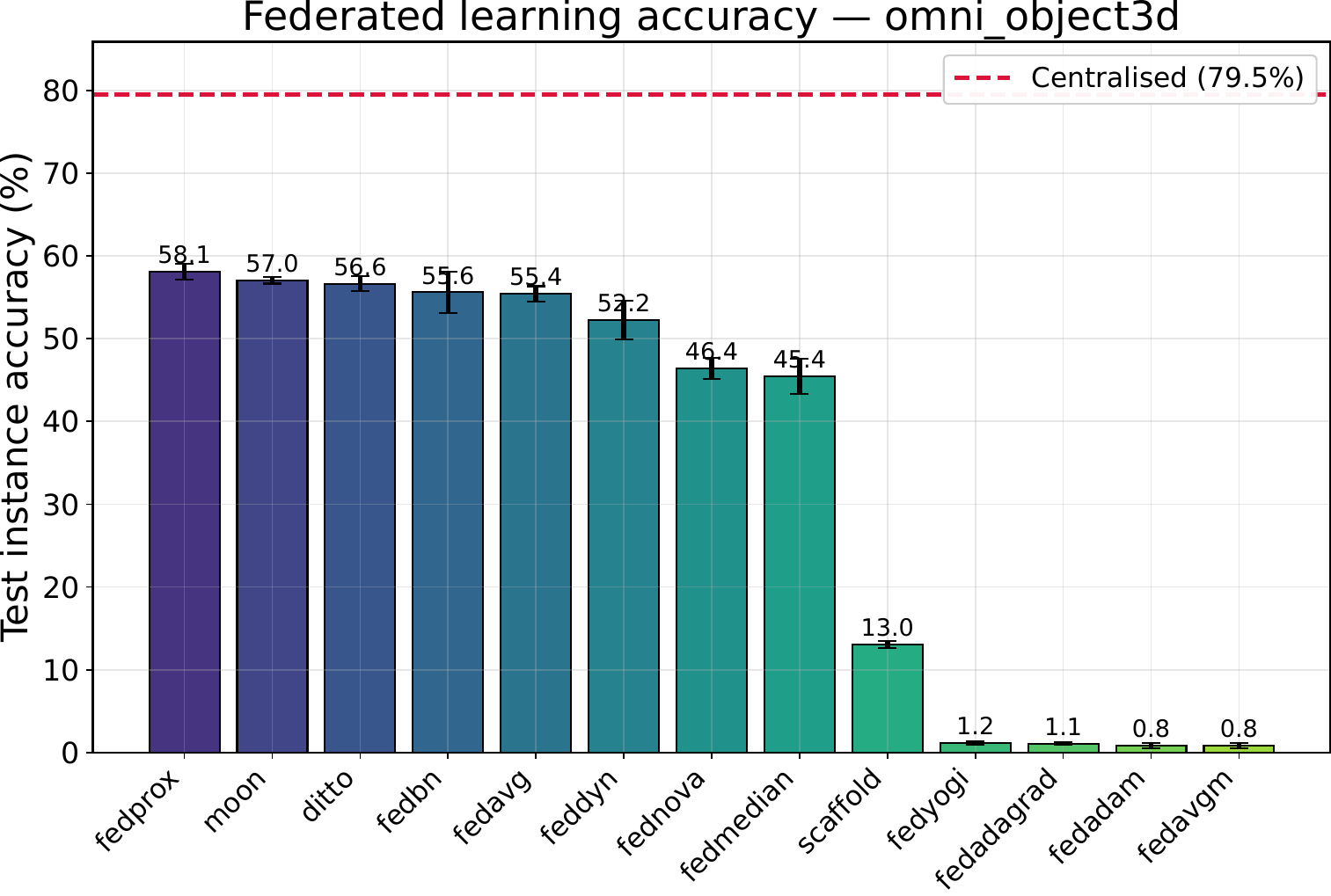}
\caption{OmniObject3D}
\label{fig:fl_acc_omni}
\end{subfigure}
\caption{\textbf{Federated strategy accuracy on two additional datasets (best per run, mean$\pm$std over three seeds).}
Ranked bar-chart view of the convergence endpoints in \Cref{fig:fl_convergence_extra}; the dashed line marks the centralized reference. \textbf{(a)} On ModelNet10 the order runs FedNova ($74.5\%$), SCAFFOLD, FedDyn, FedProx, then a middle band, down to the four server-side optimizers ($11$ to $20\%$). \textbf{(b)} On OmniObject3D the order runs FedProx ($58.1\%$), MOON, Ditto, then a middle band, down to SCAFFOLD ($13\%$) and the four server-side optimizers near zero.}
\label{fig:fl_accuracy_extra}
\end{figure}

\smallskip\noindent\textbf{Federated convergence.}
\Cref{fig:fl_convergence_extra} plots the per-round federated accuracy on both datasets and \Cref{fig:fl_accuracy_extra} the corresponding best-accuracy ranking; the pattern matches \Cref{fig:fl_convergence}. A leading group of drift-correction and robust-aggregation methods climbs steadily over the twenty rounds, whereas the server-side optimizers lag far behind: FedAvgM and FedYogi sit at the chance level on both datasets, FedAdam barely rises above it, and FedAdagrad collapses on OmniObject3D while reaching only about $20\%$ on ModelNet10. On ModelNet10 the strongest teacher (FedNova) reaches about $74\%$ instance accuracy against a $95.4\%$ centralized reference, and on OmniObject3D the strongest (FedProx) reaches about $58\%$ against a $79.5\%$ centralized reference, so the federated accuracy gap widens with task difficulty, exactly as on the two main datasets.

\begin{figure}[tb]
\centering
\begin{subfigure}{0.49\textwidth}\centering
\includegraphics[width=\linewidth]{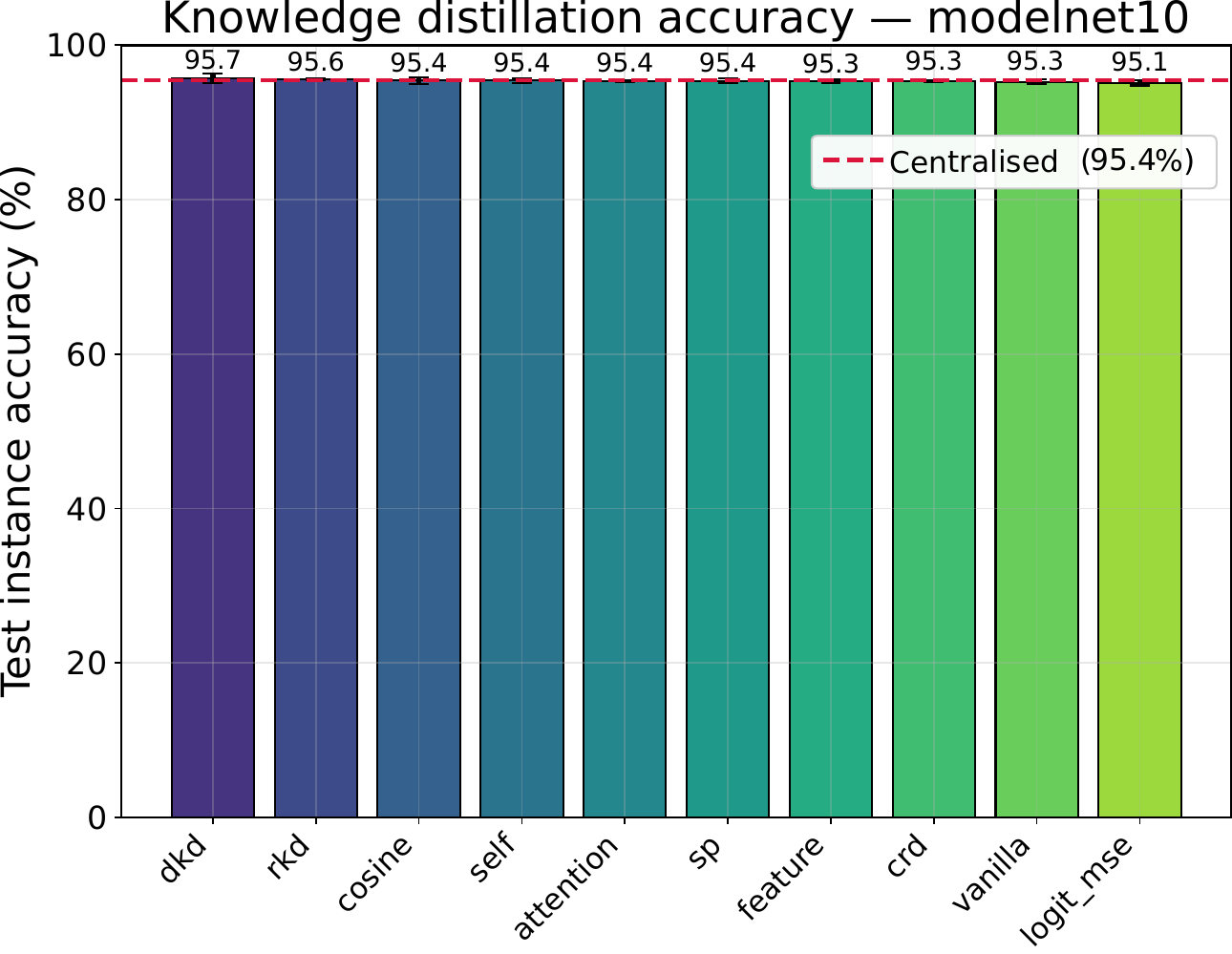}
\caption{ModelNet10}
\label{fig:kd_acc_mn10}
\end{subfigure}\hfill
\begin{subfigure}{0.49\textwidth}\centering
\includegraphics[width=\linewidth]{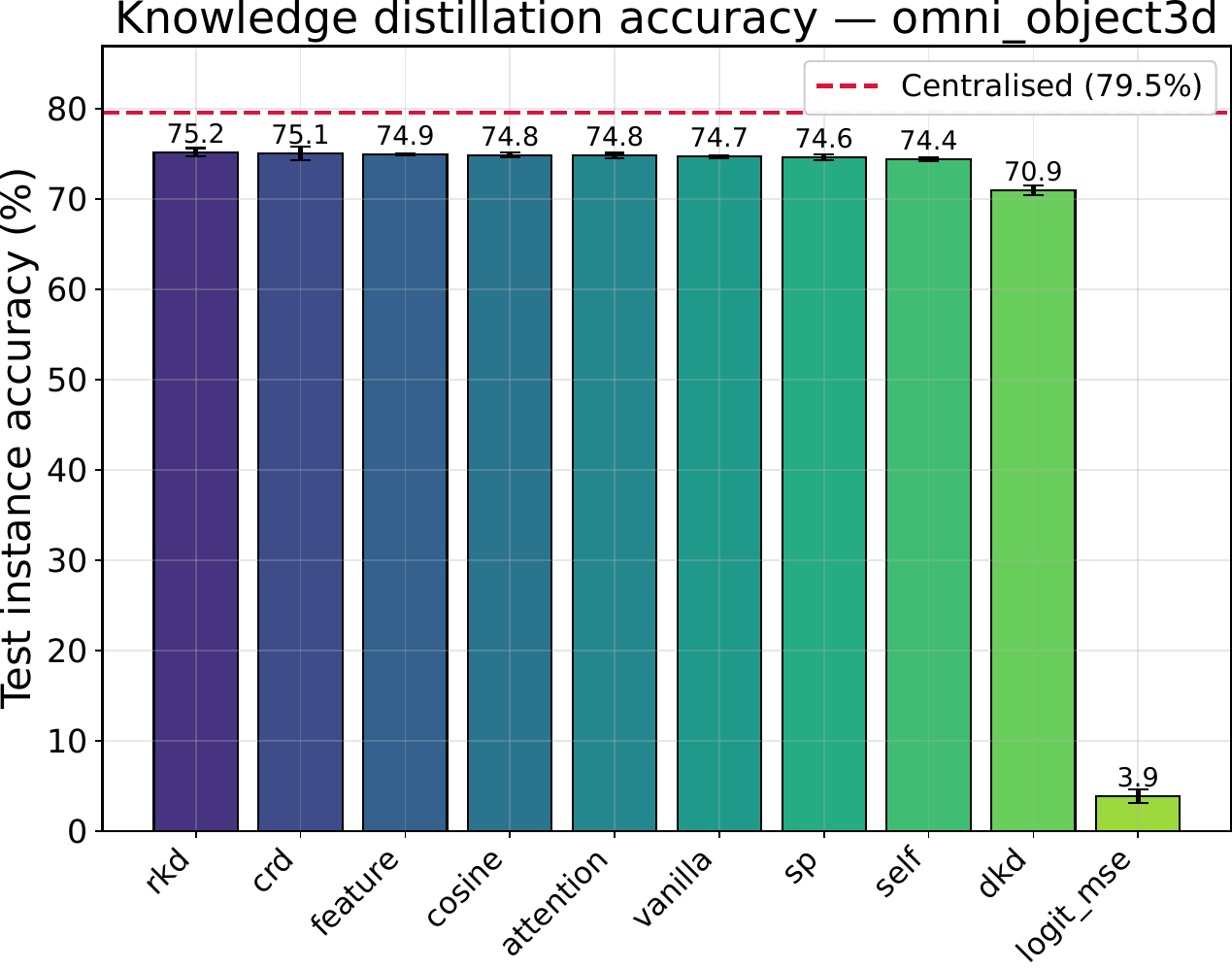}
\caption{OmniObject3D}
\label{fig:kd_acc_omni}
\end{subfigure}
\caption{\textbf{Standalone knowledge distillation on two additional datasets.}
Test instance accuracy of the ten KD objectives distilling the centralized teacher into the compact student; the dashed line marks the centralized reference. \textbf{(a)} On ModelNet10 all ten objectives reach the $95\%$ ceiling. \textbf{(b)} On OmniObject3D most objectives land near $75\%$, Decoupled KD trails at $70.9\%$, and Logit-MSE collapses to $3.9\%$, the same fragility it shows on ModelNet40.}
\label{fig:kd_accuracy_extra}
\end{figure}

\smallskip\noindent\textbf{Standalone distillation.}
\Cref{fig:kd_accuracy_extra} reports the ten distillation objectives distilling the centralized teacher into the compact student. On ModelNet10 every objective recovers the teacher almost exactly, landing between $95.1\%$ and $95.7\%$ against the $95.4\%$ centralized accuracy. On the much harder OmniObject3D, eight of the ten objectives cluster between $74.4\%$ and $75.2\%$, within roughly five points of the $79.5\%$ teacher, while Decoupled KD trails at $70.9\%$ and Logit-MSE collapses to $3.9\%$. The Logit-MSE failure echoes its weak standing on ModelNet40 (Section~4.3 of the main paper) and confirms that the raw logit-matching objective is the least robust of the ten as the label space grows.

\smallskip\noindent\textbf{Compression.}
Because the same compact student is distilled on every dataset, the compression is dataset-independent: the student keeps its ${\approx}1.4$\,MB footprint against the ${\approx}5.6$\,MB teacher and runs at roughly half the teacher's inference latency, just as on ModelNet40 and the clinical task, and on the easier ModelNet10 it gives up almost no accuracy for that reduction.

% Commented out for combining appendix with main.tex
\iffalse
\clearpage

% ---- References ----
\bibliographystyle{splncs04}
\bibliography{ref}

\end{document}
\fi

\end{document}